\documentclass[pdftex, useAMS,usenatbib]{mn2e}

\usepackage{natbib}

\usepackage{graphicx}
\usepackage[space]{grffile}
\usepackage{latexsym}
\usepackage{natbib}
\usepackage{url}
\usepackage{color}
\usepackage[utf8]{inputenc}
\usepackage{longtable,lscape}
\usepackage{fancyref}
\usepackage{times}
\usepackage{amssymb}

\usepackage{floatflt}

\def\kms{$\mbox{km s}^{-1}$}

\newcommand{\msun}{{\rm M}_\odot}
\newcommand{\vrotsigma}{V_{\rm rot} \sin{i} \ /\sigma_{\rm 1D}}

\def\aap{A\&A}
\def\aapr{A\&A Rev.}
\def\apj{ApJS}
\def\apjl{ApJL}
\def\apjs{ApJ Supp.}

\def\mnras{MNRAS}
\def\aj{AJ}
\def\nat{Nature}

\def\araa{ARA\&A}
\def\pasp{PASP}
\def\pasj{PASJ}

\def\na{New Astron.}
\def\memsai{Mem. Societa Astronomica Italiana}

\usepackage{txfonts}

\title[Multiple populations in GCs: the distinct kinematic
imprints of different formation scenarios]{Multiple populations in globular clusters: the distinct kinematic
imprints of different formation scenarios}
\author[V. H\'{e}nault-Brunet et al.]{V. H\'{e}nault-Brunet\thanks{E-mail: v.henault-brunet@surrey.ac.uk}, M. Gieles, O. Agertz, J. I. Read\\
Department of Physics, Faculty of Engineering and Physical Sciences, University of Surrey, Guildford, GU2 7XH, UK}

\begin{document}

\date{Accepted 2015 March 25.  Received 2015 March 14; in original form 2014 December 18}

\pagerange{\pageref{firstpage}--36} \pubyear{2015}
\maketitle

\label{firstpage}

\begin{abstract}
Several scenarios have been proposed to explain the presence of multiple stellar populations in globular clusters. Many of them invoke multiple generations of stars to explain the observed chemical abundance anomalies, but it has also been suggested that self-enrichment could occur via accretion of ejecta from massive stars onto the circumstellar disc of low-mass pre-main sequence stars. These scenarios imply different initial conditions for the kinematics of the various stellar populations. Given some net angular momentum initially, models for which a second generation forms from gas that collects in a cooling flow into the core of the cluster predict an initially larger rotational amplitude for the polluted stars compared to the pristine stars. This is opposite to what is expected from the accretion model, where the polluted stars are the ones crossing the core and are on preferentially radial (low-angular momentum) orbits, such that their rotational amplitude is lower. Here we present the results of a suite of $N$-body simulations with initial conditions chosen to capture the distinct kinematic properties of these pollution scenarios. We show that initial differences in the kinematics of polluted and pristine stars can survive to the present epoch in the outer parts of a large fraction of Galactic globular clusters. The differential rotation of pristine and polluted stars is identified as a unique kinematic signature that could allow us to distinguish between various scenarios, while other kinematic imprints are generally very similar from one scenario to the other.

\end{abstract}


\section{Introduction}

Once considered as quintessential simple stellar populations (i.e. no dispersion in age or metal content), globular clusters (GCs) now appear to host multiple stellar populations. This has been revealed by ubiquitous star-to-star abundance variations of light elements such as Na, O, and Al \citep[e.g.][]{GCB_2012}, indicating that a significant fraction of cluster stars \citep[$\sim30-70\%$; e.g.][]{Carretta2009} must have been polluted by material processed in high-temperature stellar interiors via proton capture reactions. It is also apparent from the multiple or extended  main sequences, subgiant and red-giant branches seen in the colour-magnitude diagrams of a growing number of Galactic GCs \citep[e.g.][]{Piotto2012}, which have been attributed to different levels of helium enrichment \citep[e.g.][]{Dantona2002, Norris2004, Pasquini_2011} and light element abundance variations \citep[e.g.][]{Marino2008, Sbordone2011} among the cluster stars.


Several models have been proposed in an attempt to explain the observed chemical abundance anomalies, with different types of stars suggested as the source of helium-enriched material. The yields from these potential enrichment sources must also explain the observed chemical abundance patterns (e.g. the Na-O anticorrelation) without producing a spread in iron abundances \citep[such a spread is absent in almost all but the most massive GCs; e.g.][]{Carretta2009Fe}. Asymptotic Giant Branch (AGB) stars \citep[e.g.][]{Ventura_2001, DErcole_2008}, fast rotating massive stars \citep[``spin-stars"; e.g.][]{Decressin_2007}, interacting massive binaries \citep{deMink_2009} and super-massive stars with $M\sim10^4 \ \msun$ \citep{denissenkov2014} have all been considered as possible polluters.

Obtaining the right chemistry for the polluting material is one aspect of the problem, but finding a way to get this material into stars is another equally important and challenging issue. Many of the models put forward invoke the formation of a second generation of stars from the processed material released by a first generation. For this reason, the terms multiple {\it generations} and multiple {\it populations} are often interchangeably used in the literature to refer to the presence of polluted and pristine stars, but note that the notion of multiple star-formation events within a single cluster is an assumption of these models for which there is no direct observational evidence.

In a scenario that has received significant attention, \citet{DErcole_2008} explored the possibility that a second generation is formed from the gas ejected by AGB stars. They used 1D hydrodynamical simulations to show that the low-velocity ejecta of AGB stars can be retained by the cluster and form a cooling flow that rapidly collects in the innermost regions. The new generation then forms a more concentrated stellar subsystem, in keeping with observations of a more centrally concentrated polluted (i.e. Na-rich, O-poor) population in a number of Galactic GCs \citep[e.g.][]{Lardo_2011}. This scenario was further studied by different authors \citep[e.g.][]{ConroySpergel2011, Bekki2011}, many of which stressed the importance of accreting pristine interstellar gas onto the cluster for it to work. In particular, it was suggested that the total mass of accreted pristine gas must be comparable to that of AGB ejecta in order to reproduce the observed Na-O anticorrelation \citep[e.g.][]{DErcole2010, Ventura2013}. How such pristine interstellar material can remain free of pollution by supernova ejecta from the first generation (as required by the lack of an iron abundance spread) however remains unclear. In addition to the above model based on AGB stars, the main scenario invoking spin-stars as the source of pollution also works under the assumption that the slow ejecta from these stars form the basis of a new stellar generation \citep[e.g.][]{Decressin_2007}. \citet{Charbonnel2014} even postulated that, within this framework, all low-mass stars present in globular clusters today could in fact be second-generation stars.

The models that require multiple star-formation events come with many potential shortcomings \citep[for an overview see][]{Renzini_2008, Bastian_2013, cabrera2015}. For example, they generally suffer from an ``internal mass budget" problem, as the amount of processed material returned by either AGB stars or spin-stars is insufficient to account for the large observed fractions of polluted stars in GCs. To circumvent this issue, these models either invoke a non-standard initial mass function for the first or second generation \citep[e.g.][]{Decressin_2007} and a 100\% star-formation efficiency for the second-generation stars \citep[e.g.][]{DAntona2013}, or imply that GCs were initially $\sim10-100$  times more massive than observed at the present time and that a large fraction of first-generation (i.e. pristine) stars have since been lost \citep[e.g.][]{DErcole_2008, Schaerer2011, Bekki2011}. The latter possibility appears to be in conflict with the ``external mass budget" constraints imposed by observations of GCs and field stars in nearby dwarf galaxies (Fornax, WLM, IKN), which imply that GCs in these systems could not have been more than about five times more massive initially \citep{Larsen2012, Larsen2014}. These models are also at odds with observations of young massive clusters forming at the present epoch, for which there is currently no conclusive evidence for age spreads, ongoing star formation, or reservoirs of gas and dust from which a new generation could form \citep[e.g.][]{Kudrya2012, BastianSilvaVilla2013, BastianCabrera2013, Cabrera2014, BastianStrader2014, BastianHollyhead2014}.

To address these problems, \citet{Bastian_2013} proposed a new scenario in which processed material shed from interacting massive binaries during the first $\sim$10~Myr of the cluster's life is accreted onto the circumstellar discs of low-mass stars (still on the pre-main sequence), ultimately polluting those stars. It must be said that the suggestion that pollution could originate from the accretion of ejecta from higher mass stars was first made by \citet{Dantona1983}, who at the time considered that the observed abundance patterns were reflecting surface contamination. More recent evidence however indicates that abundance anomalies are present throughout the star, as evolved red giant branch stars (mixed through convection) show the same Na-O anti-correlations as main-sequence stars \citep{Cohen2002}. The variation proposed by \citet{Bastian_2013} addresses this by polluting stars when they are on the pre-main sequence and convective, and it also maximises the efficiency of accretion by invoking gas capture by discs. 

In the so-called ``early disc accretion" scenario, the polluted stars originate from the same generation as the pristine stars, with no need for multiple star-formation events. Only the stars that pass through the cluster core to sweep up gas are polluted, as this is where high-mass stars and their ejecta are expected to be preferentially located, either because they form there \citep[e.g.][]{bb1998} or because they segregate soon after their formation \citep[e.g.][]{SPZ2010}. Depending slightly on the velocity structure and the gas distribution, this leaves about half of the low-mass stars, which did not cross the core, with normal abundances. The model also predicts an enriched population that is more centrally concentrated than the pristine population. Given that a large fraction of high-mass stars are in binary systems that will interact at some point in their evolution \citep{Sana2012, Sana2013}, these interacting systems are promising sources of polluting material. It has been argued that their ejecta can display the required chemical properties and also satisfy the internal mass budget constraints \citep{deMink_2009} without requiring GCs to have been significantly more massive initially. 

Among the caveats of the early disc accretion model, an important one is the uncertainty on the expected lifetime of circumstellar discs around low-mass pre-main sequence stars in the dense environment of a globular cluster. These discs need to survive for $5-10$~Myr to sweep up enough material for this scenario to work, and Bondi-Hoyle accretion is expected to be inefficient due to the large velocity dispersions of massive clusters. \citet{DAntona2014} also pointed out that the structure of the seed stars may not remain fully convective for the whole duration of the accretion phase. If this is true, the full mixing of the helium-enriched material required down to the stellar centre would not be achieved. Thermohaline mixing may offer a way around this, but the same authors suggested that this could destroy any lithium surviving in the envelope, which may be in contradiction with observations \citep[e.g.][]{Shen2010}.

Both the better-studied multiple generations model \citep[with AGB stars as the source of pollution; e.g.][]{DErcole_2008} and the newer early disc accretion model \citep{Bastian_2013} have their difficulties, but neither has been conclusively ruled out to date. In the present paper, we therefore compare two main scenarios for how enriched material makes its way into stars: one in which a second generation is formed from gas that collects at the centre of the cluster (hereafter referred to as the {\it multiple generations} scenario), and another one in which pollution is due to accretion onto low-mass stars from the same generation when they pass through the core of the cluster (hereafter referred to as the {\it accretion} scenario). Instead of focusing on the chemistry as is often the case in studies of multiple populations, we approach this problem from another angle and explore the fossil kinematic imprints that may follow from these two formation scenarios after a Hubble time of dynamical evolution. Our results are not tied to a specific source of polluting material. They apply to any ``multiple generations" scenario that proceeds as suggested by \citet{DErcole_2008}, whether AGB stars are the polluters or not. Similarly, they would be relevant for any ``accretion" scenario that proceeds as outlined by \citet{Bastian_2013}, regardless of the source of pollution or whether accretion proceeds via discs or not.

In \S \ref{prev_dyn}, we start by reviewing previous work on the dynamics of multiple populations, which was mainly set in the context of models invoking multiple generations. \S \ref{nbody} describes the $N$-body simulations that we performed and the different initial conditions that we chose to capture the distinct kinematic properties implied by the two self-enrichment models considered. We then present in \S \ref{results} the results of these simulations, with an emphasis on the identification of a unique kinematic signature that allows to distinguish between the two models. In \S \ref{observables}, we discuss how this kinematic signature could be observed and we identify a large subsample of Galactic GCs in which it is expected to have survived to the present epoch. We present our conclusions in \S \ref{conclusions}.

\section{Previous studies of the dynamics of multiple populations}
\label{prev_dyn}

As mentioned previously, the internal mass budget problem of models invoking multiple generations can in principle be avoided by assuming that GCs were initially much more massive. A significant fraction of first-generation stars ($\sim90\%$ or more) must then be lost (i.e. unbound) by dynamical evolution while the more concentrated second generation remains relatively unaffected. \citet{Decressin2008} showed with $N$-body simulations that two-body relaxation alone cannot cause such a strong preferential loss of first-generation stars, and suggested that a more efficient mechanism such as primordial gas expulsion is needed to unbind the first-generation stars located in the outer regions of the cluster on a short timescale.

\citet{Decressin2010} found that, under favourable circumstances, the expulsion of gas left over from star formation can lead to a cluster containing 60\% of second-generation stars. The radial distribution of the two populations is still expected to be different after gas expulsion, so two-body relaxation could further increase the fraction of second-generation stars. Note however that the importance of early gas expulsion in young massive clusters has recently been questioned by hydrodynamical simulations \citep{Kruijssen2012} and kinematic observations \citep[e.g.][]{HB2012_sigma, cottaar2012} suggesting that newly formed clusters are relatively gas-poor at their centre and in virial equilibrium from a young age (a few Myr). 

\citet{DErcole_2008} studied the effect of mass loss from type~II supernova ejecta (without including primordial gas left over from star formation) and suggested that this alone could trigger rapid early expansion of the cluster and lead to a strong preferential loss of first-generation stars. In any case, it seems that the initial conditions and mass lost in this early phase need to be fine-tuned to explain that the mass ratio of the observed populations is always of order unity. One may otherwise expect GCs experiencing weak tidal fields to contain a much larger fraction of pristine stars. This does not seem to be the case for the remote and massive cluster NGC 2419 ($R_{\rm G} = 94.7$~kpc, $M_V = -9.5$). Multi-band photometric observations of this cluster indeed suggest that a significant fraction of its stars belong to a helium-enriched population \citep{Beccari2013}.

In the following sections, we will nevertheless ignore these caveats about early mass loss and follow the long-term dynamical evolution of multiple populations starting from a cluster in virial equilibrium with a similar number of polluted and pristine stars. This approach was also adopted by \citet{Vesperini_2013}, who studied the long-term evolution of the relative spatial distribution of first- and second-generation stars (with spherically symmetric distributions) in the context of the scenario presented by \citet{DErcole_2008}. Starting with a second generation that is more centrally concentrated than the first generation, they explored how the timescale for spatial mixing of the two populations depends on the initial concentration of the second generation. They found that complete spatial mixing is expected only for dynamically evolved clusters which have lost at least $60-70\%$ of their mass due to two-body relaxation (i.e. ignoring the rapid early loss of first-generation stars), irrespective of the initial concentration of the second generation. In \S\ref{observables}, we comment on the fraction of Galactic GCs expected to have evolved to that stage. 

In addition to having different spatial distributions, stars belonging to different populations could also display distinct kinematic properties. \citet{Bekki_2010, Bekki2011} simulated the formation of second-generation stars from AGB ejecta within a cluster and showed that if the first generation formed with some small net angular momentum, the second generation can show considerable rotation and even adopt a flattened distribution initially. This is a simple consequence of conservation of angular momentum, along with dissipative processes driving the polluted material to higher densities towards the centre of the cluster. Such properties are reminiscent of the multiple stellar populations (with different ages and chemical properties) observed in nuclear star clusters, the youngest of which often form centrally concentrated stellar discs \citep[e.g.][]{Seth2006}. \citet{Bekki_2010, Bekki2011} also showed that the second-generation stars are expected to initially have a lower velocity dispersion than the first-generation stars.

\citet{Mastrobuono_Battisti_2013} were the first to study the long-term dynamical evolution of a GC with nested structures while also taking into account rotation and a flattened spatial distribution for the more concentrated population. Their $N$-body simulations, tailored to $\omega$~Cen, trace the evolution of a relatively low-mass ($10^5 \ \msun$) cold disc component embedded in a massive GC ($2.5\times10^6 \ \msun$). They showed that these initial conditions can leave behind signatures even after 12 Gyr of relaxation-driven dynamical evolution\footnote{Although note that the half-mass relaxation time of a cluster like $\omega$~Cen is comparable to a Hubble time or larger.}. For example, while the disc stars do become more isotropically distributed with time, they do not attain a completely relaxed spherical shape at the end of the simulations. Also, as these disc stars exchange angular momentum with the other stars, the whole cluster can become slightly flattened. Another prediction of this study is a lower velocity dispersion and a more radially anisotropic velocity distribution for the polluted (disc) stars compared to the pristine stars. We will address the uniqueness of these kinematic signatures in \S \ref{results} when comparing the long-term evolution of the accretion and multiple generations scenarios.

Before moving to our own $N$-body simulations, we note that the work presented here is also motivated by recent observational results. For example, from HST proper motion measurements in 47 Tuc, \citet{Richer_2013} found that the stars on the bluer side of the main sequence (presumably more helium-rich/polluted) exhibit the largest proper-motion anisotropy, with preferentially radial orbits, while the presumably pristine stars on the redder side of the main sequence display no velocity anisotropy (in the plane of the sky). The larger radial anisotropy of the polluted stars was interpreted by these authors as a possible effect of two-body relaxation, with the more centrally concentrated polluted population slowly diffusing outward. Kinematic differences were also observed between the stellar populations of 47 Tuc from radial velocity measurements. \citet{Kucin2014} indeed reported that the line-of-sight velocity dispersion of their subsample of more enriched stars is lower than that of their subsample of presumably pristine stars by about 1~\kms.

Based on radial velocities and Na abundance measurements in 20 Galactic GCs, the study of \citet{bellazzini_2012} is the only one to date that has explored the possible connection between kinematics and multiple chemical populations in a large sample of clusters. Their data generally did not reveal an obvious correlation between either velocity dispersion or systemic cluster rotation and Na abundance (a convenient tracer of self-enrichment). Three of their clusters (NGC~6388, NGC~6441, NGC~2808) show tentative evidence for a drop in the velocity dispersion with increasing Na abundance, but this will need to be verified with larger and more evenly distributed samples. While it was reported that the Na-rich and Na-poor subsamples display similar rotational patterns in NGC 6441 and NGC 6388, there is some indication that Na-poor stars display a larger rotational amplitude than Na-rich stars in NGC 2808 ($\simeq2.25$~\kms \ and 1.0 \kms, respectively\footnote{\citeauthor{bellazzini_2012} actually quote values of $A_{\rm rot}\simeq4.5$~\kms \ and 2.0 \kms, but they define $A_{\rm rot}$ as two times the mean rotational amplitude.}). Two other clusters (NGC 6171 and NGC 7078) show hints of Na-poor stars having a larger rotational amplitude by a few \kms, albeit from small samples.

\section{$N$-body simulations}
\label{nbody}

\subsection{Initial conditions for the pristine and polluted populations}

\label{IC_section}

We studied the dynamical evolution of polluted and pristine stars in the context of the {\it accretion} (\S \ref{iso}) and {\it multiple generations} (\S \ref{mgen}) scenarios by means of $N$-body simulations. More details about the aspects common to all of our simulations are summarised in the next subsection. Here we start by describing the initial conditions of the different populations and how these depend on the pollution scenario considered.

\subsubsection{Accretion scenario}
\label{iso}

To capture the initial conditions of an accretion scenario like the one proposed by \citet{Bastian_2013}, we set up an isochrone model \citep{Henon1959}. The isochrone model is a convenient choice because the orbits in this potential are analytic, and we can easily flag as polluted the stars on an orbit crossing the core of the cluster. The potential of the isochrone model is given by

\begin{equation}
\Phi(r) = -\frac{G M}{r_0 + (r_0^2 + r^2)^{1/2}},
\end{equation}

\noindent{where $M$ is the cluster mass, $r$ is the distance to the centre and $r_0$ is the scale radius (we shall refer to it as the core radius from now on), which is related to the half-mass radius of this model by $r_{h} = 3.06 \ r_0$. The density profile, with a constant density core followed by a $r^{-4}$ fall off (in 3D), is reminiscent of what is observed in nearby young massive clusters \citep{EFF1987}.

To set up our systems, positions and velocities of the stars were sampled randomly from an isotropic isochrone model, and the masses of the particles were sampled from a Kroupa mass function \citep[][see also \S \ref{runs}]{kroupa2001}, with no correlation between stellar mass and position.

We assumed that the enriched material - expelled by massive stars in the model proposed by \citet{Bastian_2013} - is located within the core radius of the cluster and we split the stars into two categories: stars that do not enter the core of the cluster are considered as pristine stars, while stars on an orbit with a pericentre smaller than the core radius are assumed to accrete and are flagged as members of the polluted population. This yields a cluster containing about 45\% of polluted stars at the start of the simulation, and these are by construction more centrally concentrated than the pristine stars.

Note that both in \citet{Bastian_2013} and in the present work, the fraction of polluted stars is determined by the initial orbit distribution of low-mass stars. Early segregation (either primordial or dynamical) of the most massive stars is simply invoked as a way to get a source of polluted material in the central regions of the cluster. We do not set up our clusters with primordial mass segregation, but dense clusters that are not primordially mass segregated will quickly segregate by two-body interactions anyway \footnote{ A massive star of mass $m$ sinks towards the centre of the cluster by dynamical friction on a timescale $t_s = \langle m\rangle / m  \times t_{\rm rl}$ \citep{Spitzer1969}, where $t_{\rm rl}$ is the relaxation time of the region of interest, and $ \langle m\rangle$ is the mean stellar mass. The timescale $t_s$ can actually be shorter than a few Myr for dense clusters \citep{pz1999}.}.

Not including primordial mass segregation could potentially affect the early dynamical evolution of the cluster, but we argue below that it does not change our results about the long-term evolution. Early mass loss from centrally concentrated massive stars can lead to a stronger expansion than the same mass loss distributed throughout the body of the cluster. For tidally limited and strongly mass segregated clusters, \citet{Vesperini2009} showed with N-body simulations that rapid dissolution may occur as a consequence of this early expansion and the associated larger flow of mass over the tidal boundary. However, they also showed that segregated clusters initially underfilling their Roche lobe survive the early expansion because much of this expansion occurs within the tidal radius and thus does not cause a significant loss of stars. In this case, the subsequent evolution, lifetime, and mass-loss rate of the cluster do not differ significantly from those of an initially non-segregated cluster. The main notable difference is that long-lived initially segregated clusters will tend to have a less concentrated structure, delaying core collapse. Note that the early mass-loss driven expansion of a mass segregated cluster is not homologous, as the massive (segregated) population loses more mass than the lower mass stars in the outer parts. The expansion is therefore more important in the core and has much less severe effects in the outer parts. 

Because all our simulated clusters are initially tidally underfilling (as most GCs probably start) and because we are interested in long-term kinematic signatures that are more easily observed (and that more easily survive) in the outer parts of clusters, we conclude from the arguments above that our results would not be significantly affected if we had included primordial mass segregation.

In the scenario proposed by \citet{Bastian_2013}, low-mass stars may accrete a significant fraction of their final mass, which could modify the mass function of the polluted population. However, we did not take this into account when building our N-body models for the accretion scenario, and instead assumed a standard mass function for both the polluted and the pristine population initially. This is partly for the sake of simplicity, but also because currently there are no clear predictions or robust observational constraints on the mass function of the polluted population following this gas accretion phase.


We only considered isochrone models with an isotropic velocity distribution. \citet{Bastian_2013} showed that when considering a radially anisotropic velocity distribution with the Osipkov-Merritt prescription \citep{Osipkov1979, Merritt1985} and adopting an anisotropy radius equal to the half-mass radius, the fraction of stars that spend very little or no time in the core is in good agreement with the isotropic case. A radially anisotropic velocity distribution in the outer parts may represent more realistic initial conditions if clusters form via mergers of smaller clumps, but we will show in the next section that our simulated clusters rapidly develop radial anisotropy in the outer parts anyway.

We computed various models with different amounts of kinetic energy in rotation initially. This is motivated by studies revealing that rotation with a typical amplitude of a few \kms \ is common among Galactic GCs \citep{Cote1995, Anderson2003, vdB2006, Lane2009, Lane2010a, Lane2010b, bellazzini_2012, Fabricius2014}. Strong rotation has also been found in young and intermediate-age massive clusters \citep{HB2012, Mackey2013}, while theoretical work suggests that cluster formation through violent relaxation in the presence of an external tidal field can lead to significant internal differential rotation \citep{VespeVarri2014}. To quantify the initial amount of rotation of our models, we use the dimensionless spin parameter introduced by \citet{Peebles1969}: 

\begin{equation}
\lambda = \frac{J \ |E|^{1/2}}{G \ M^{5/2}},
\end{equation}

\noindent{where $J$ is the total angular momentum, $E$ the total energy and $M$ the mass of the system. For each scenario, we modelled clusters with $\lambda=0$, 0.091, and 0.129. In the specific case of an isochrone potential, these values of $\lambda$ correspond to fractions of respectively 0, 10, and 20\% of the total kinetic energy of the cluster in rotation. These are reasonable values considering the typical rotational amplitudes and velocity dispersions of GCs \citep[$0\lesssim\vrotsigma\lesssim0.5$; e.g.][]{MeylanHeggie1997}. For all our models with net rotation, the adopted direction of the rotation axis (the positive z-axis) is such that the cluster is corotating with its orbit around the centre of the galaxy.

To consider clusters with different amounts of angular momentum, we added rotation to the isochrone model by flipping the z-component of the angular momentum vector of a fraction of randomly selected stars having negative z-angular momentum within the cluster (which preserves virial equilibrium). The rotation curve initially has the same shape as the velocity dispersion profile, but as two-body relaxation proceeds it looks similar to what is obtained from self-consistent models including rotation \citep{VarriBertin2012}, with a peak located near the half-mass radius of the cluster. We set up three systems (\tt ACC0}, {\tt ACC1}, {\tt ACC2}) with increasing values of $\lambda$ matching those used for the initial conditions of the multiple generations scenario (see Table~\ref{models} and \S \ref{mgen}). The stability of these systems was verified by checking that their total angular momentum, their half-mass radius, and their global virial ratio stayed constant for several crossing times after the start of the simulation (for test runs without stellar mass loss).

\subsubsection{Multiple generations scenario}
\label{mgen}

To emulate the initial conditions implied by the model of \citet{DErcole_2008} and the hydrodynamical simulations of \citet{Bekki_2010, Bekki2011}, we set up self-gravitating models with a flattened and rotating stellar distribution embedded in a more extended spherical cluster. We preferred a purely dynamical approach over hydrodynamical simulations, partly to avoid the complications inherent to modelling the star formation process, but also because this approach is fast and gives us more control over the initial conditions.

We used the code {\tt magalie} \citep{Boily2001} included in the {\tt NEMO}\footnote{http://bima.astro.umd.edu/nemo/} Stellar Dynamics Toolbox \citep[][version 3.3.2]{teuben1995}. {\tt magalie} can create multi-component (e.g. disc+halo) stellar systems in near equilibrium using a method adapted from the {\tt BUILDGAL} procedure of \citet{Hernquist1993}. When embedding a disc within a spherical component, the method starts with the exact spatial density profiles and then approximates the phase space distribution function of all components (and hence the velocity field) using moments of the collisionless Boltzmann equation. For all our initial conditions generated with {\tt magalie} (see below), we verified the stability of the constructed systems by checking that the angular momentum in both populations, their half mass radii and the global virial ratio stayed constant for several crossing times and revolution periods of the disc after the start of the simulation (for test runs without stellar mass loss).

\begin{figure}
\begin{center}
\includegraphics[trim = 3mm 15mm 10mm 20mm, clip, width=\columnwidth]{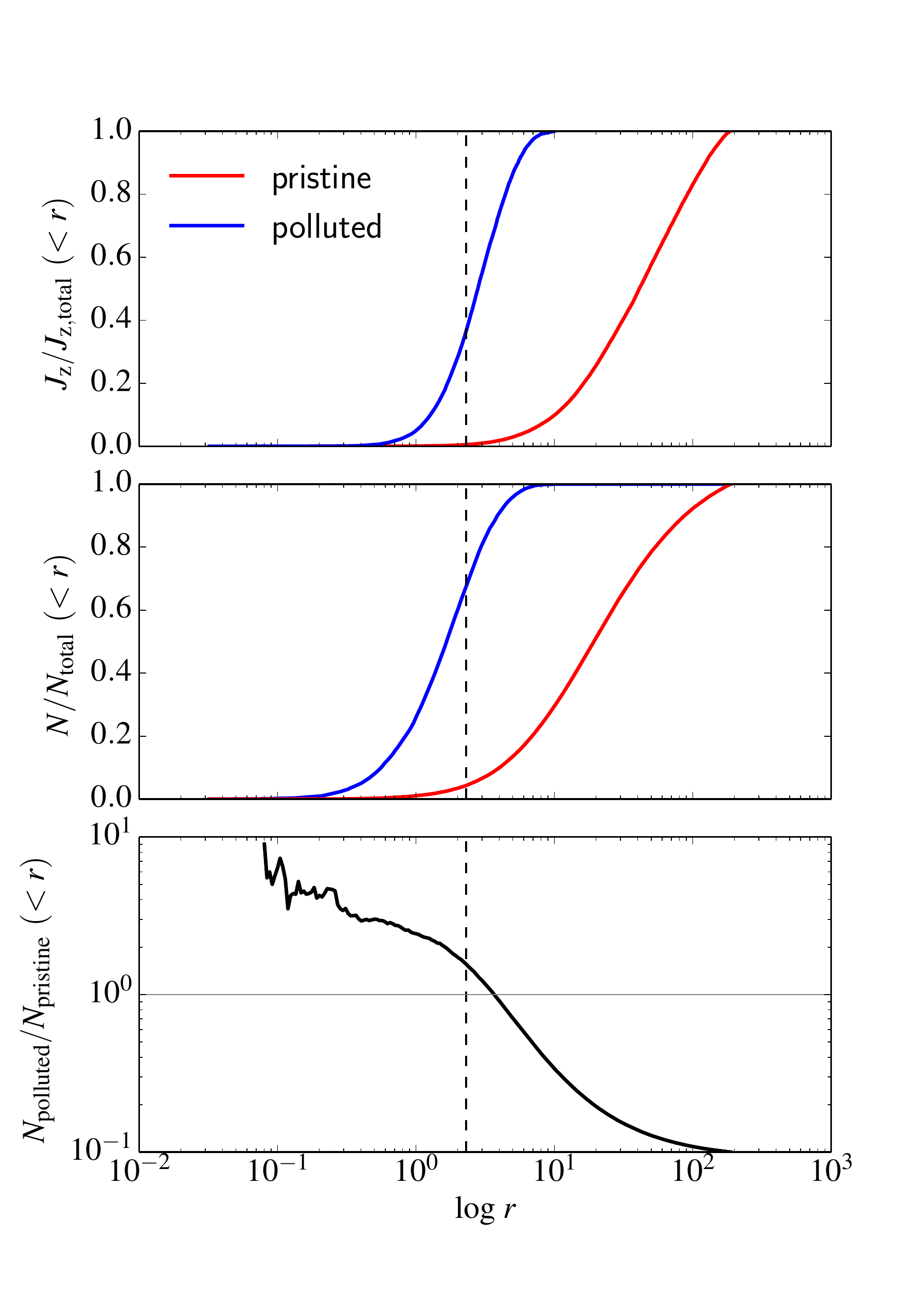}
\caption{\textbf{\label{J_budget}} 
Cumulative number and angular momentum distributions of polluted (blue curves) and pristine stars (red curves), as a function of radius, in an equilibrium ``disc+halo" configuration with $\lambda=0.091$, $M_{\rm halo}=10\ {M}_{\rm disc}$, and $J_{\rm z, halo}=20\ {J}_{\rm z, disc}$. This captures the configuration of the multiple generations scenario before early violent mass loss. The vertical dashed lines indicate the radius beyond which 90\% of the cluster stars reside. {\it Top panel:} Cumulative distribution of angular momentum (z-component) for the polluted (disc) and pristine (halo) stars as a function of radius, normalised to the total angular momentum in each component. {\it Middle panel:} Cumulative number distribution of the polluted and pristine stars as a function of radius, normalised to the total number of stars in each component. {\it Bottom panel}: Ratio of the number of polluted and pristine stars within radius $r$.}
\end{center}
\end{figure}

We considered two distinct populations in our toy models, namely a polluted and a pristine one. For models with net angular momentum, we represented the pristine stars (first generation) by a spherically symmetric halo following a Hernquist profile\footnote{The density of the Hernquist model as a function of radius is given by $\rho(r) = \frac{\rho_0}{(r/a) \ (1+r/a)^3}$, where $a$ is the scale length.} \citep{Hernquist1990} and the embedded polluted population (second generation) by an exponential disc with a scale height equal to 20\% of its scale length. Like the isochrone model, the Hernquist model has a power-law density profile with a $r^{-4}$ fall off in the outer parts, which describes young massive clusters very well \citep{EFF1987} and therefore represents a good choice for realistic initial conditions.

We first explored how much angular momentum is expected to be lost from the first generation if $\sim90\%$ of the initial cluster mass is rapidly removed from the outer parts. To do so, we set up systems in which the mass of the halo is 10 times larger than the mass of the disc (i.e. before mass loss). We assumed that half of the mass of the disc is made of ejecta from the first generation, and the other half comes from accretion onto the cluster of diluting (pristine) gas with zero net angular momentum. From conservation of angular momentum, we thus require that the total angular momentum of the halo is 20 times larger than that of the disc. The amount of rotation in the halo is controlled by flipping the z-component of the angular momentum vector of a fraction of randomly selected stars having negative z-angular momentum. A system satisfying the above constraints and having $\lambda=0.091$ is obtained by flipping the z-momentum of 65\% of the halo stars having a negative z-momentum and by setting the ratio between the scale length of the Hernquist halo ($a_{\rm halo}$) and the scale length of the exponential disc ($R_{\rm disc}$) to $a_{\rm halo}/R_{\rm disc} = 9.37$.

Figure \ref{J_budget} shows the cumulative number and angular momentum distribution as a function of radius (in 3D) for the pristine and polluted populations of such a system. The vertical dashed line in each panel indicates the radius beyond which 90\% of the cluster stars reside. If we assume that the stars beyond this radius are removed instantaneously, we can see that this would not only preferentially remove the pristine stars (as required in this scenario, leaving the number ratio of the two populations close to unity), but it would also remove a large fraction of the angular momentum of the pristine population. In this particular setup, the resulting cluster would have a larger total angular momentum in the second generation than in the first generation by a factor of $\sim3$.

We also looked at similar configurations but with different values of $\lambda$ and a larger initial mass ratio between the first and second generations (a mass ratio of 10 is a lower limit if the internal mass budget problem is to be avoided in scenarios with multiple generations). For a fixed value of $\lambda$, configurations with a larger mass ratio  
have a disc that is less compact with respect to the halo (i.e. smaller $a_{\rm halo}/R_{\rm disc}$). In this case, the preferential loss of first-generation stars is less important when the outermost stars are removed. This can produce a system in which the total angular momentum of the first generation is still larger than that of the second generation, but it may also yield a fraction of second-generation stars that is too low (see Figure \ref{Jz_loss}). As $\lambda$ is increased, $a_{\rm halo}/R_{\rm disc}$ also gets smaller and it becomes increasingly difficult to remove a large fraction of pristine stars without also losing a large fraction of polluted stars. If we take the simple dynamical considerations above at face value and if GCs can form with a large amount of angular momentum, a large initial mass ratio between the polluted and pristine populations ($M_{\rm halo} \gtrsim 20\ {M}_{\rm disc}$) may be difficult to reconcile with the observed fraction of polluted stars in GCs.

\begin{figure}
\begin{center}
\includegraphics[width=\columnwidth]{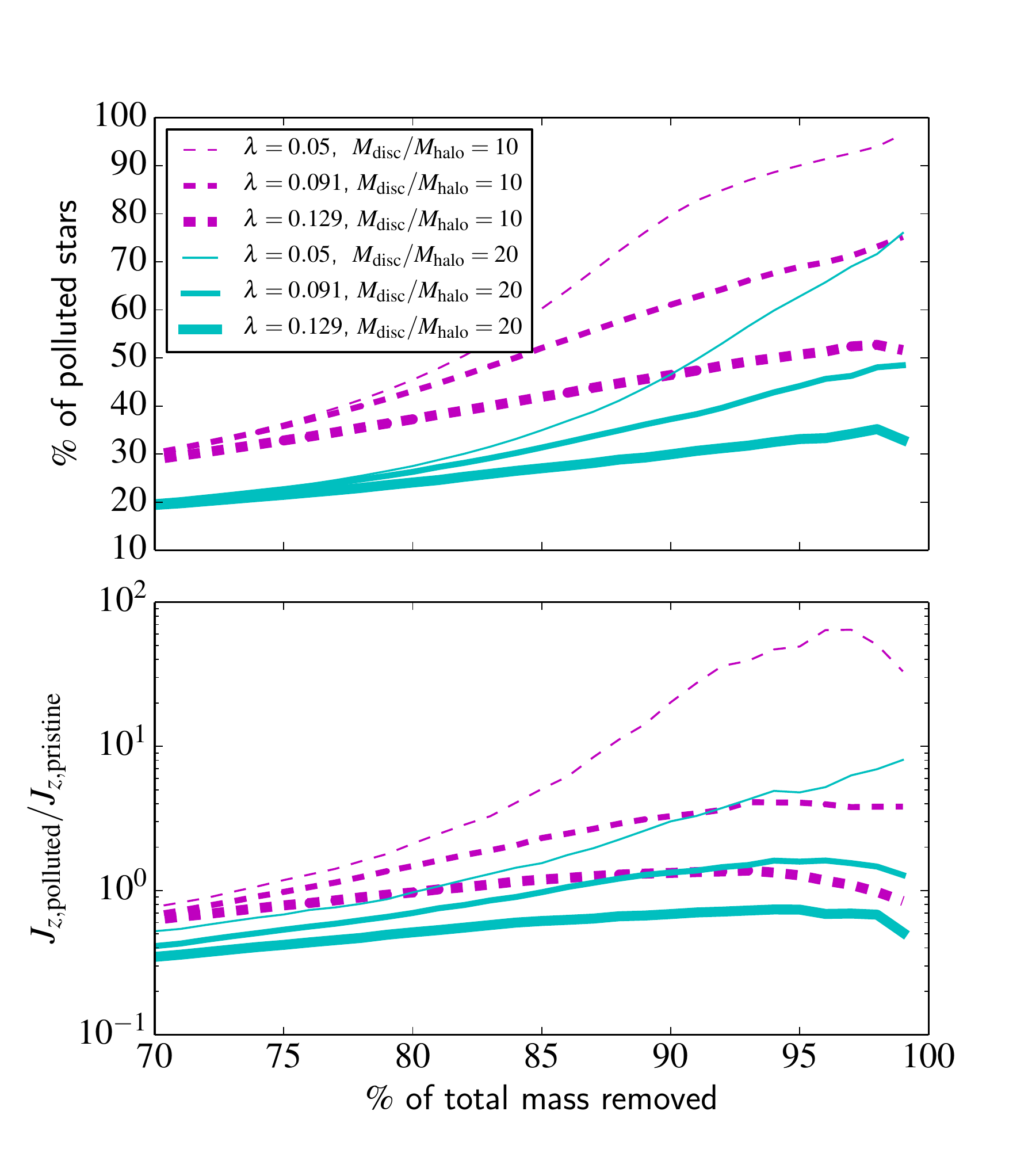}
\caption{\textbf{\label{Jz_loss}} {\it Top panel:} Percentage of polluted stars in the cluster after instantaneous removal of all the stars beyond a certain radius (in the multiple generations scenario), as a function of the percentage of the total mass removed. {\it Bottom panel:} Ratio of total angular momentum in the remaining polluted and pristine stars, as a function of the percentage of the total mass removed from the cluster. In both panels, the different curves illustrate the effect of different initial conditions for the mass ratio between the two populations and the total amount of angular momentum (before mass removal).}
\end{center}
\end{figure}

\begin{table*}
\caption{Summary of the initial conditions of our suite of $N$-body simulations. We also list the fraction of the initial mass and number of stars left after 10.75~Gyr for all the models with stellar evolution. The model without stellar evolution was followed until complete dissolution. More details on the parameters common to all models are presented in the main text.}
\label{models}
\begin{tabular}{l c c c c c c c c }
\hline
Model & Density profile &   stellar& $\lambda$ & $N_{\rm polluted}/N_{\rm pristine}$& $J_{z,{\rm pristine}}/J_{z,{\rm polluted}}$& $a_{\rm halo}/R_{\rm disc}$ & $M_{\rm f}/M_0$ & $N_{\rm f}/N_0$  \\
& &evolution & & & & \\
\hline
{\tt MGEN1} & Exp. disc + Hernquist & $\checkmark$ & 0.091 & 1& 0& 3.38 & 0.48 & 0.86  \\
{\tt MGEN1-nosev}  & Exp. disc + Hernquist  & & 0.091 & 1& 0& 3.38 &\ldots &\ldots \\
{\tt MGEN2}  & Exp. disc + Hernquist & $\checkmark$ & 0.129 & 1& 0& 1.36 & 0.50 & 0.90 \\
\hline
\hline 
Model & Density profile  &   stellar& $\lambda$ & $N_{\rm polluted}/N_{\rm pristine}$& $J_{z,{\rm pristine}}/J_{z,{\rm polluted}}$& $a_{\rm pristine}/a_{\rm polluted}$& $M_{\rm f}/M_0$ & $N_{\rm f}/N_0$  \\& &evolution & & & &  \\
\hline
{\tt MGEN0a} & Hernquist + Hernquist & $\checkmark$ & 0  & 1& \ldots& 1.36 & 0.46 & 0.84\\
{\tt MGEN0b} &Hernquist + Hernquist & $\checkmark$ & 0 & 1& \ldots & 3.38 & 0.45 & 0.82 \\
{\tt MGEN0c} & Hernquist + Hernquist &$\checkmark$ & 0 & 1& \ldots & 10.0 & 0.42 & 0.75 \\
\hline
\hline
Model & Density profile  &   stellar & $\lambda$ & $N_{\rm polluted}/N_{\rm pristine}$  & $J_{z,{\rm pristine}}/J_{z,{\rm polluted}}$   &$r_{\rm 0}/r_{\rm h}$ & $M_{\rm f}/M_0$ & $N_{\rm f}/N_0$ \\
& &evolution & & & &  \\
\hline
{\tt ACC0} & Isochrone & $\checkmark$ & 0 & 0.81 &\ldots & 0.33 & 0.50 & 0.90 \\
{\tt ACC1} & Isochrone &$\checkmark$ & 0.091 & 0.81 & 5.53& 0.33 & 0.50 & 0.90 \\
{\tt ACC2} & Isochrone &$\checkmark$ & 0.129 &0.81 &5.66& 0.33 & 0.50 & 0.90 \\
\hline
\end{tabular}

\end{table*}

In what follows, we focus on systems for which removing a large fraction of stars from the outer parts can leave the cluster with at least $\sim50\%$ of polluted stars. In these cases, the total angular momentum in the remaining polluted stars is always larger than the total angular momentum in the remaining pristine stars (Figure~\ref{Jz_loss}). For simplicity, in all our models mimicking the multiple generations scenario, we therefore assumed that the net angular momentum of the cluster is dominated by polluted stars and that the pristine population has no net angular momentum initially ($J_{z,{\rm pristine}}/J_{z,{\rm polluted}} = 0$). We also attributed an equal number of stars to each of the two populations. This captures the initial conditions after the early mass loss phase implied by this scenario. In the remainder of this paper, when we refer to mass loss, we are only concerned with the mass lost during the long-term evolution driven by two-body relaxation.

The initial conditions of our different simulations are summarised in Table~\ref{models}. For a subset of these (models {\tt MGEN1}, {\tt MGEN2}, {\tt MGEN1-nosev}), we set up an exponential disc embedded in a spherical Hernquist halo. The spatial extent of the disc with respect to the halo is then dictated by the choice of $\lambda$. One of these simulations ({\tt MGEN1-nosev}), with $\lambda=0.091$, is the one for which we did not include stellar evolution and which we followed until complete dissolution (see \S \ref{runs}).

For comparison, we also considered another subset of initial conditions where the cluster has no net angular momentum and the second generation is represented by a spherically symmetric Hernquist model embedded in a more extended Hernquist halo. Unlike the rotating models above, the spatial extent of the second generation in these cases is not influenced by rotational support and does not depend on the properties of the first generation. Ultimately, it depends on the star formation process for this second generation. We thus decided to consider a few different concentrations for the second generation. In two cases (models {\tt MGEN0a} and {\tt MGEN0b}), the ratio of the radial scales of the two populations ($a_{\rm pristine}/a_{\rm polluted}$) was taken to be the same as the $a_{\rm halo}/R_{\rm disc}$ values of our rotating models. We also considered a model in which the second generation is more concentrated ({\tt MGEN0c}; see Table~\ref{models}).

\subsubsection{Comparison of the initial conditions of the two scenarios}
\label{comp}

\begin{figure*}
\begin{center}
\includegraphics[width=7.0in]{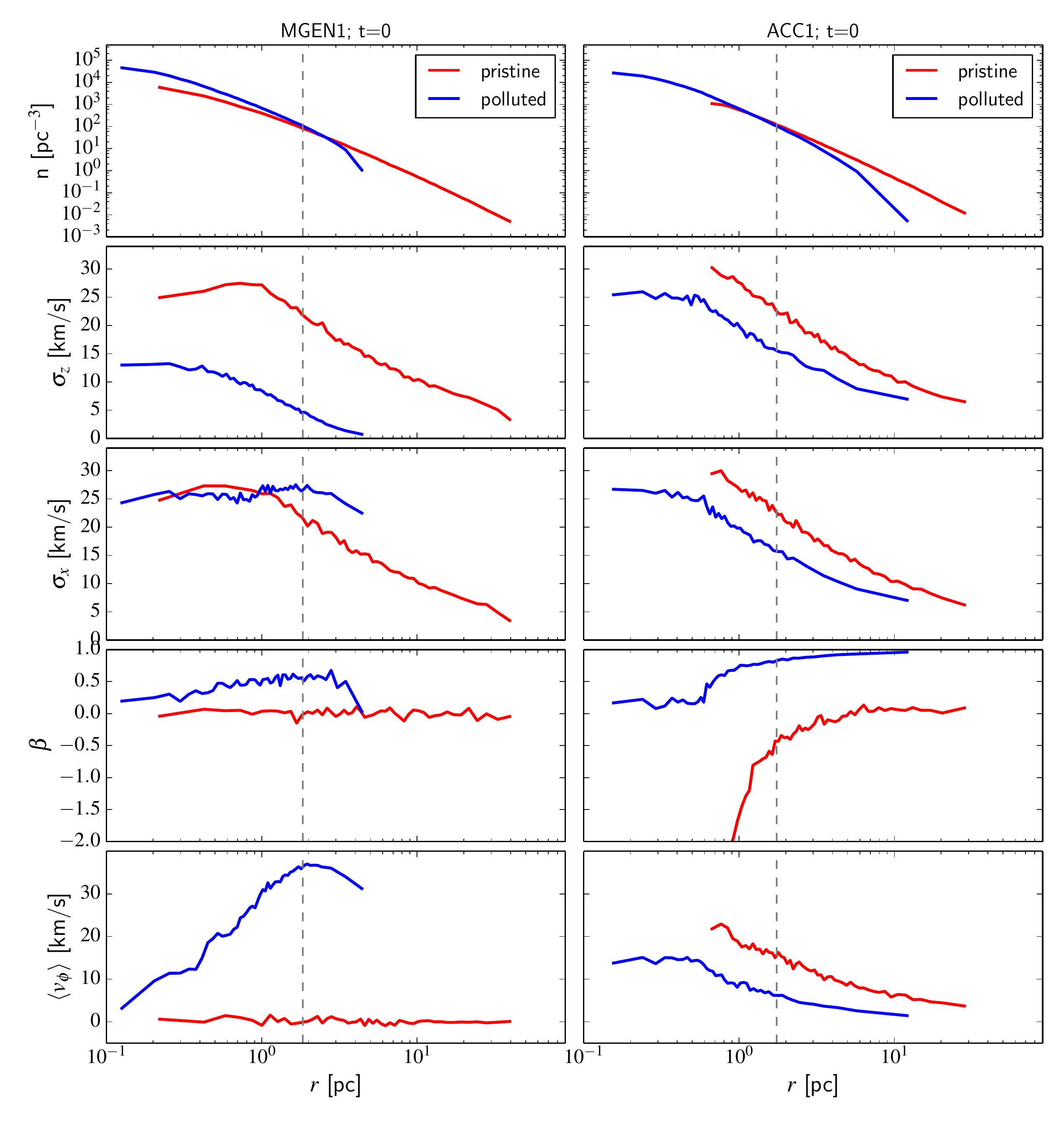}
\caption{\label{IC_snapshots} From top to bottom: number density, velocity dispersion in the $z$ direction, velocity dispersion in the $x$ direction, velocity anisotropy, and mean azimuthal velocity as a function of radius (in 3D) for polluted (blue) and pristine (red) stars. The initial conditions for model {\tt MGEN1} are shown in the left panels, and those for model {\tt ACC1} are shown in the right panels. The half-mass radius is indicated by a grey dashed line.}
\end{center}
\end{figure*}

As the polluted stars in the accretion model are assumed to be the ones crossing the core of the cluster, they are expected to be preferentially on radial orbits initially. In the presence of net angular momentum, this preference for radial orbits among the polluted stars also means that the mean rotational amplitude of these stars (at a given radius) will be lower than the mean rotational amplitude of the pristine stars. This is opposite to what is expected from the multiple generations scenario, where the angular momentum in the polluted population and the mean rotational velocity of the polluted stars is expected to be larger (Figure \ref{IC_snapshots}).

Note that we have neglected the effect of gas accretion on the orbits of stars in our simulations. In the early disc accretion scenario, the enriched gas is released by massive stars near the centre of the cluster before being quickly swept up (within a few Myr). As these stars located near the centre have low orbital angular momentum compared to the rest of the cluster stars, the enriched material that they expel would also have low specific angular momentum. The specific angular momentum of a star would thus typically be reduced following accretion, and its orbit would shrink (through conservation of energy and angular momentum). If anything, taking into account the effect of gas accretion would make the difference in the mean rotational velocity (at a given radius) between the polluted and pristine populations even larger, enhancing the difference already present between the initial conditions of the multiple generations and accretion scenarios. 

In Figure \ref{IC_snapshots}, we illustrate the similarities and differences in the initial conditions of the two scenarios by showing the number density, velocity dispersion ($z$ and $x$ components), velocity anisotropy, and mean azimuthal velocity (around the $z$-axis) profiles at time $t=0$ for the polluted and pristine populations of models {\tt MGEN1} and {\tt ACC1}. Similar plots are shown in section~\ref{appendix_ICs} of the appendix to compare the initial conditions of models {\tt MGEN0c} and {\tt ACC0} (Figure~\ref{IC_snapshots_AGB0C_EDA0}) and models {\tt MGEN2} and {\tt ACC2} (Figure~\ref{IC_snapshots_AGB2_EDA2}).

The 3D number density profile is already very similar in both of the scenarios considered at the beginning of the simulation. The global density profile of all our models will evolve towards a King-like profile \citep{King1966} at late times, and any memory of the specific model used to generate the initial conditions will gradually be erased as the density profile is reshaped by two-body relaxation and the tidal field.

The initial conditions of both scenarios are also generally characterised by a dynamically cooler and more centrally concentrated polluted population. Because the velocity dispersion of the cluster is higher towards the centre of the cluster, it may appear counter-intuitive that the more centrally concentrated population has a lower velocity dispersion. This is however a natural consequence of the more rapid fall off of the density distribution of this population \citep[see e.g.][]{vdb1999}. Another way to see this is that because the polluted population is preferentially located in the central regions, it contains fewer stars on wide orbits that cross the inner regions of the cluster at high velocity. 

For models {\tt MGEN1} and {\tt MGEN2}, note however that the $z$ component of the velocity dispersion is by definition initially smaller for the disc population than for the spherical (pristine) population. The only case where the velocity dispersion of the polluted population can be larger than that of the pristine population is when looking at the component of the velocity dispersion in the plane of the disc ($\sigma_x$ or equivalently $\sigma_y$) for these two models, from about the half-mass radius and beyond.

The anisotropy parameter is defined in the usual way as $\beta = 1 - \frac{\sigma_{\rm t}^2}{2 \sigma_{\rm r}^2}$, where $\sigma_{\rm t}$ and $\sigma_{\rm r}$ are the tangential and radial velocity dispersions. The velocity distribution is thus radially anisotropic for $\beta > 0$. For all models with net rotation, the anisotropy profiles of the accretion and multiple generations scenarios are qualitatively similar. In both scenarios, the polluted population has a radially anisotropic velocity distribution, while the pristine population is either isotropic or even tangentially anisotropic in the inner regions for the early disc accretion scenario. Our accretion model without rotation ({\tt ACC0}) displays the same behaviour, while both populations are (by construction) fully isotropic in the initial conditions of the multiple generations models without rotation ({\tt MGEN0a}, {\tt MGEN0b}, {\tt MGEN0c}).

\begin{figure*}
\begin{center}
\includegraphics[width=6.0in]{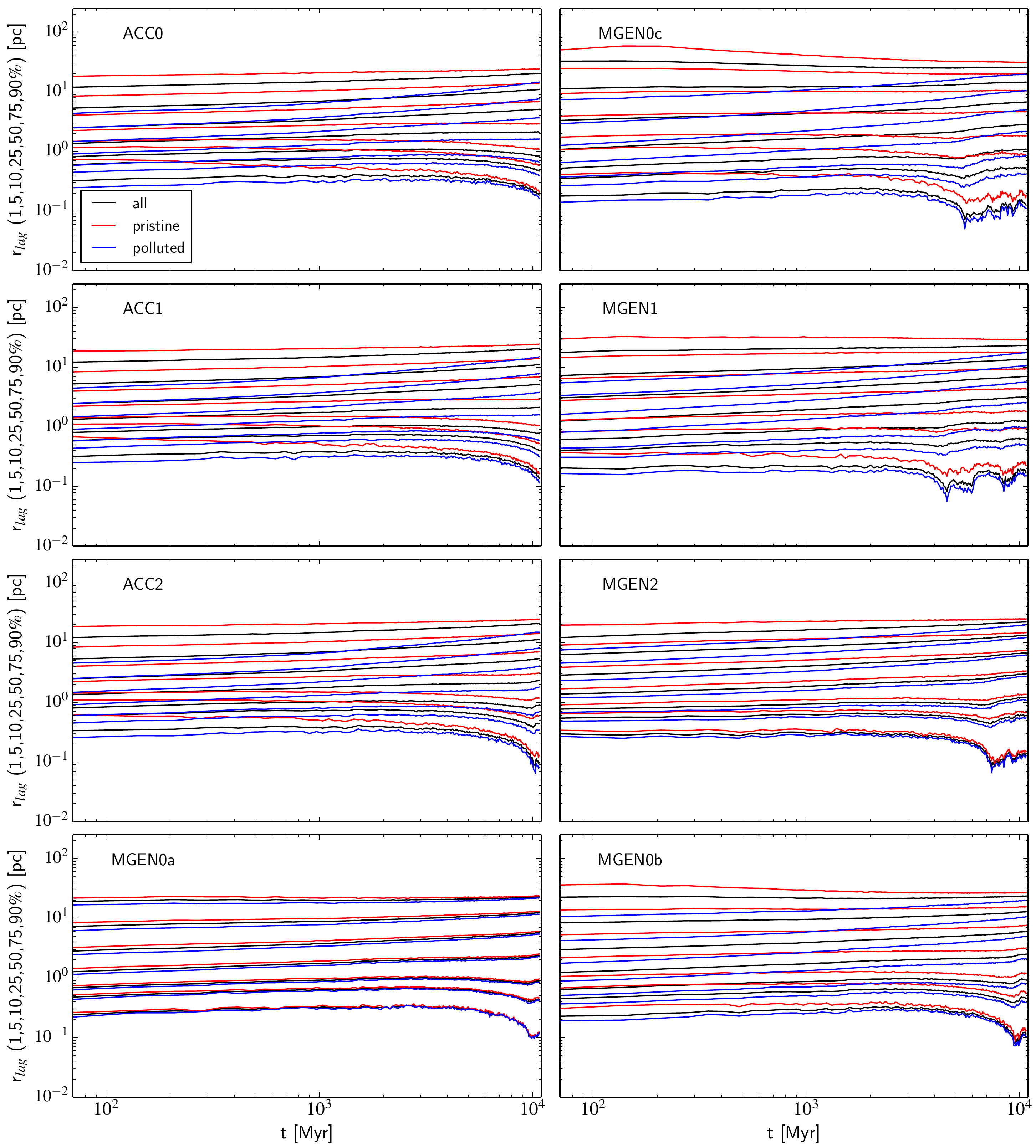}
\caption{\label{rlag} Time evolution of the 1\%, 5\%, 10\%, 25\%, 50\%, 75\%, and 90\% Lagrangian radii for the polluted (blue lines), pristine (red lines) and all stars (black lines) for all of our models with stellar evolution (see Table~\ref{models}).}
\end{center}
\end{figure*}

The mean azimuthal velocity profile (also referred to as the ``rotation curve" throughout this paper) is where the difference between the initial conditions of the accretion and multiple generations scenarios is the most striking. Apart from models without rotation where the mean azimuthal velocity profile of both populations is obviously flat and zero, all of our initial conditions imply significant differential rotation between the polluted and pristine stars. At a given radial distance from the centre of the cluster, the mean azimuthal velocity is always larger for the pristine stars in the accretion scenario, while it is always larger for the polluted stars in the multiple generations scenario. The latter follows from our choice of $J_{z,{\rm pristine}}/J_{z,{\rm polluted}} = 0$ for the initial conditions of the multiple generations scenario. It would however remain true even with $J_{z,{\rm pristine}}/J_{z,{\rm polluted}} = 1$ because the polluted population would still be more centrally concentrated and would thus need to have a larger rotational velocity amplitude for its total angular momentum to be the same as the more extended pristine population.

\subsection{Description of the runs}
\label{runs}

To follow the long-term dynamical evolution of the clusters, we used the code {\tt NBODY6}, a fourth-order Hermite integrator with \citet{1973JCoPh..12..389A} neighbour scheme \citep{1992PASJ...44..141M, 1999PASP..111.1333A, 2003gnbs.book.....A} and accelerated force calculation on NVIDIA Graphical Processing Units (GPUs) \citep{2012MNRAS.424..545N}.

All of our modelled clusters were placed on a circular orbit around the Galaxy, for which the gravitational potential was assumed to be a singular isothermal sphere. Unbound stars were removed from the simulation once they reached a distance of $2 r_{\rm J}$ from the cluster centre, with $r_{\rm J}$ (the Jacobi radius) given by

\begin{equation}
r_{\rm J} = \left(\frac{R_{\rm G}}{V_{\rm G}}\right)^{2/3} \left( \frac{G M}{2}\right)^{1/3},
\end{equation}

\noindent{where $R_{\rm G}$ is the galactocentric distance, $V_{\rm G}$ is the circular velocity around the centre of the galaxy (we adopt $V_{\rm G}=220$~\kms), $G$ is the gravitational constant, and $M$ is the mass of the cluster within radius $r_{\rm J}$. 

As we are interested in modelling mixing processes for which the evolution is driven by two-body relaxation \citep[see][]{Vesperini_2013}, we did not include primordial binaries. This is a reasonable simplification because the post-core collapse expansion of the cluster and the accompanying evolution of the two-body relaxation timescale is expected to behave in the same way irrespective of the details of the source of energy driving the expansion \citep[e.g. primordial binaries or mass loss from stellar evolution;][]{gieles2011, Giersz_Heggie_2011}.

With one exception (see below), all of our models include stellar and binary evolution (and associated mass loss\footnote{The mass lost due to stellar evolution is assumed to be instantly removed from the cluster.}) following the prescriptions of \citet{hurley2000} and \citet{hurley2002}. All stars are given a metallicity of $Z=0.0035$, but we ignore the effect of a spread in helium abundance among cluster stars on their evolution. For all our models with stellar evolution, we adopted a \citet{kroupa2001} initial mass function (IMF) between 0.1 and 100~$\msun$, with a power-law slope of -1.3 below 0.5~$\msun$ and -2.3 above.

For the multiple generations scenario (\S\ref{mgen}), including stars up to 100 $\msun$ may seem rather extreme given that we follow the evolution of the cluster after the second generation has formed and the cluster has reached virial equilibrium, in which case high-mass stars ($>8\  \msun$) from the first generation should have already exploded as supernovae. By ignoring that first-generation massive stars have already exploded, we actually overestimate the early mass loss from supernovae by only $\sim10\%$. This may lead to an overestimate of the early expansion of the cluster by $\sim10\%$ \citep[e.g.][]{Hills1980}, which is likely not very important when addressing whether signatures of the initial conditions can survive for a timescale of the order of the relaxation time.

For the model without stellar evolution, stellar masses were sampled from a double power-law IMF between 0.1 and 1.2 $\msun$ (to mimic the present-day mass spectrum of GCs), with a break at 0.5 $\msun$ and power-law slopes of $-1.3$ and $-2.0$.

All our models were computed with $N^*=10^5$ stars initially. When including stellar evolution, the two-body relaxation timescale is tied to the timescale of mass loss from stellar evolution, which fixes the physical timescale of the model. In these cases, the models with $N^*=10^5$ stars can be scaled to represent a more massive cluster with a larger number of stars by demanding that the relaxation time is the same for the full-scale cluster with $N$ stars as for the smaller-scale model. This is achieved when the radial scales of the two models are related in the following way \citep[e.g.][]{HG2008}

\begin{equation}
\frac{r_{v,*}}{r_v} = \left(\frac{N}{N_*}\right)^{1/3} \left(\frac{\log{(\gamma N_*)}}{\log{(\gamma N)}} \right)^{2/3},
\end{equation}

\noindent{where $r_{v,*}$ and $r_v$ represent the virial radius of the scaled and full-scale cluster, and $\gamma = 0.02$ \citep[e.g.][]{GHH2008}. The model with $N^*$ stars therefore has a larger radius than the full-scale cluster. We choose to scale our models to the properties of a typical massive GC like 47 Tuc, although we do not attempt to precisely reproduce the observed characteristics of this cluster. Our initial conditions are based on values from \citet{Giersz_Heggie_2011}, who used Monte Carlo models to obtain a set of initial parameters providing a satisfactory match to the kinematic and photometric data of 47 Tuc after a Hubble time of dynamical evolution. At $t=0$, our scaled models represent a cluster with $N=2\times10^6$ stars with a mean mass of $\langle m\rangle=0.6377 \ \msun$, a total mass of $M=1.275\times10^6 \ \msun$, a virial radius of $r_v = 2.38$ pc and a Jacobi radius of $r_J=86$~pc, corresponding to a galactocentric radius of $R_{\rm G}=3.3$~kpc for a singular isothermal sphere potential and a circular velocity of $V_{\rm G}=220$~\kms. The scaled parameters used were $N_* = 10^5$, total mass $M_*=6.377\times10^4 \ \msun$, $r_{v,*} = 5.2$ pc and $R_{{\rm G},*} = 47.98$ kpc.

We evolved these clusters up to an age of 10.75 Gyr, which is close to the latest age determination of 47 Tuc from the properties of its white-dwarf cooling sequence \citep{Hansen2013}. At the end of the simulations, the modelled clusters are all reasonably consistent with the present-day characteristics of 47~Tuc \citep[see][]{Giersz_Heggie_2011}: half-mass radii between 5 and 7 pc, central line-of-sight velocity dispersions between 10 and 13 \kms, Jacobi radii between 64 and 69 pc. Table~\ref{models} lists the fraction of the initial mass ($M_{\rm f}/M_0$) and number of stars ($N_{\rm f}/N_0$) left at the end of the simulation for all these models with stellar evolution.

For the model without stellar evolution, a cluster with $10^5$~stars, a virial radius of 2~pc and a mean mass $\langle m\rangle=0.3458 \ \msun$ was placed on a circular orbit with $R_{\rm G} = 10$~kpc. Because the physical timescale of this model is not set by the stellar evolution timescale, we are free to scale both the mass and radius of the system to represent clusters with different relaxation times and galactocentric radii \footnote{Note that the ratio of the half-mass radius to the Jacobi radius ($r_{\rm h}/r_{\rm J}$) is fixed.}.
We ran this model until complete dissolution of the cluster, which allows us to explore the late stages of the dynamical evolution of multiple populations without being tied to a specific scaling.

The initial half-mass relaxation time is 0.9~Gyr for all our simulations with stellar evolution and 0.3~Gyr for our model without stellar evolution, using the equation from \citet{spitzer1987}, which assumes an equal-mass cluster and spherical symmetry. One should however be cautious when directly comparing the age of the systems to this initial half-mass relaxation time, first because the half-mass relaxation time is not well defined for systems with a disc embedded in a spherical halo. We should expect two-body relaxation to be more efficient in systems in which stars are moving coherently (e.g. the disc component of our multiple generations systems), but this is not captured by the equation used. Our clusters also have a stellar mass spectrum and the mass function changes with time as a result of stellar and dynamical evolution. This will have an effect on the two-body relaxation time at any given point of the evolution.

\section{Long-term kinematic imprints} \label{results}

\subsection{Phase-space mixing}

\label{psmix}

\begin{figure}
\begin{center}
\includegraphics[width=\columnwidth]{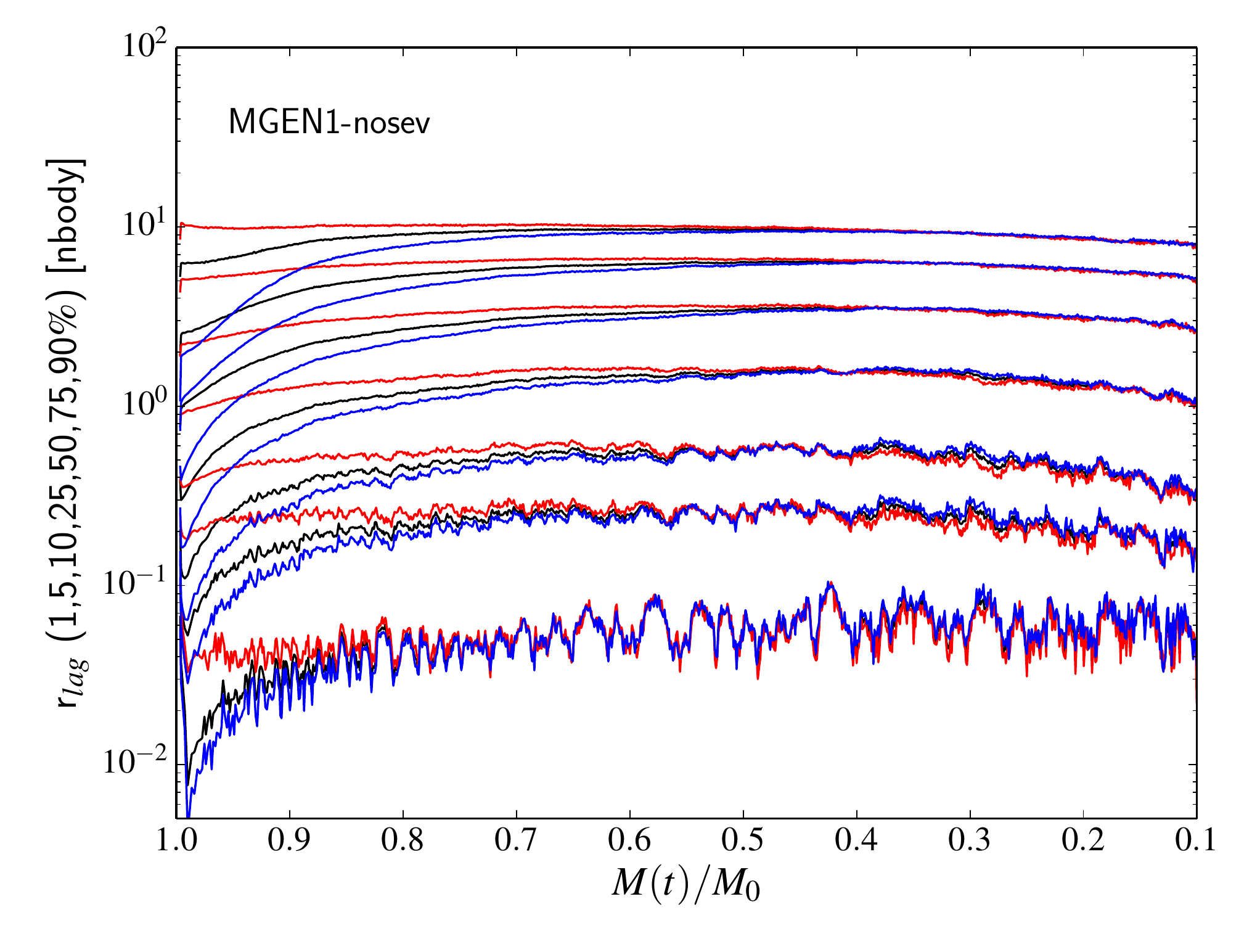}
\caption{\label{rlag_nosev} Evolution of the 1\%, 5\%, 10\%, 25\%, 50\%, 75\%, and 90\% Lagrangian radii for the polluted (blue lines), pristine (red lines) and all stars (black lines) as a function of the fraction of the initial mass left in the cluster for model {\tt MGEN1-nosev} (see Table~\ref{models}). The radius scale of this scalable model is expressed in $N$-body units.}
\end{center}
\end{figure}

We now turn to the long-term dynamical evolution of clusters from the initial conditions described in the previous section. After 10.75~Gyr, our 47 Tuc-like models with stellar evolution have lost between 50\% and 58\% of their initial mass ($\sim10-25\%$ of their stars) through a combination of stellar mass loss and tidal evaporation. At this stage, preferential loss of pristine stars has increased the ratio of polluted to pristine stars to $N_{\rm polluted}/N_{\rm pristine}\approx0.85$ for the accretion models and $1.03 < N_{\rm polluted}/N_{\rm pristine} <1.42$ for the multiple generations models.

Figure~\ref{rlag} shows the time evolution of the Lagrangian radii of polluted and pristine stars (along with those of the whole cluster) for all the models with stellar evolution. Overlap of all Lagrangian radii for the two populations would indicate complete spatial mixing, but for none of the models in Figure~\ref{rlag} are the polluted and pristine stars fully mixed at the end of the simulation. Only in model {\tt MGEN0a} are both populations almost mixed, but this is the model in which the polluted population started the least concentrated with respect to the pristine population. Interestingly, the effect of the gravogyro instability in accelerating core collapse \citep[e.g.][]{Ernst2007} is seen when comparing models {\tt ACC0}, {\tt ACC1} and {\tt ACC2}, for which the initial conditions only differ in their total amount of angular momentum.

\begin{figure}
\begin{center}
\includegraphics[width=\columnwidth]{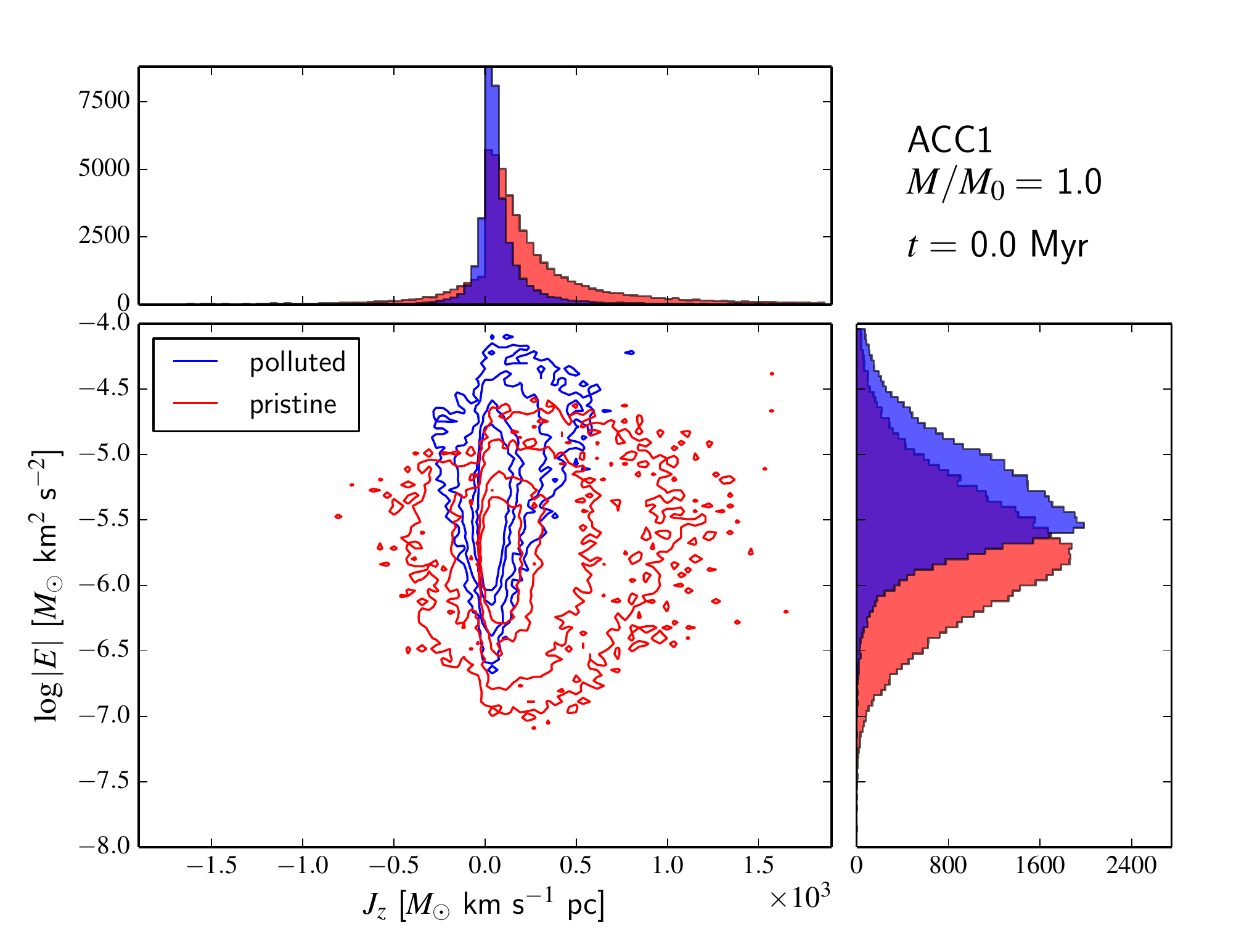}
\includegraphics[width=\columnwidth]{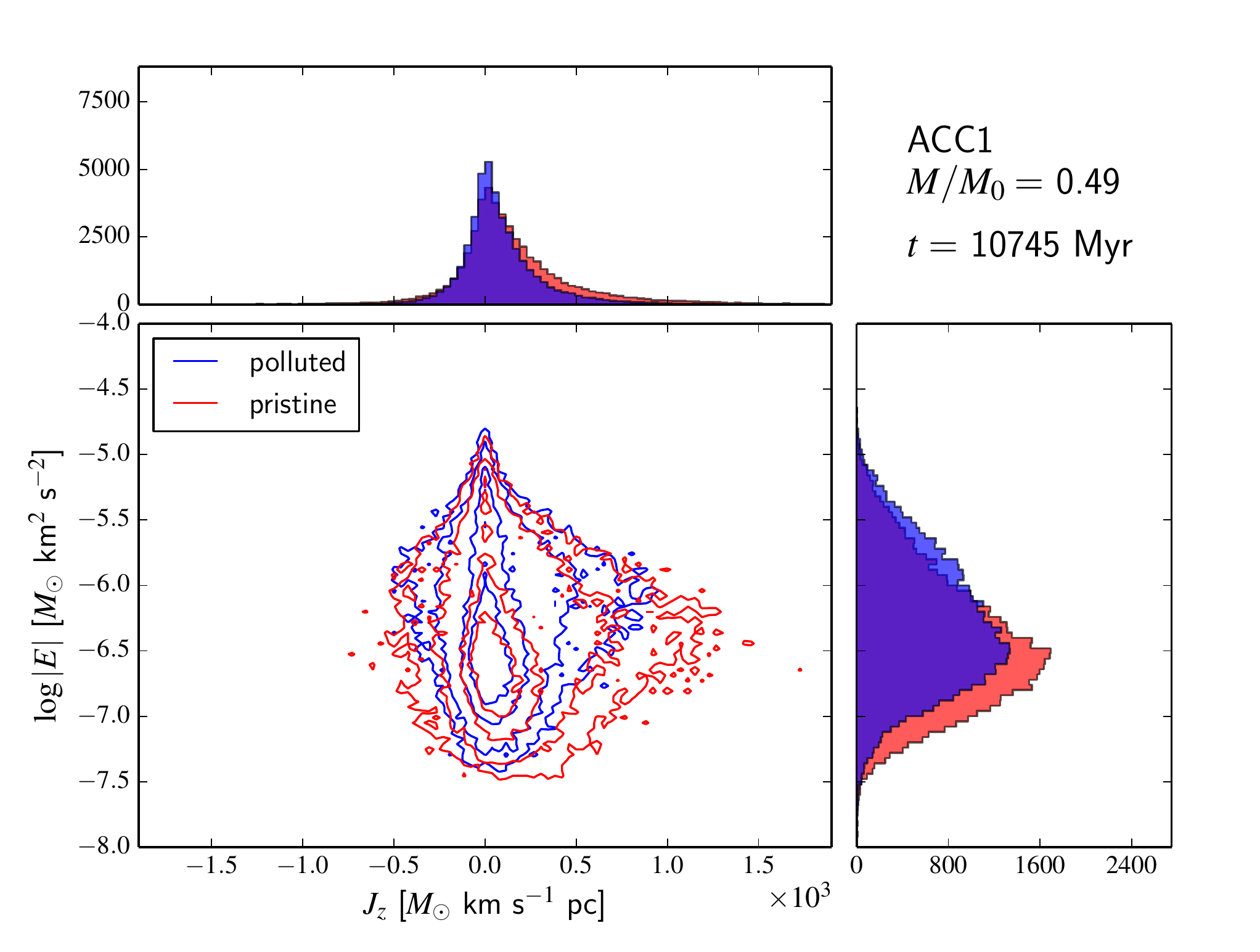}
\caption{\label{EDA1_mix} Distribution of energy and z-angular momentum for the polluted (blue) and pristine (red) stars of model {\tt ACC1} at $t=0$ (top panel) and at the end of the simulation (bottom panel). The solid lines represent isodensity contours in the $|E|$ vs. $J_z$ plane for each population.}
\end{center}
\end{figure}

\begin{figure}
\begin{center}
\includegraphics[width=\columnwidth]{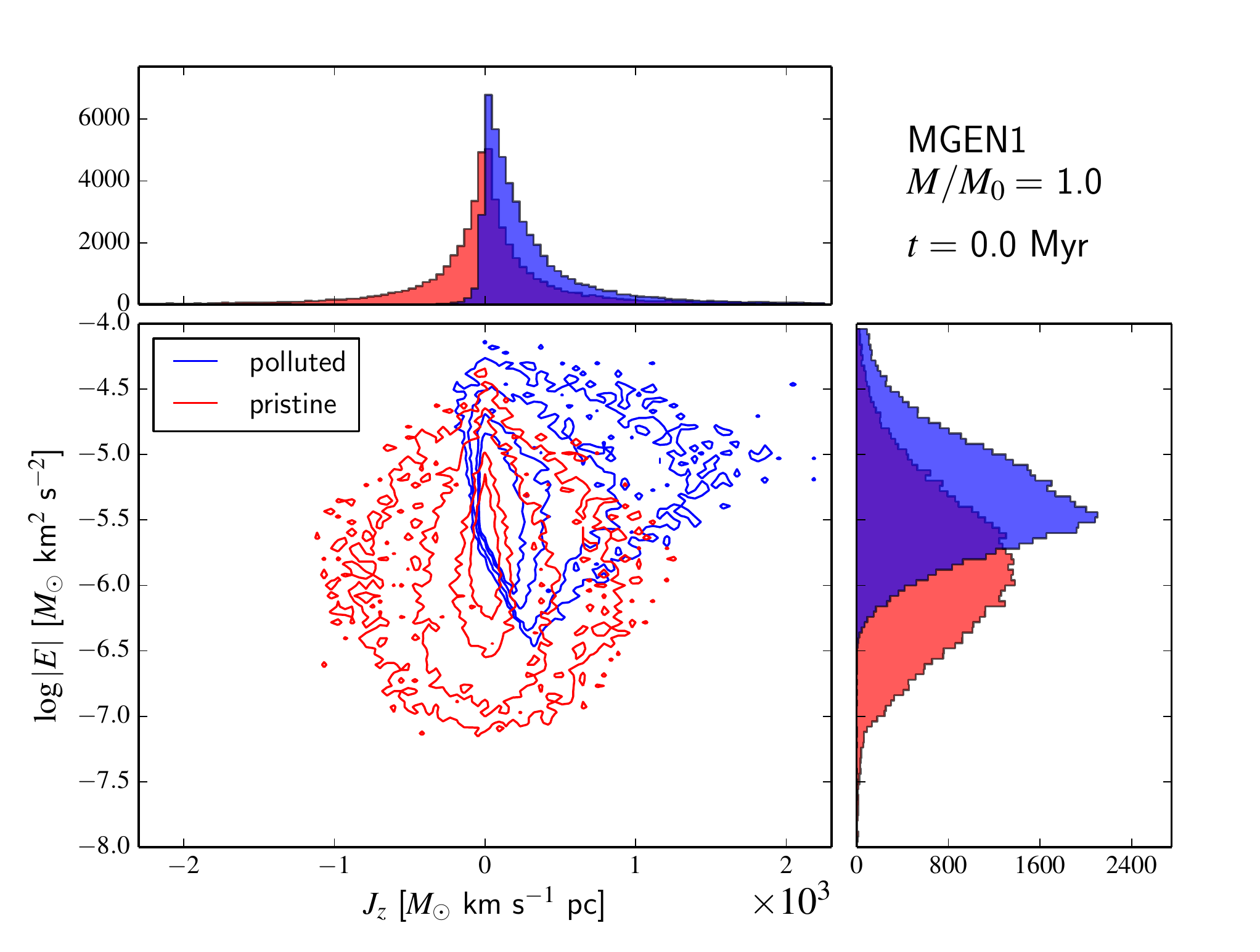}
\includegraphics[width=\columnwidth]{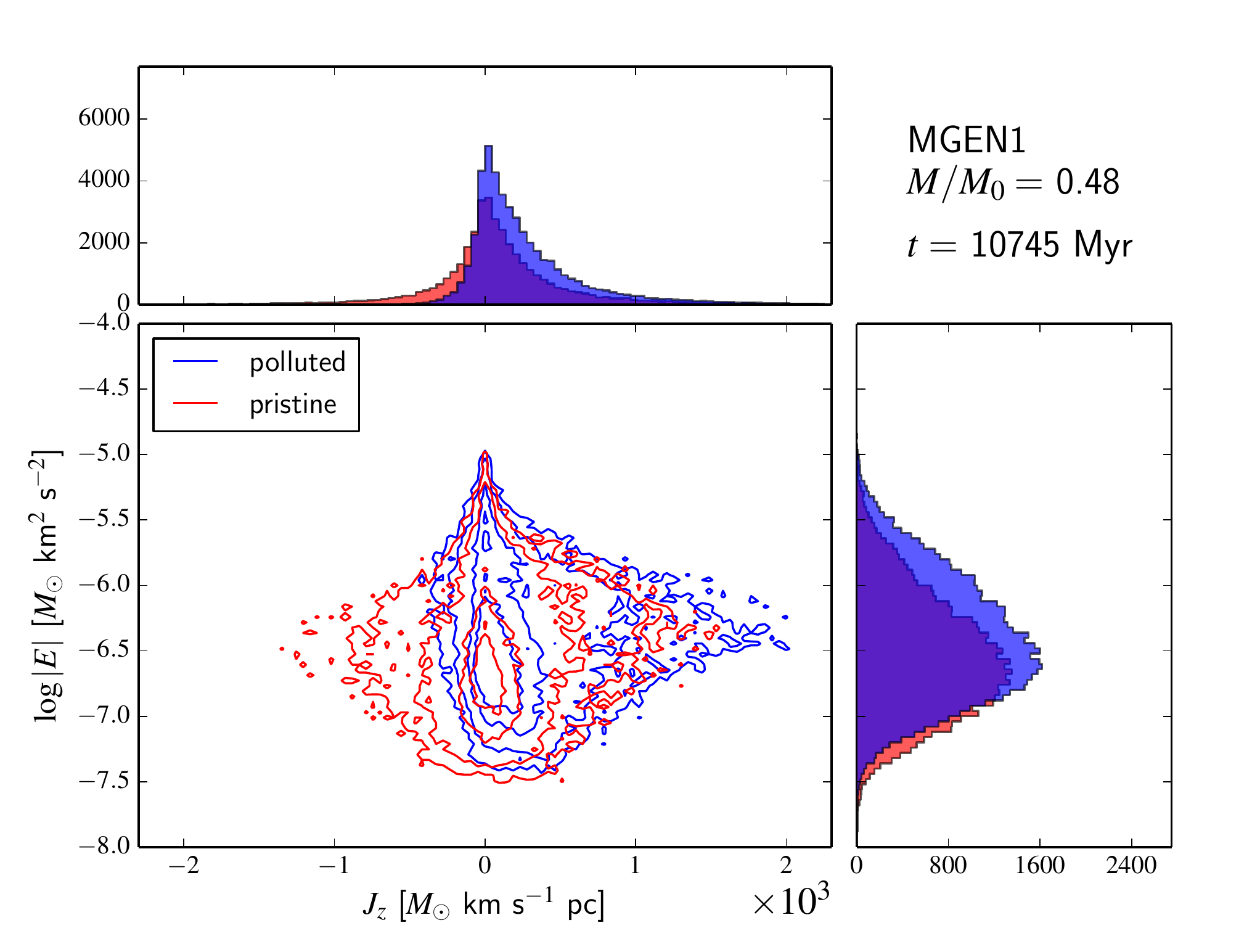}
\caption{\label{AGB1_mix} Distribution of energy and z-angular momentum for the polluted (blue) and pristine (red) stars of model {\tt MGEN1} at $t=0$ (top panel) and at the end of the simulation (bottom panel). The solid lines represent isodensity contours in the $|E|$ vs. $J_z$ plane for each population.}
\end{center}
\end{figure}

\begin{figure}
\begin{center}
\includegraphics[width=0.9\columnwidth]{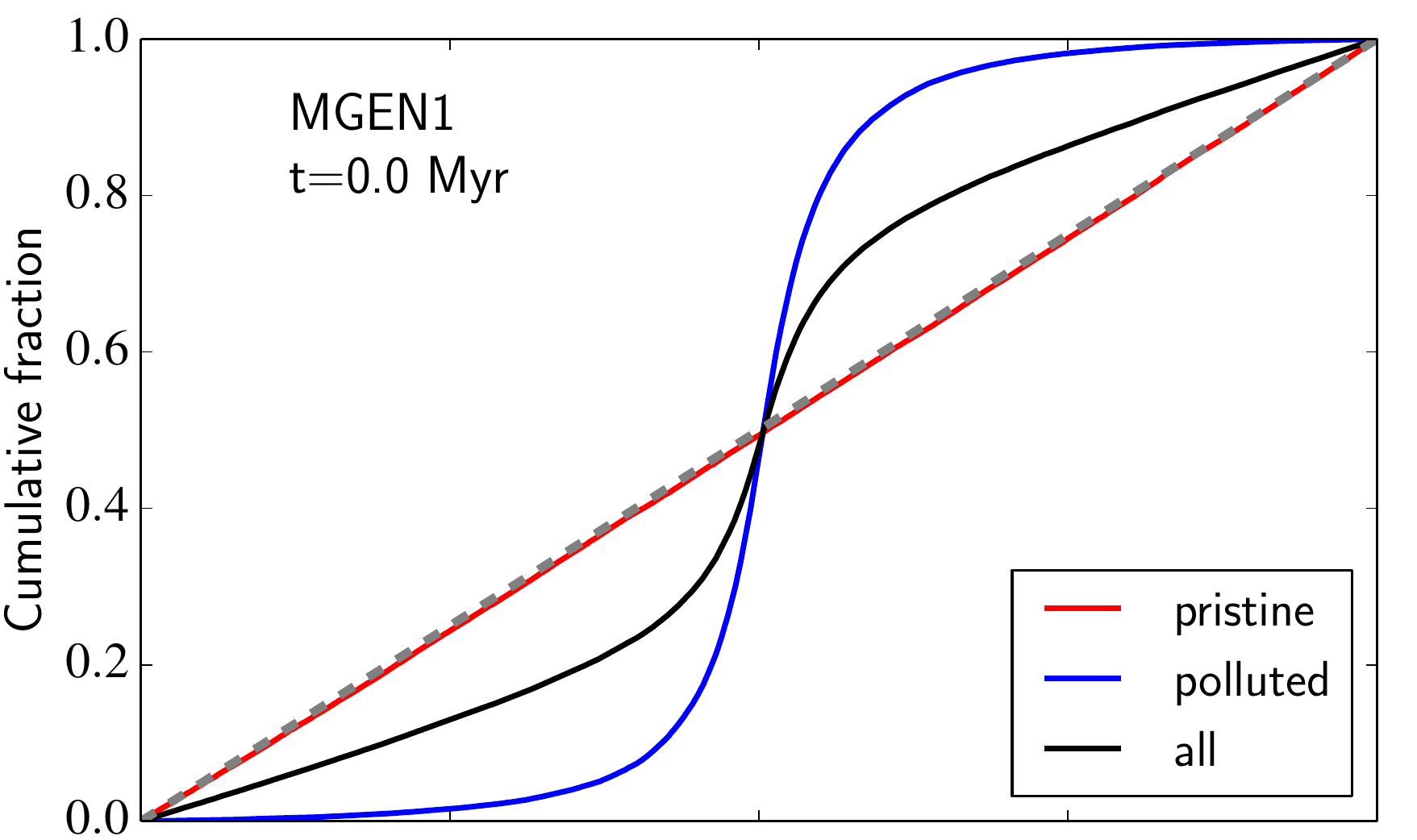}
\includegraphics[width=0.9\columnwidth]{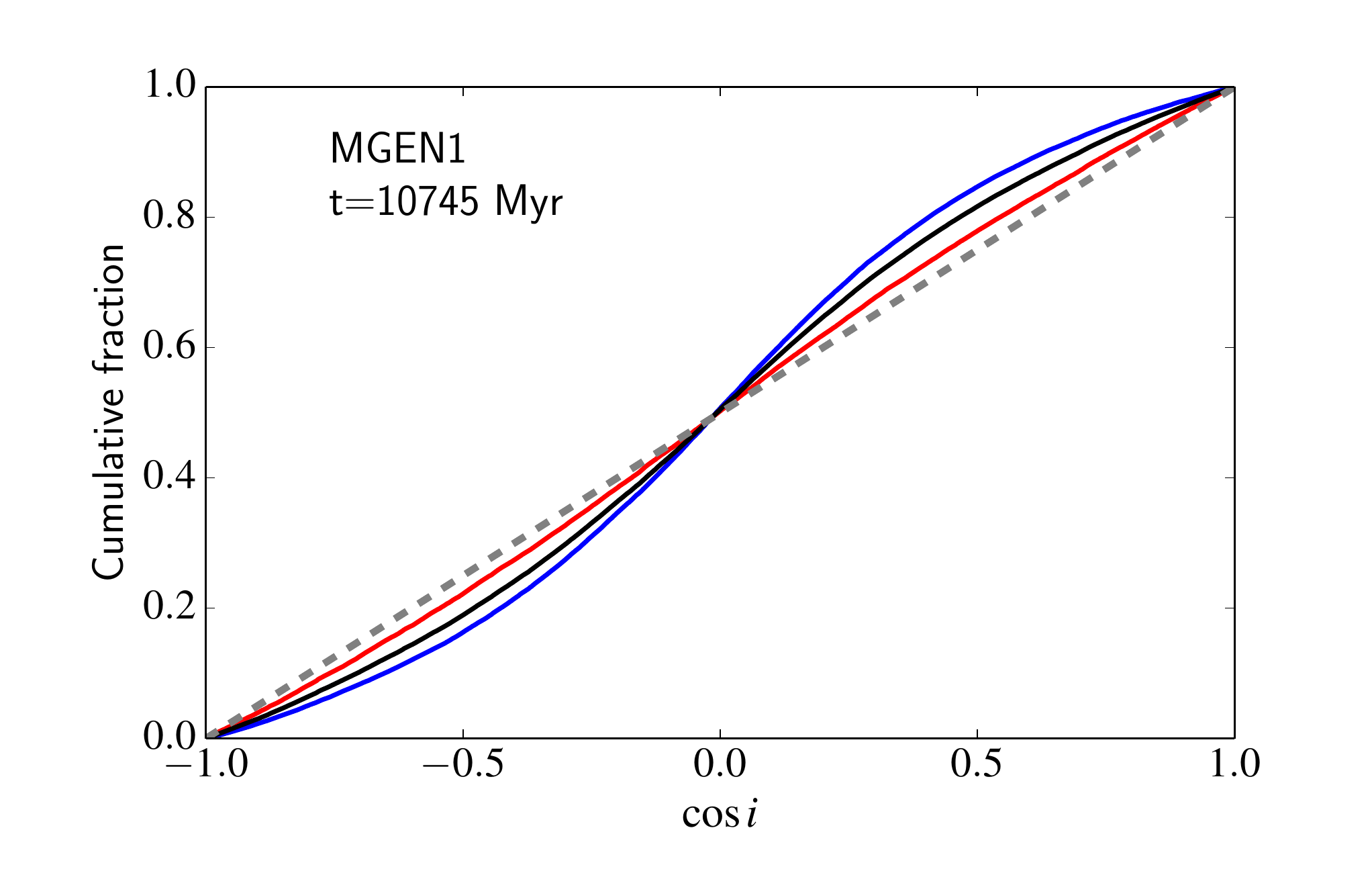}
\caption{\label{cosi} Cumulative fraction of stars as a function of $\cos{i}$ (where $i$ is the position angle of the star with respect to the positive z-axis) for the polluted (blue), pristine (red) and all stars (black) of model {\tt MGEN1} at $t=0$ (top panel) and at the end of the simulation (bottom panel). Stars in the plane of the disc (the $x-y$ plane) have $\cos{i}=0$. The dashed grey line represents an isotropic spatial distribution. }
\end{center}
\end{figure}

\begin{figure}
\begin{center}
\includegraphics[width=0.98\columnwidth]{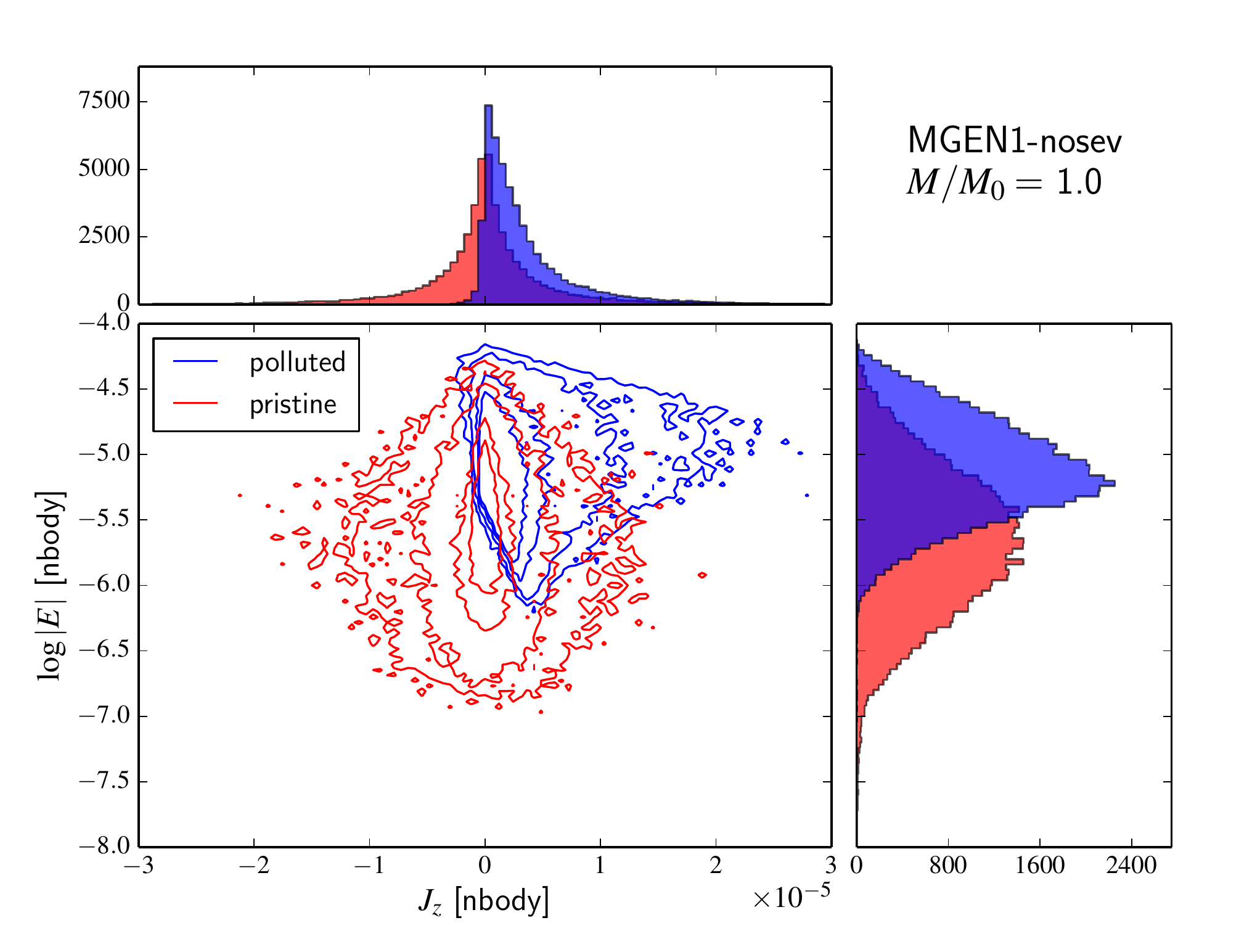}
\includegraphics[width=0.98\columnwidth]{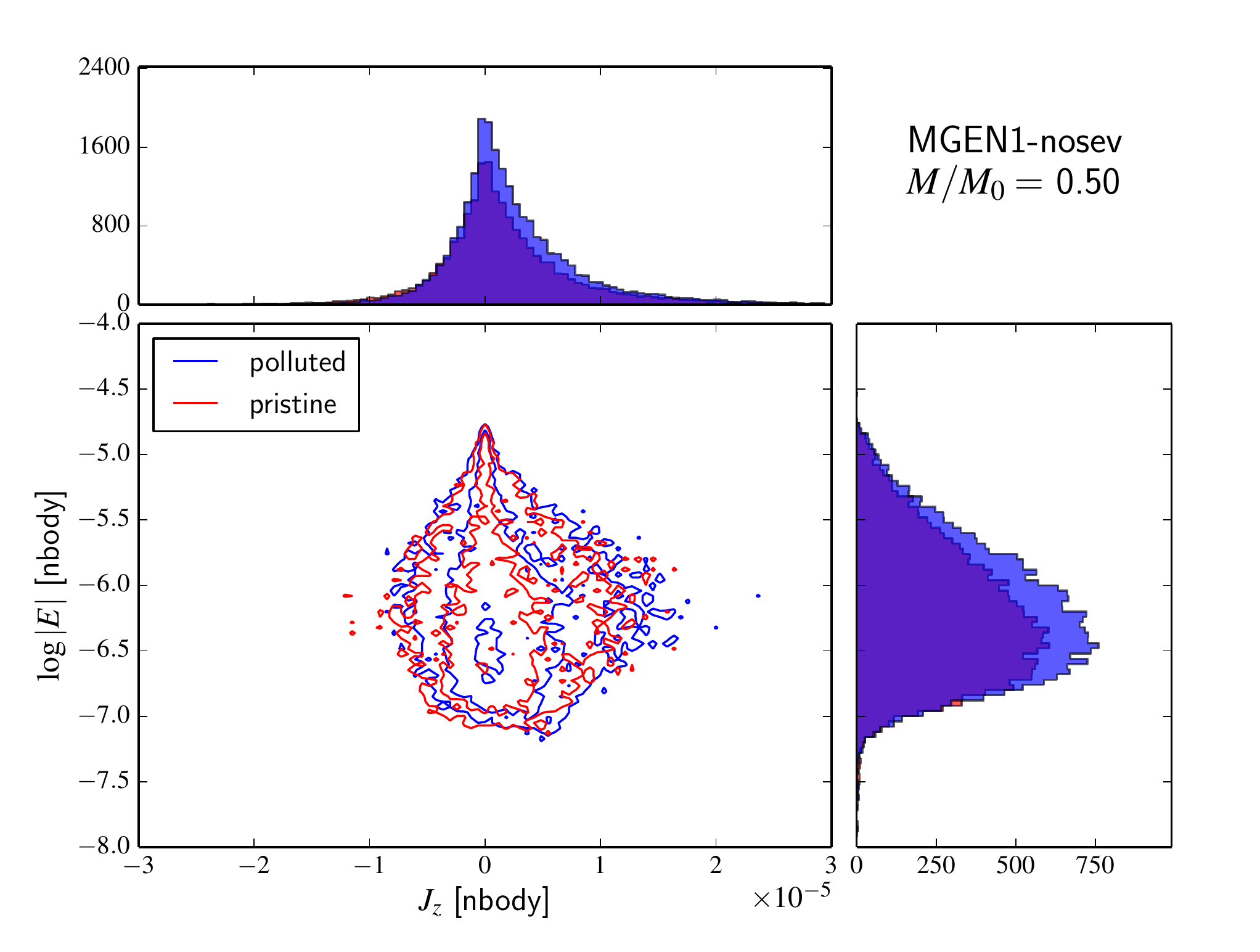}
\includegraphics[width=0.98\columnwidth]{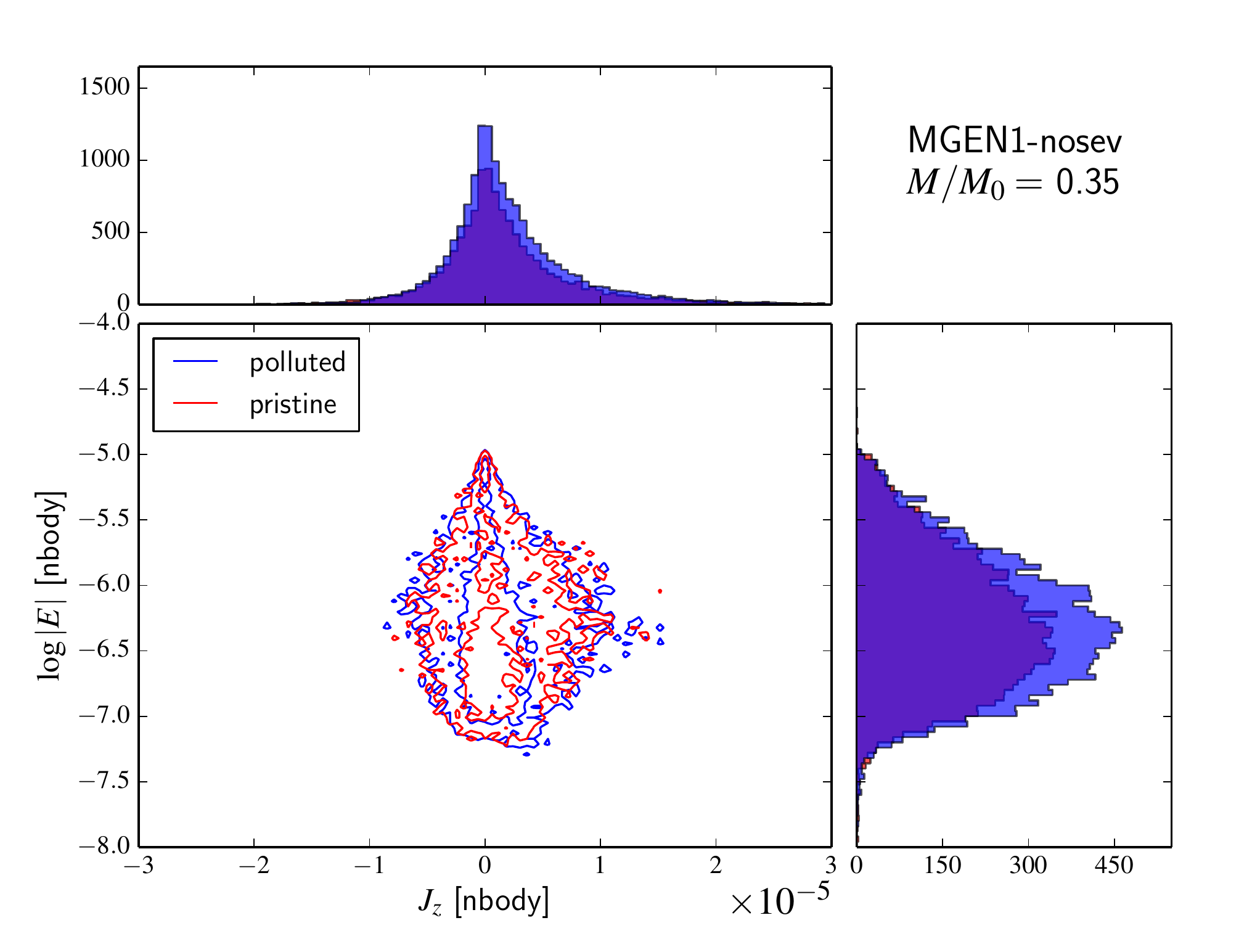}
\caption{\label{EJ_nosev} Distribution of energy and z-angular momentum for the polluted (blue) and pristine (red) stars of model {\tt MGEN1-nosev} at $t=0$ (top panel), when $M/M_0=0.50$ (middle panel) and when $M/M_0=0.35$ (bottom panel). The solid lines represent isodensity contours in the $|E|$ vs. $J_z$ plane for each population.}
\end{center}
\end{figure}

\citet{Vesperini_2013} showed that both internal two-body relaxation and mass loss drive the evolution of the mixing of multiple populations. The relaxation timescale is longer in the outer regions of the cluster, making two-body relaxation (and mixing) less efficient at these larger distances from the centre. This is amplified by the expansion of the cluster which slows down the rate of dynamical ageing with time. Mass loss acts to slow the growth of the relaxation time, which eventually starts to decrease as the remaining mass and the radius of the cluster decrease. The Jacobi radius also decreases as the cluster mass goes down, and the less mixed outer layers are gradually stripped away, which accelerates the evolution towards complete spatial mixing.

Regardless of initial differences in the concentration of the polluted population, \citet{Vesperini_2013} found that clusters always approach a state of complete mixing after losing $60-70\%$ of their mass. This is also what we find in model {\tt MGEN1-nosev}, which was followed until complete dissolution. Figure~\ref{rlag_nosev} shows the evolution of the Lagrangian radii of the polluted and pristine stars for this model as a function of the fraction of the initial mass left ($M/M_0$). All Lagrangian radii for the two populations overlap when $M/M_0 \approx 0.35$.

In the present work, we are mainly interested in the kinematic properties of multiple populations as a way to distinguish between the proposed scenarios. It is therefore important to verify if these kinematic imprints are erased on the same timescale as full spatial mixing occurs. In their study of the long-term dynamical evolution of globular clusters with two stellar populations (one of them being more concentrated initially), \citet{Decressin2008} found that the loss of information on the stellar orbital angular momentum occurs on the same timescale as spatial homogenisation. To check this, we looked at the evolution of the two populations of our models in energy and angular momentum space.

Figure~\ref{EDA1_mix} shows the distribution of energy and z-angular momentum for the two populations of model {\tt ACC1} at the beginning and at the end of the simulation. The energy of a star is defined as $E = \frac{1}{2} m v^2 + m \phi_{\rm c}$, where $\phi_{\rm c}$ - the potential energy due to all other cluster stars - excludes the contribution of the tidal field. One can see the obvious differences in the initial conditions of the two populations for the accretion scenario. Polluted stars typically having a lower energy (i.e. more negative - they are more tightly bound to the cluster due to their smaller distance from the centre and their low velocity). The net angular momentum of this model and the larger angular momentum of the pristine population are also apparent from the distribution of $J_z$. After 10.75 Gyr, differences are still present both in the energy and angular momentum distributions of the two populations.

Figure~\ref{AGB1_mix} shows the same quantities but for model {\tt MGEN1}. Again the polluted stars typically have a lower energy. The angular momentum distribution of the pristine stars is initially symmetric around $J_z=0$, while the polluted stars have an angular momentum distribution skewed towards positive values of $J_z$ for this multiple generations model with net angular momentum. After 10.75 Gyr, the two populations are again not fully mixed in energy and angular momentum space. Some exchange of angular momentum has occurred between the polluted and pristine stars, as the final $J_z$ distribution of pristine stars is slightly skewed towards positive values\footnote{Although we note that tidal effects can also induce some mild rotation. We will come back to this briefly in \S \ref{rot_section}}.

As pointed out by \citet{Mastrobuono_Battisti_2013}, partial relaxation and the exchange of angular momentum between the polluted and pristine populations can have consequences on the spatial morphology of the two populations (and the cluster as a whole) when the polluted population starts as a disc/flattened structure. We illustrate this in Figure~\ref{cosi} by showing the cumulative fraction of stars as a function of $\cos{i}$ (where $i$ is defined as the position angle of the star with respect to the positive z-axis) for the polluted, pristine, and all stars of model {\tt MGEN1} at the beginning and at the end of the simulation. The polluted population has not relaxed to a completely spherical shape (the isotropic distribution is represented by the dashed grey line in Figure~\ref{cosi}) by the end of the simulation. Slight flattening of the pristine population (which started isotropic) and of the cluster as a whole also follows from the exchange of angular momentum between the two populations, although the pristine population is obviously still less flattened that the polluted one. Studying the ellipticity profiles of multiple populations in globular clusters may reveal important clues about their formation. To establish whether this would allow to distinguish between different scenarios, one should however also consider accretion models in which the cluster is initially flattened, but this is beyond the scope of the present work.

Like the two examples shown in Figures \ref{EDA1_mix} and \ref{AGB1_mix}, none of our 47 Tuc-like models are fully mixed in energy and angular momentum space after 10.75 Gyr (see figures in section~\ref{appendix_mixing} of the appendix). To estimate when full mixing occurs, we turn to model {\tt MGEN1-nosev} again. Figure \ref{EJ_nosev} shows the distribution of angular momentum and energy of polluted and pristine stars in this model at three different moments of the evolution of the cluster: when $M/M_0= 1.0$, 0.5, and 0.35. After the cluster has lost 50\% of its initial mass, the distribution of angular momentum of the polluted stars is still skewed towards larger positive values of $J_z$ compared to that of the pristine stars, similar to what was found for model {\tt MGEN1}. The energy distribution of the polluted stars is also skewed towards lower (more negative) values compared to the pristine stars. When $M/M_0=0.35$, the shapes of the $|E|$ and $J_z$ distributions are essentially the same for polluted and pristine stars (up to a scaling factor due to the different number of stars in the two populations). This suggests that kinematic differences between polluted and pristine stars are erased on the same timescale as full spatial mixing occurs.

In the next subsections, we look in more details into the specific kinematic signatures that are expected from each scenario when a cluster is not fully mixed.

\subsection{Velocity dispersion}

\label{sigma_section}

\begin{figure*}
\begin{center}
\includegraphics[width=6.0in]{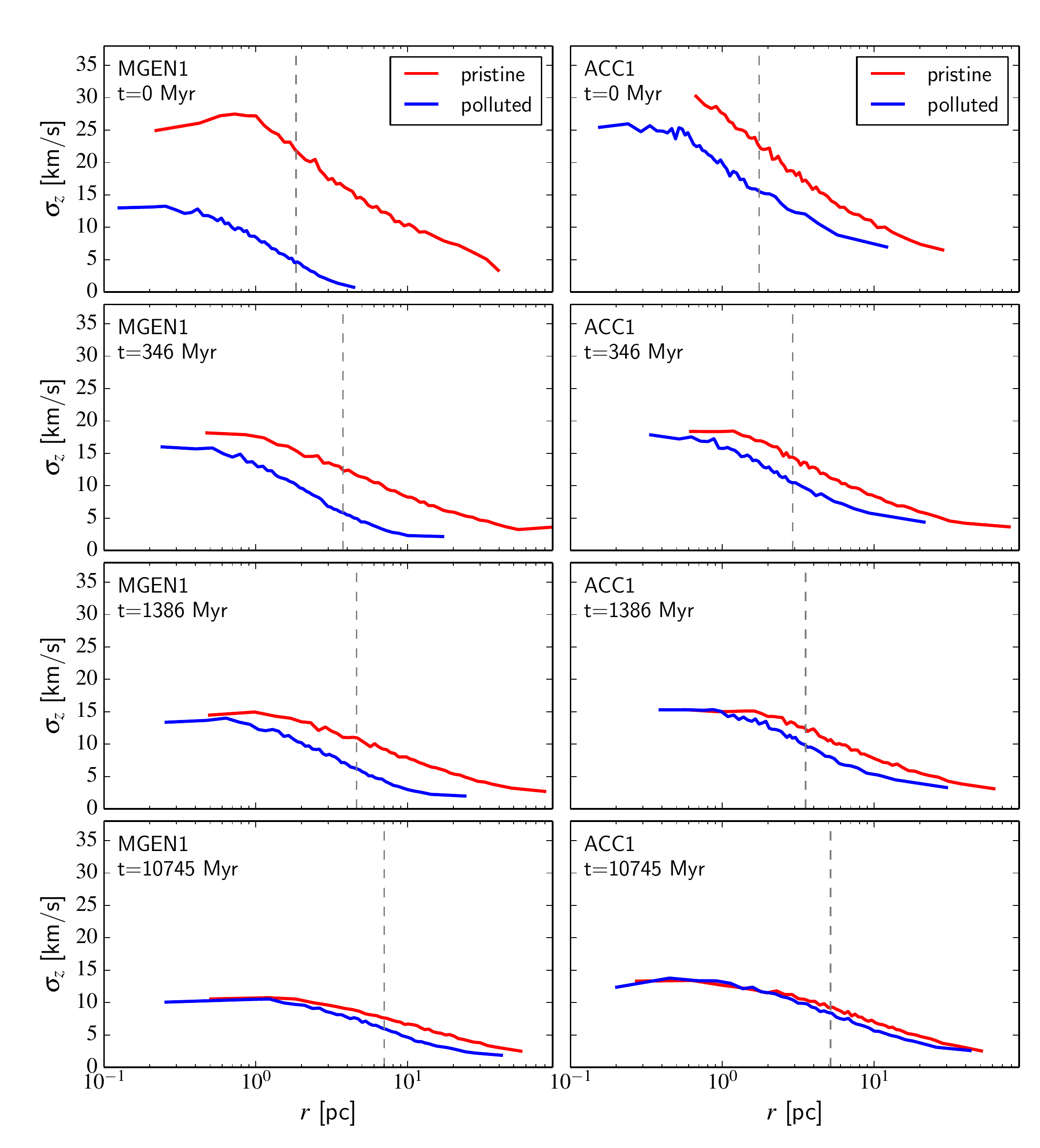}
\caption{\label{sigmaz_evol} Velocity dispersion (z-component) as a function of radius (in 3D) for the polluted (blue) and pristine (red) stars of model {\tt MGEN1} (left panels) and model {\tt ACC1} at different times in the evolution of the cluster. The half-mass radius is indicated by grey dashed lines.}
\end{center}
\end{figure*}

We now briefly discuss the evolution of the velocity dispersion profile of the two populations for our models with stellar evolution. Figure~\ref{sigmaz_evol} shows the time evolution of the velocity dispersion (the $z$-component) profile of the polluted and pristine stars for models {\tt MGEN1} and {\tt ACC1}.

The variation of the two-body relaxation time as a function of radius and time is clearly seen in Figure~\ref{sigmaz_evol}. After only about 1~Gyr, the central $z$-velocity dispersion is the same for the polluted and pristine stars in both models. Nearly 10 Gyr later, the region within which the velocity dispersion profile of the two populations is identical extends further out, but differences of the order of 1 or 2 km~s$^{-1}$ persist in the outer parts of the cluster where the relaxation time is longer. The region where these differences persist appears to extend further in for model {\tt MGEN1}, but when also considering simulations with other values of $\lambda$ and different concentrations of the polluted population initially (see Figures~\ref{sigmaz_evol_0} and \ref{sigmaz_evol_2} in the Appendix), this is not always apparent. It does not appear to be an unambiguous signature that would allow to distinguish between the scenarios.

Note that for a dynamically young system, differences in the velocity dispersion of the polluted and pristine stars may persist even in the inner regions of the cluster. This could explain why the core velocity dispersions of the first and second generation stars are still different after 12~Gyr in the simulations of  \citet{Mastrobuono_Battisti_2013}, as their disc+halo model is tailored to $\omega$~Cen.

As discussed by \citet{Bekki_2010, Bekki2011} and \citet{Mastrobuono_Battisti_2013}, a lower velocity dispersion for the polluted population is one kinematic signature that can be left in a scenario with multiple generations, especially when looking at the component perpendicular to the initial plane of the disc. This is however clearly not unique to such a model, as we find the same signature in the accretion scenario, regardless of the initial value of $\lambda$. Therefore, the lower velocity dispersion reported for the enriched stars in 47 Tuc \citep{Richer_2013, Kucin2014} cannot be used to favour or discard one of the two scenarios considered here.

As mentioned in \S \ref{comp}, the initial $x$ (or $y$) velocity dispersion (i.e. the component in the rotation plane of the disc) of the polluted population can be larger than that of the pristine population in a multiple generations model with net angular momentum. Figure~\ref{sigmax_evol} shows the time evolution of the velocity dispersion (the $x$ component) profile of the polluted and pristine stars for models {\tt MGEN1} and {\tt ACC1} (see Figures~\ref{sigmax_evol_0} and \ref{sigmax_evol_2} in the Appendix for cases with different values of $\lambda$ initially).

At $t=10.75$~Gyr, there is almost no trace of the initially larger $x$ velocity dispersion of the polluted population in model {\tt MGEN1}, as both the polluted and pristine populations have essentially the same $x$ dispersion profile. In model {\tt MGEN2} (Figure~\ref{sigmax_evol_2}), a larger $x$ velocity dispersion for the polluted population is however still apparent at the end of the simulation in the outer parts of the cluster, but the difference is very small ($<1$~\kms).

For the accretion models and the multiple generation models without a disc (i.e. no net angular momentum), which are all spherically symmetric, the $x$ and $y$ components of the velocity dispersion are of the same magnitude as the $z$ component, so the arguments above about the $z$ dispersion still hold for these systems. The velocity dispersion (any component) is always smaller for the polluted (and more concentrated) population in these cases.

\begin{figure*}
\begin{center}
\includegraphics[width=6.0in]{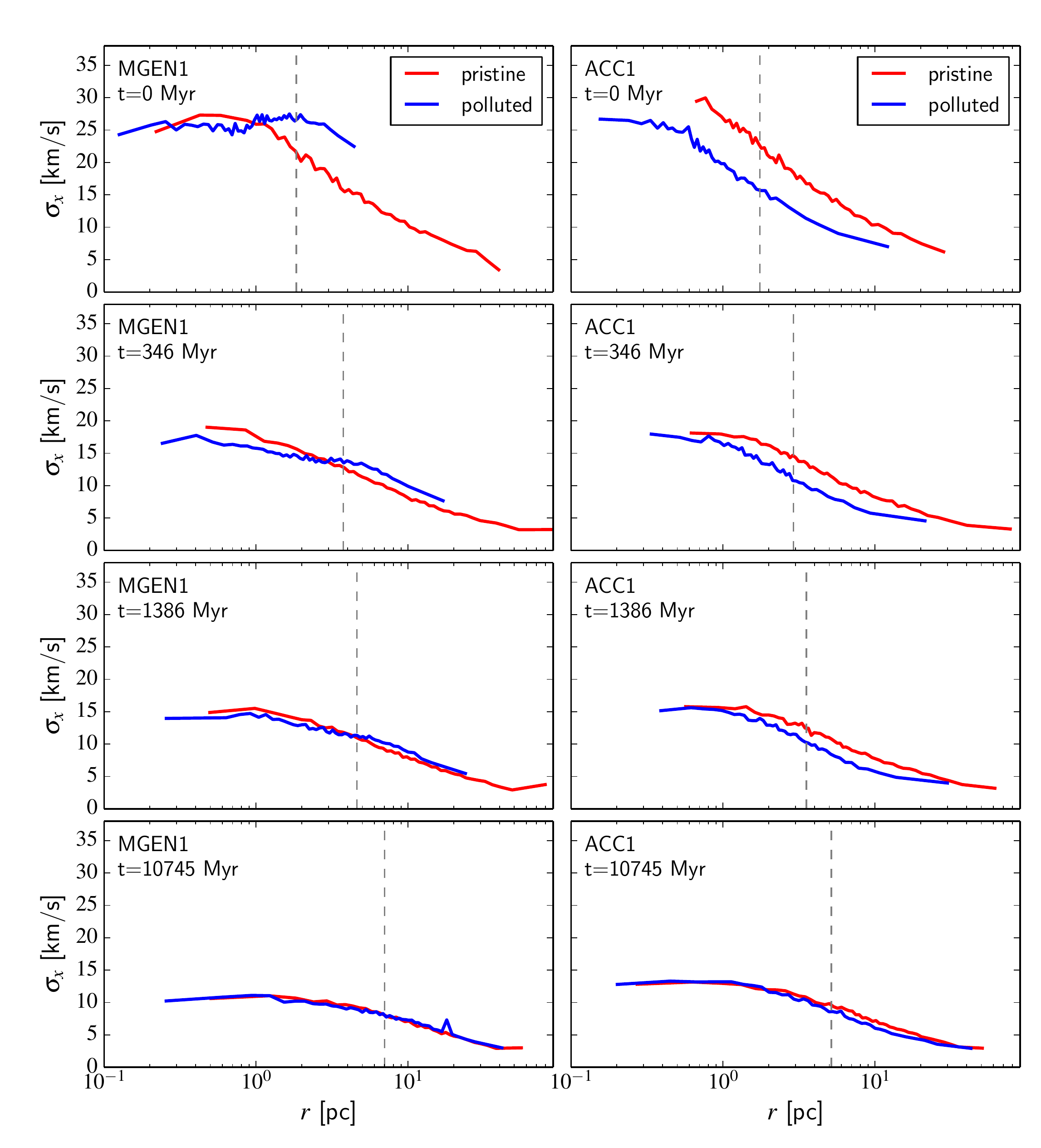}
\caption{\label{sigmax_evol} Velocity dispersion (x-component) as a function of radius (in 3D) for the polluted (blue) and pristine (red) stars of model {\tt MGEN1} (left panels) and model {\tt ACC1} at different times in the evolution of the cluster. The half-mass radius is indicated by grey dashed lines.}
\end{center}
\end{figure*}

\subsection{Velocity anisotropy}

\begin{figure*}
\begin{center}
\includegraphics[width=6.0in]{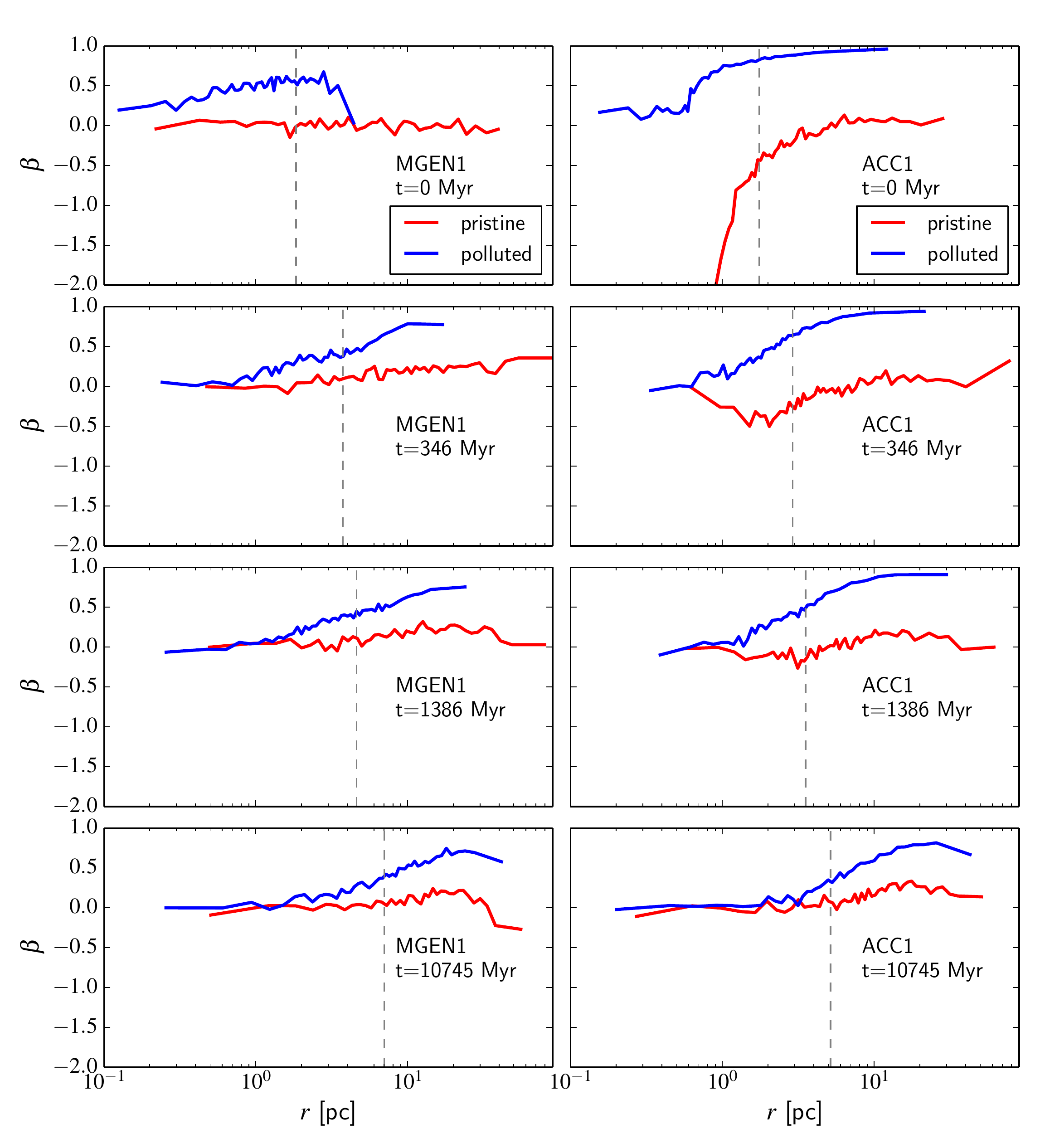}
\caption{\label{beta_evol_1} Velocity anisotropy ($\beta$) as a function of radius (in 3D) for the polluted (blue) and pristine (red) stars of model {\tt MGEN1} (left panels) and model {\tt ACC1} at different times in the evolution of the cluster. The half-mass radius is indicated by grey dashed lines.}
\end{center}
\end{figure*}

\begin{figure*}
\begin{center}
\includegraphics[width=6.0in]{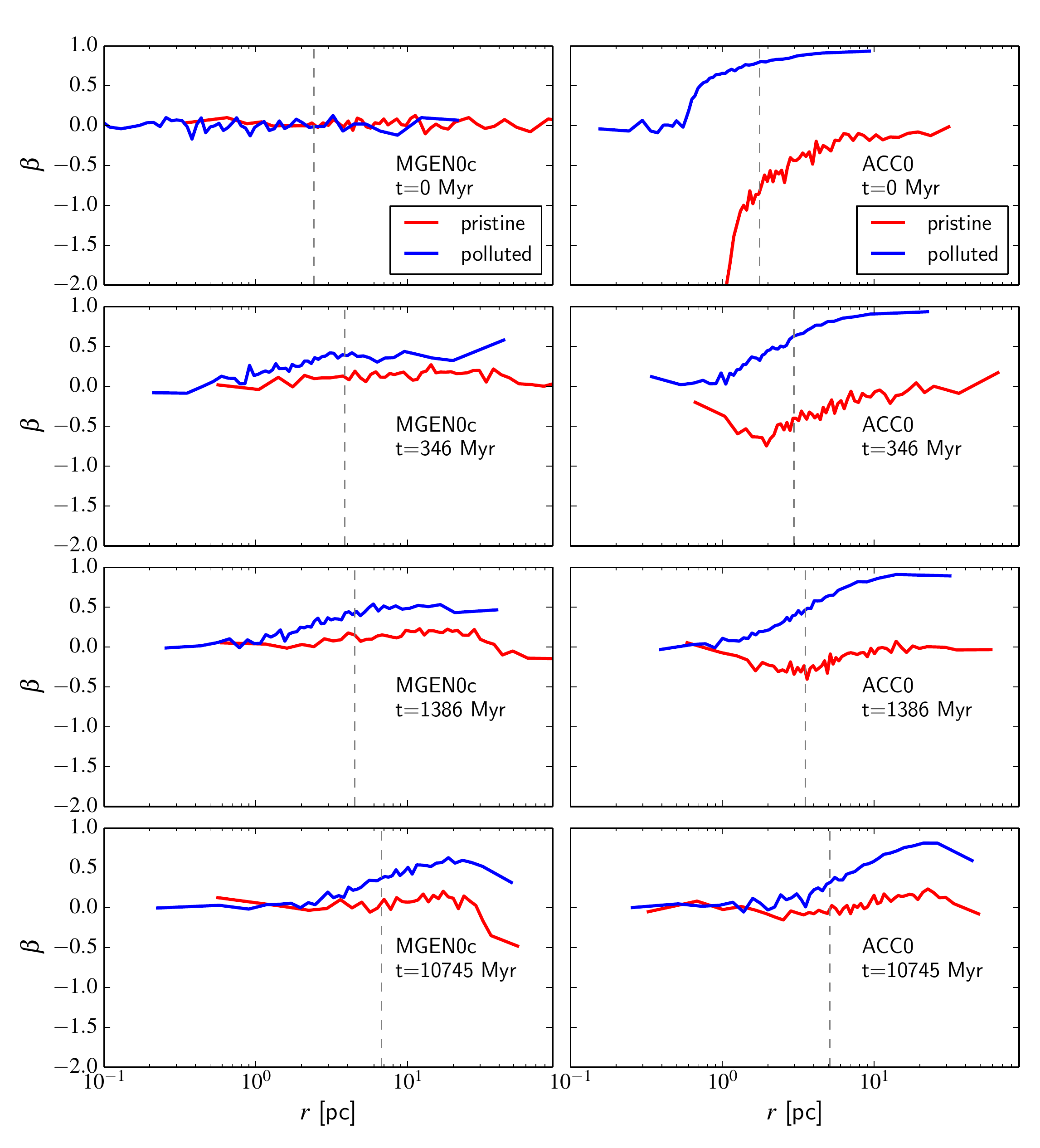}
\caption{\label{beta_evol_0} Velocity anisotropy ($\beta$) as a function of radius (in 3D) for the polluted (blue) and pristine (red) stars of model {\tt MGEN0c} (left panels) and model {\tt ACC0} at different times in the evolution of the cluster. The half-mass radius is indicated by grey dashed lines.}
\end{center}
\end{figure*}

Motivated by the finding of a radially anisotropic polluted population in 47 Tuc \citep{Richer_2013}, we now look at the evolution of the velocity anisotropy profile of the polluted and pristine stars in the accretion and multiple generations scenarios. Figure~\ref{beta_evol_1} shows the time evolution of this profile for each of the two populations in models {\tt MGEN1} and {\tt ACC1}.

In both models, the polluted population stays more radially anisotropic than the pristine population in the outer parts of the cluster. The anisotropy profile of the polluted stars is characterised by an inner isotropic core followed by radial anisotropy that increases with radius. In the late stages, the radial anisotropy peaks at an intermediate distance from the centre and then decreases outwards, as expected when the effect of tides becomes important and stars on radial orbits are preferentially lost \citep[e.g.][]{bm2003, fuku2000} and gain angular momentum due to the tides \citep{OhLin1992}. After 10.75 Gyr, the pristine population is also isotropic in the inner parts (over a larger radial extent than the polluted population) and shows some very mild radial anisotropy at larger radii, followed by a decrease of $\beta$ and isotropy or mild tangential anisotropy in the outermost regions.

These conclusions do not change in models with different values of $\lambda$. Even when $\lambda=0$ and both populations of a model with multiple generations start fully isotropic across the whole cluster, the more rapidly expanding (and initially more centrally concentrated) polluted population rapidly develops stronger radial anisotropy in the outer parts compared to the pristine population. The kinematic signature at the end of the simulation in this case is remarkably similar to all of our other models (Figure~\ref{beta_evol_0}). A more radially anisotropic polluted population is therefore not a unique signature of either scenarios, but another natural consequence of starting with a more centrally concentrated polluted population.

\subsection{Rotation}
\label{rot_section}

\begin{figure*}
\begin{center}
\includegraphics[width=6.0in]{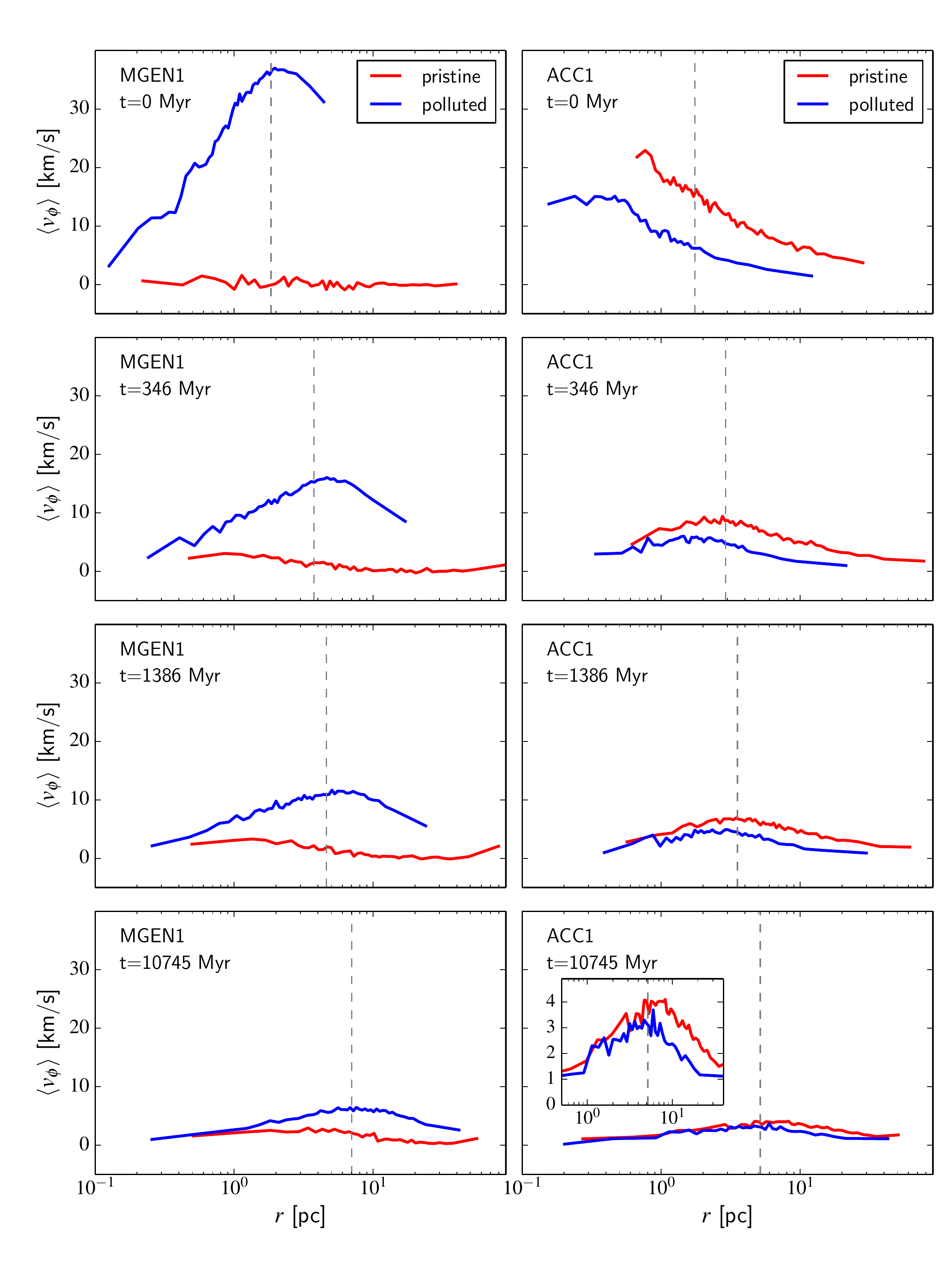}
\caption{\label{vphi_evol_1} Mean azimuthal velocity as a function of radius (in 3D) for the polluted (blue) and pristine (red) stars of model {\tt MGEN1} (left panels) and model {\tt ACC1} at different times in the evolution of the cluster. The half-mass radius is indicated by grey dashed lines. The inset plot in the bottom right panel shows a zoomed view of a portion of the rotation curve for the final snapshot of model {\tt ACC1}.}
\end{center}
\end{figure*}

We now look at the long-term evolution of the bulk rotation of pristine and polluted stars, a promising diagnostic given the very distinct initial conditions of the accretion and multiple generations scenarios for this kinematic property (Figure~\ref{IC_snapshots}). To illustrate the differences between the two scenarios as a function of time, Figure~\ref{vphi_evol_1} shows the evolution of the mean azimuthal velocity profile for the two populations in models {\tt MGEN1} and {\tt ACC1}.

As the clusters evolve due to two-body relaxation and lose mass, angular momentum is transported outwards and the total angular momentum decreases. The peak of the rotation curve (located near the half-mass radius) also moves outward and its amplitude decreases. In all models with net rotation, as mixing progresses, the mean azimuthal velocity of the pristine stars eventually matches that of the polluted stars in the inner regions. The rotational amplitude in these inner regions is very low after 10.75 Gyr, but a small velocity gradient (of the order of 1~km\,s$^{-1}$\,pc$^{-1}$) is still present across the core of the cluster, even in post-core collapse clusters (for example model {\tt MGEN1}, see Figure~\ref{rlag}). This suggests that observational evidence for a velocity gradient/rotation in the inner parts of a cluster cannot be used to argue that the cluster is in the pre-core collapse phase of its evolution \citep[c.f.][]{Fabricius2014}.

In all our models with net rotation, the differential rotation of the pristine and polluted stars is maintained in the outer parts of the cluster after 10.75 Gyr of dynamical evolution. For the multiple generations scenario, the polluted stars rotate faster than the pristine stars, while the opposite is true for the accretion scenario. While the rotation curves of the two populations become indistinguishable in the inner regions of the clusters, differences of the order of $1-2$~km~s$^{-1}$ or more persist in the outer parts\footnote{As discussed in \S \ref{iso}, taking into account the effect of gas  accretion on individual stellar orbits would likely enhance the amplitude of the differential rotation between pristine and polluted stars in the accretion scenario.}. A larger initial value of $\lambda$ yields larger differences in the rotation curves of pristine and polluted stars (see Figure~\ref{vphi_evol_2} in the appendix).

In models {\tt MGEN1} and {\tt MGEN2} (Figures~\ref{vphi_evol_1} and \ref{vphi_evol_2}), the exchange of angular momentum from the polluted to the pristine population is seen through the subtle growth of the rotational amplitude of the pristine population (which has zero angular momentum initially) in the inner parts of the clusters. The rotation curve of the pristine stars also rises very slightly in the outer parts of these clusters as they evolve, showing a gain of weak prograde rotation at the level of $\sim1$~km~s$^{-1}$. This effect is also seen in our models with no initial rotation (Figure~\ref{vphi_evol_0}), but the effect remains small and is always confined to regions beyond the tidal radius of the cluster. Tidal effects are expected to cause a preferential loss of stars on prograde orbits \citep{Henon1970, Keenan1975, OhLin1992, fuku2000}, so we are most likely witnessing the stripping of stars on prograde orbits when seeing an increasing prograde rotation outside of the tidal radius. In any case, for our models with initial rotation, such a tidally induced rotational signature never compromises the clear differential rotation signal between the polluted and pristine stars.

\section{Prospect of detecting the identified signatures}
\label{observables}

Based on the results of the previous section, comparing the velocity dispersion of the polluted and pristine populations may offer a way to distinguish between the scenarios kinematically, although the expected signal is weak. A larger velocity dispersion for the polluted population in the plane of rotation can be explained by a multiple generations model with net angular momentum, but not by the accretion model nor any model without net rotation initially.  Observationally, this component of the velocity dispersion is best probed with radial velocities if the rotation axis is perpendicular to the line of sight, but best probed with proper-motion data if the rotation axis is along the line of sight.

If a larger velocity dispersion is measured for the polluted population in a given dataset, the multiple generations model would be favoured and the same data should reveal a larger rotational amplitude for this polluted population. On the other hand, if a lower velocity dispersion is measured for the polluted population, it could mean that the accretion model prevails but could also be because the inclination of the rotation axis is not favourable enough to probe the velocity dispersion in the plane of rotation. In that sense, the velocity dispersion is perhaps not the most powerful probe of the formation scenario. On the other hand, it is clear that the differential rotation of polluted and pristine stars represents a unique kinematic signature that can help to constrain scenarios for the formation of multiple populations in GCs.

The expected strength of the signal in a cluster like 47 Tuc with a moderate amount of angular momentum initially is a few km~s$^{-1}$ or less. This could be observable with precise ground-based radial velocity measurements from large high-resolution spectroscopic samples of RGB stars (bearing in mind inclination effects). As discussed in \S \ref{prev_dyn}, tantalising results in that respect have been obtained by \citet{bellazzini_2012}, but they will need to be confirmed with larger datasets or at least with a more in-depth statistical analysis. Interestingly, in the three clusters for which these authors report different rotational amplitudes for the Na-rich and Na-poor samples, the pristine (Na-poor) stars always display a larger rotational amplitude, as expected in the accretion scenario.

The Gaia satellite will also allow to obtain proper motions with a precision better than 2 km~s$^{-1}$ for RGB stars in the outskirts of GCs, which could be sufficient to see a difference in the rotation of polluted and pristine stars in some clusters. Gaia will not be helpful for proper motions in the dense cores of GCs, but for the application discussed here, this is not a problem because the signal is expected to be stronger in the outer parts of clusters, where the two-body relaxation timescale is longer. Proper motions from HST \citep[e.g.][]{Richer_2013} may also be useful to uncover the expected signal. In this case, the analysis would be made easier by the differential nature of the signal, which eliminates the need for determining an absolute reference frame and avoids the associated complications, especially when the field-of-view of the observations is small \citep[e.g.][]{Anderson2003}.

\begin{figure}
\begin{center}
\includegraphics[width=\columnwidth]{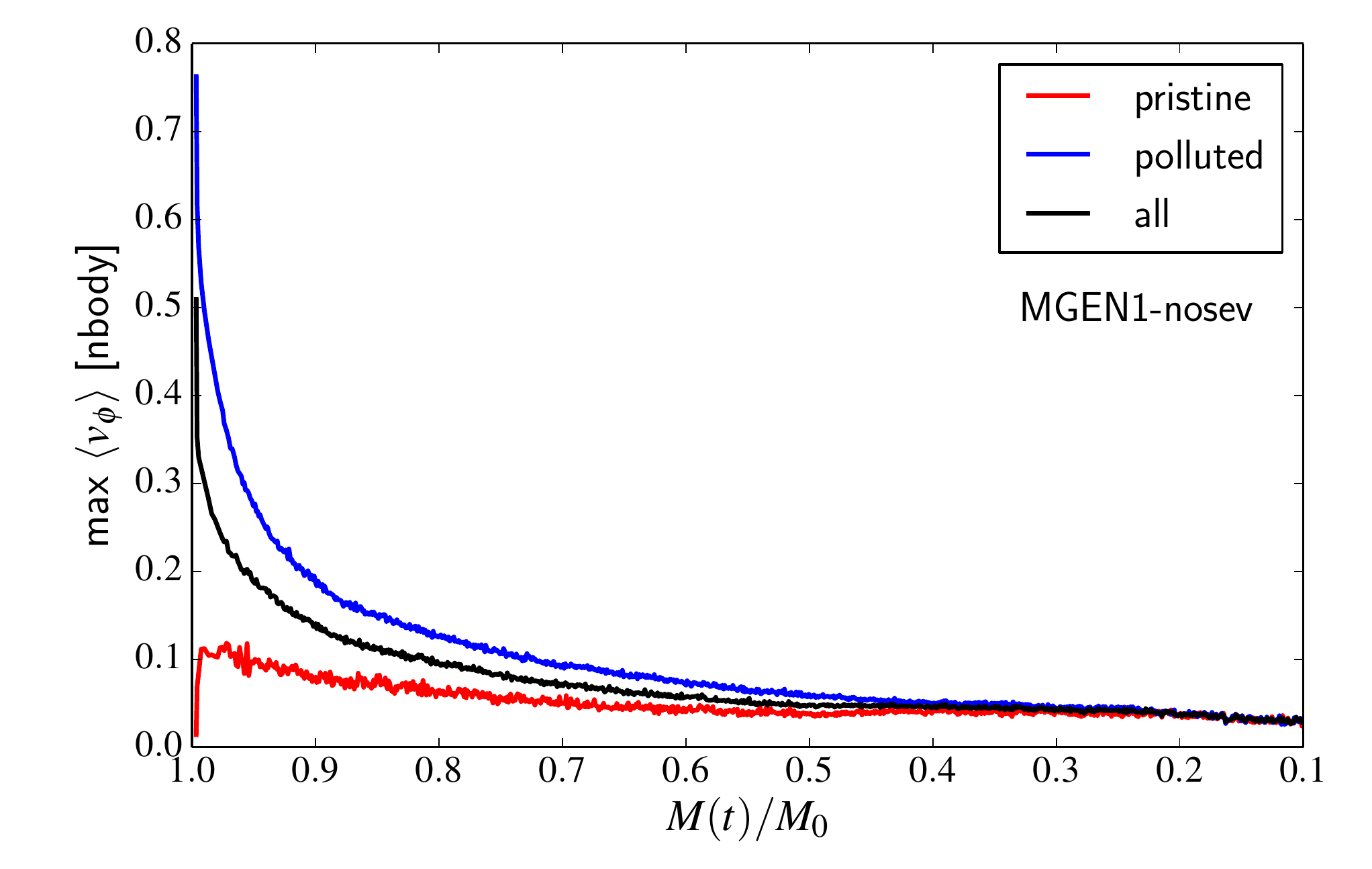}
\caption{\label{max_rot_M} Amplitude of the rotation curve (i.e. peak height of the mean azimuthal velocity profile) for the polluted (blue lines), pristine (red lines) and all stars (black lines) as function of the fraction of the initial mass left in the cluster for model {\tt MGEN1-nosev}.}
\end{center}
\end{figure}

\begin{figure}
\begin{center}
\includegraphics[width=\columnwidth]{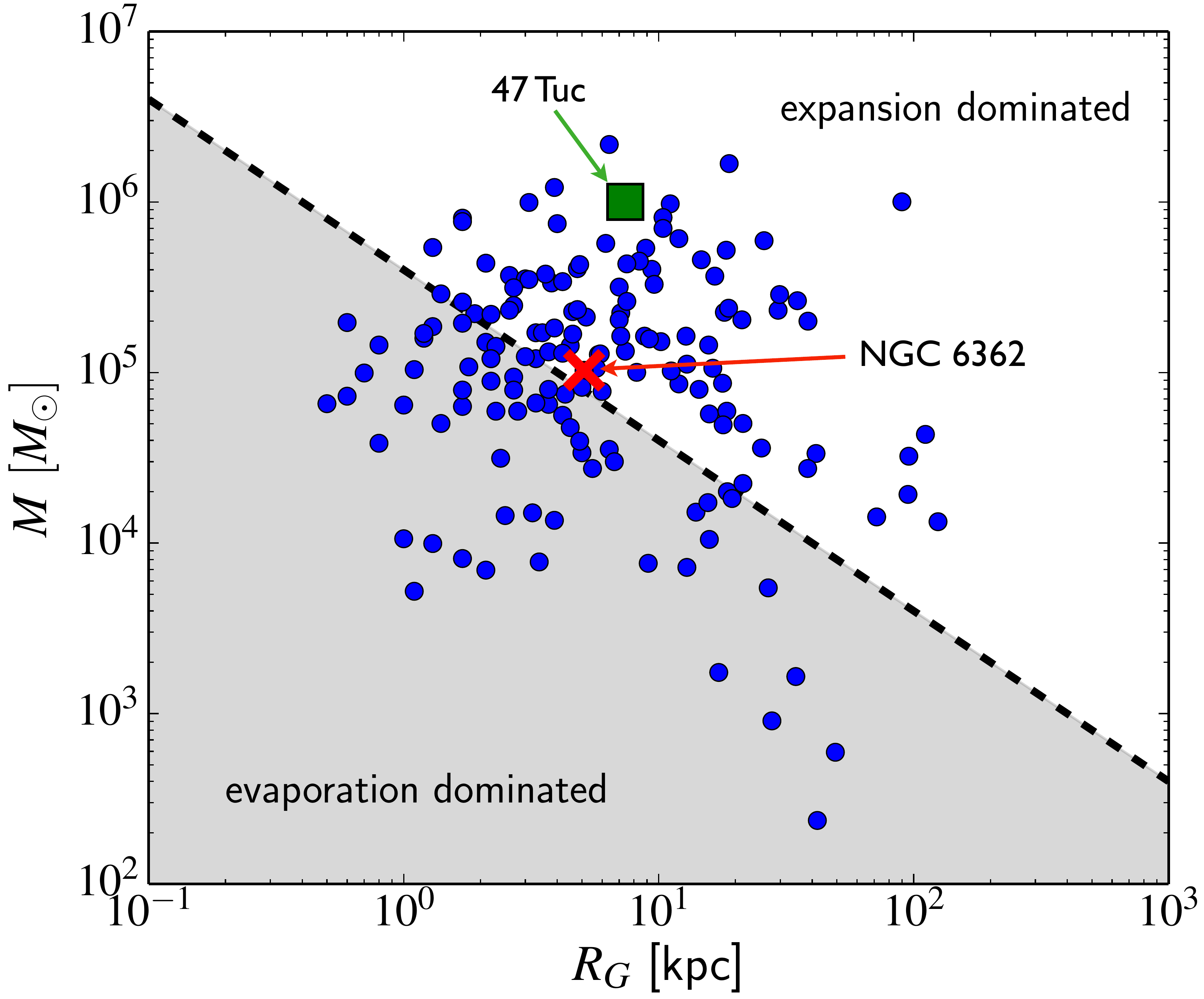}
\caption{\textbf{\label{harris}} Distribution of stellar mass and galactocentric radius of Galactic GCs from the 2010 version of the Harris catalogue. Clusters above the dashed line are presumably in the expansion-dominated phase of their evolution and not fully mixed, while clusters below the line are expected to be in the evaporation-dominated phase. 47~Tuc is represented by a green square, and the fully spatially mixed cluster NGC\,6362 by a red cross.}
\end{center}
\end{figure}

To estimate the fraction of the Galactic population of GCs in which we can hope to detect differential rotation of the pristine and polluted stars, we must identify the clusters for which the different populations are not expected to be fully mixed. As we argued previously, these are the clusters which have lost less than $\sim60-70\%$ of their initial mass (during the long-term evolution driven by two-body relaxation). Figure~\ref{max_rot_M} shows that the amplitude of the rotation curve is the same for the polluted and pristine stars of model {\tt MGEN1-nosev} when the cluster has lost 70\% of its mass or more, supporting the idea that the differential rotation signature would be erased beyond that stage.

To identify the parameters of the clusters which are more likely to have lost a large fraction of their mass, we follow \citet{gieles2011}, who presented a simple description of the life cycle of initially compact clusters in a tidal field. They showed that the half-mass radius of a cluster increases during (roughly) the first half of its evolution (the so-called `expansion dominated' phase) and decreases during the second half (the `evaporation dominated' phase). Clusters in the evaporation-dominated phase have lost about 50\% of their mass or more. From their evolution equations and adopting a few simplifying assumptions (e.g. circular cluster orbits in a constant Milky Way potential), \citet{gieles2011} showed that clusters satisfying the relation

\begin{equation}
M \gtrsim 10^5 \ {\rm M}_{\odot} \left( \frac{4 \ {\rm kpc}}{R_{\rm G}} \right)
\end{equation}

\noindent{should (to first order) be in the expansion-dominated phase, in which case we do not expect them to be fully mixed. In Figure~\ref{harris}, we show clusters in the $M - R_{\rm G}$ plane using data from the 2010 version of the Harris catalogue\footnote{http://www.physics.mcmaster.ca/Globular.html} \citep{Harris1996}, where we adopted a mass-to-light ratio of 2 to convert luminosities to masses \citep{MvM2005}. There is a large sample of Galactic GCs (roughly half of the data points) with large enough masses and galactocentric radii to belong to the group of expansion-dominated clusters. This is where the kinematic signature identified in the present work should be looked for. 47 Tuc is found in this region of the $M - R_{\rm G}$ plane, which is consistent with the kinematic differences reported between its pristine and polluted stars \citep{Richer_2013, Kucin2014}. On the other hand, the lower-mass cluster NGC~6362, the first GC found to be fully spatially mixed \citep{Dalessandro2014}, is interestingly very close to the boundary between the expansion-dominated and evaporation-dominated regimes.

\section{Conclusions}
\label{conclusions}

We focused on identifying a unique long-term kinematic imprint that would allow to distinguish between different scenarios for how enriched material makes its way into stars in the context of multiple stellar populations in GCs. To do so, we presented a suite of $N$-body simulations of the long-term dynamical evolution of globular clusters, with initial conditions chosen to capture the distinct kinematic properties of two main pollution scenarios.

The first scenario (the {\it multiple generations} scenario) is one in which gas collects in a cooling flow into the core of the cluster, where a new generation of stars is eventually formed. In the other scenario (the {\it accretion scenario}), a fraction of the stars accrete material when passing through the core of the cluster, with no need for multiple star-formation events. While in the current paradigm these scenarios are naturally associated to the models put forward by \citet{DErcole_2008} and \citet{Bastian_2013} and the specific source of polluting material favoured by these authors (AGB stars and massive interacting binaries, respectively), our study does not concern the chemistry and our results would apply to scenarios that proceed in the same way dynamically, regardless of the source of pollution. 

We showed that the two scenarios imply different initial conditions for the kinematics of the various populations. The most striking difference between the initial conditions of the accretion and multiple generations scenarios arises in the presence of cluster rotation. As a result of dissipative processes driving the polluted material to higher densities towards the centre of the cluster and conservation of angular momentum, an initially larger rotational amplitude for the polluted stars compared to the pristine stars is expected from the multiple generations scenario. In the accretion scenario, the polluted stars (the ones crossing the core) are preferentially on radial orbits and their initial rotational amplitude is expected to be smaller than that of the pristine stars.

We showed that initial differences in the kinematics of pristine and polluted stars can survive for a Hubble time of relaxation-driven dynamical evolution in a cluster with properties similar to those of 47 Tuc, especially in the outer parts of the cluster where the two-body relaxation timescale is longer. In particular, some differential rotation of the pristine and polluted stars is preserved, which means that this unique signature can potentially be used to distinguish between the pollution scenarios considered. The expected strength of the differential rotation signal is typically a few km~s$^{-1}$ in the outer parts (the exact value depends on the specific scenario, the initial amount of angular momentum of the cluster, the radius at which this is measured, and on the degree of mixing - i.e. the dynamical age of the cluster). This is challenging to measure but within the reach of current and future surveys of the internal kinematics of GCs (either through radial velocities or proper motions).

Initial differences in the velocity dispersion and velocity anisotropy profiles of polluted and pristine stars are also expected to survive for a Hubble time in a 47 Tuc-like cluster. These kinematic imprints are however remarkably similar in the accretion and multiple generations scenarios. A lower velocity dispersion and a more radially anisotropic velocity distribution for the polluted stars (common signatures of both scenarios) can therefore not be used to distinguish the two scenarios. That said, the multiple generations scenario could be favoured if a larger velocity dispersion is measured for the polluted population. This would stem from the initially larger velocity dispersion of this population in the plane of rotation.

We argued that full kinematic mixing of the two populations happens on the same timescale as full spatial mixing. It is achieved when $\sim60-70\%$ of the initial mass of the cluster has been lost. We identified a large subsample of Galactic GCs having large enough masses and galactocentric radii to still be in the expansion-dominated phase of their evolution \citep{gieles2011}. Clusters in this phase are still expanding within their tidal boundary and have not suffered substantial tidal evaporation. They are thus good candidates for observing the signatures of incomplete spatial and kinematic mixing, in particular the differential rotation of pristine and polluted stars.

We should mention that the present work does not encompass all suggested frameworks for the formation of multiple populations. For example, \citet{maxwell2014} suggested that GCs formed in the inner regions of dwarf galaxies can accrete gas on each passage through the centre of these galaxies. The accreted matter, eventually forming a second generation within the cluster, would in this case be a mix of pristine material from the intergalactic medium and surrounding dwarf, plus moderate velocity AGB ejecta retained by the potential well of the galaxy. These GCs would be pushed out onto larger orbits as star formation progresses, and may be stripped during mergers. While it avoids some of the problems of the other models, this scenario cannot explain the presence of multiple stellar populations in the Galactic GCs thought to form {\it in situ} \citep[e.g.][]{Forbes2010, Leaman2013}. It would however be interesting to see if this scenario can leave a characteristic kinematic signature. Similarly, we have not investigated the implications of another scenario in which second-generation stars are proposed to form in the decretion discs of fast rotating massive stars \citep{Krause2013}.

Finally, there is definitely scope to improve the toy models that we have set up to study the kinematics of multiple populations in different scenarios. For example, the effect of gas accretion on the orbits of stars could be included in a self-consistent way. It would also be interesting to consider the effect of the different initial helium content of the polluted and pristine stars on their evolution, and indirectly on the dynamical evolution of the cluster.

\section*{Acknowledgements}
We thank the anonymous referee for a constructive report. We also thank Nate Bastian for useful comments, and Florent Renaud for pointing us to the code {\tt magalie}. VHB is grateful to the ``Fonds the recherche du Qu\'ebec - Nature et technologies" (FRQNT) for financial support through a postdoctoral fellowship grant. MG thanks the Royal Society for financial support in the form of a University Research Fellowship (URF) and the European Research Council (ERC-StG-335936, CLUSTERS).

\bibliographystyle{mn2e}


\appendix
\section{Supplementary figures (initial conditions)}
\label{appendix_ICs}

This section presents two additional figures to complement \S \ref{IC_section} and Figure~\ref{IC_snapshots} and illustrate initial conditions for models without net angular momentum initially ({\tt MGEN0c} and {\tt ACC0}; Figure~\ref{IC_snapshots_AGB0C_EDA0}), and for models with a larger initial amount of angular momentum ({\tt MGEN2} and {\tt ACC2}; Figure~\ref{IC_snapshots_AGB2_EDA2}).

\begin{figure*}
\begin{center}
\includegraphics[width=7.0in]{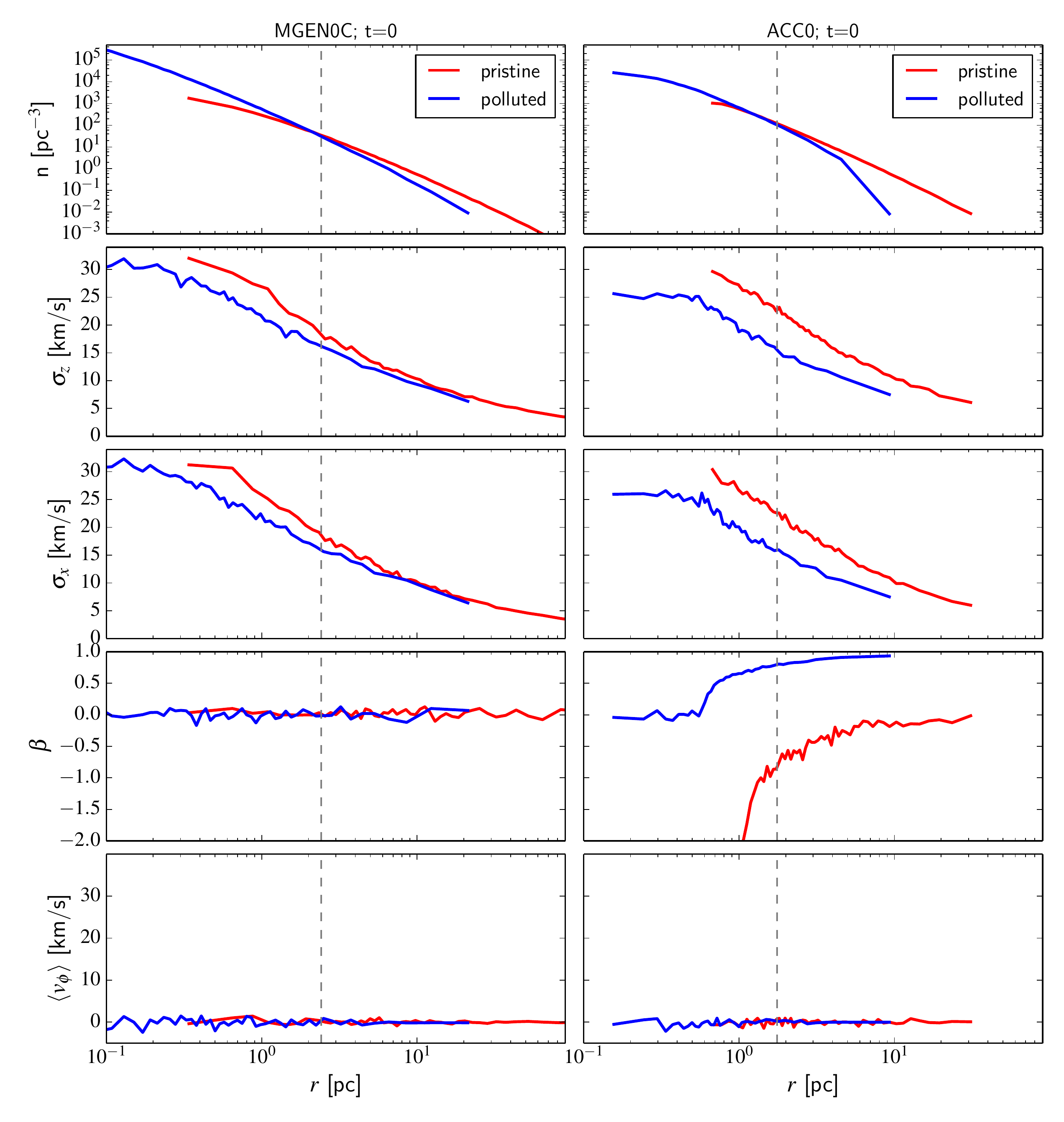}
\caption{\label{IC_snapshots_AGB0C_EDA0} Same as Figure~\ref{IC_snapshots}, but comparing the initial conditions of models {\tt MGEN0c} and {\tt ACC0}.}
\end{center}
\end{figure*}

\begin{figure*}
\begin{center}
\includegraphics[width=7.0in]{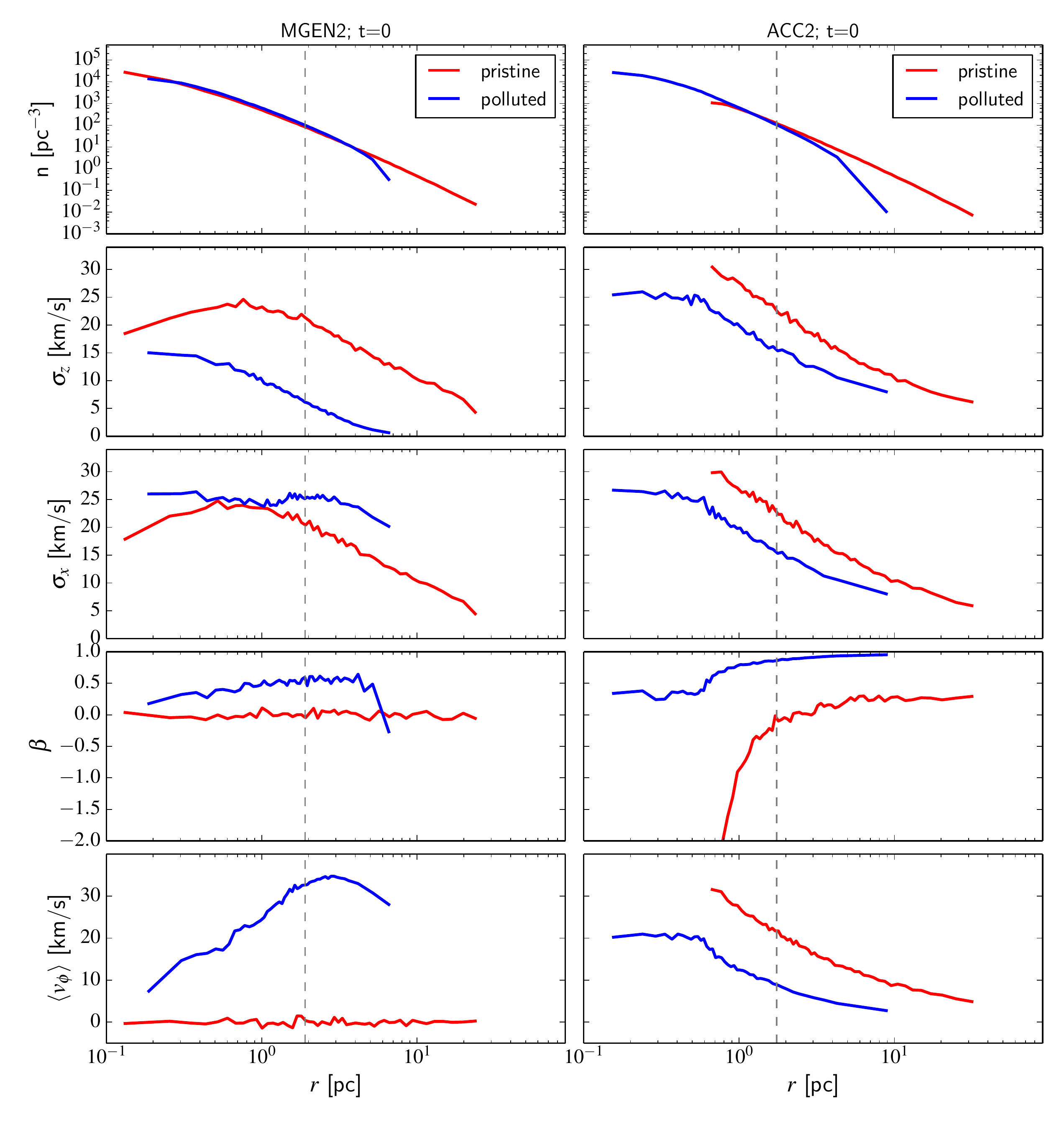}
\caption{\label{IC_snapshots_AGB2_EDA2} Same as Figure~\ref{IC_snapshots}, but comparing the initial conditions of models {\tt MGEN2} and {\tt ACC2}.}
\end{center}
\end{figure*}

\clearpage

\section{Supplementary figures (phase-space mixing)}
\label{appendix_mixing}

The additional figures presented in this section show, for the models not illustrated in \S \ref{psmix}, the initial and final energy/z-angular momentum distributions of the pristine and polluted populations.

\begin{figure}
\begin{center}
\includegraphics[width=\columnwidth]{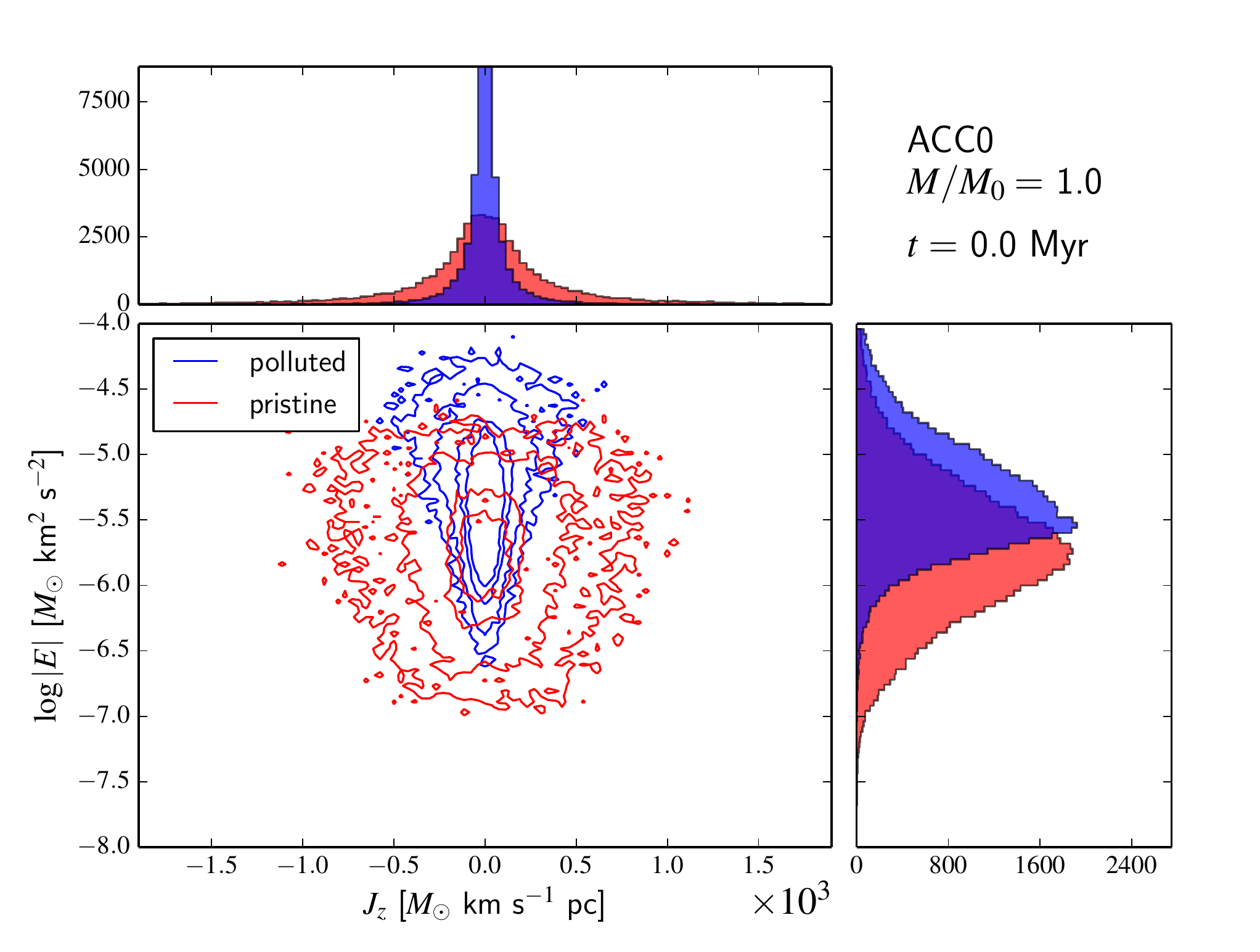}
\includegraphics[width=\columnwidth]{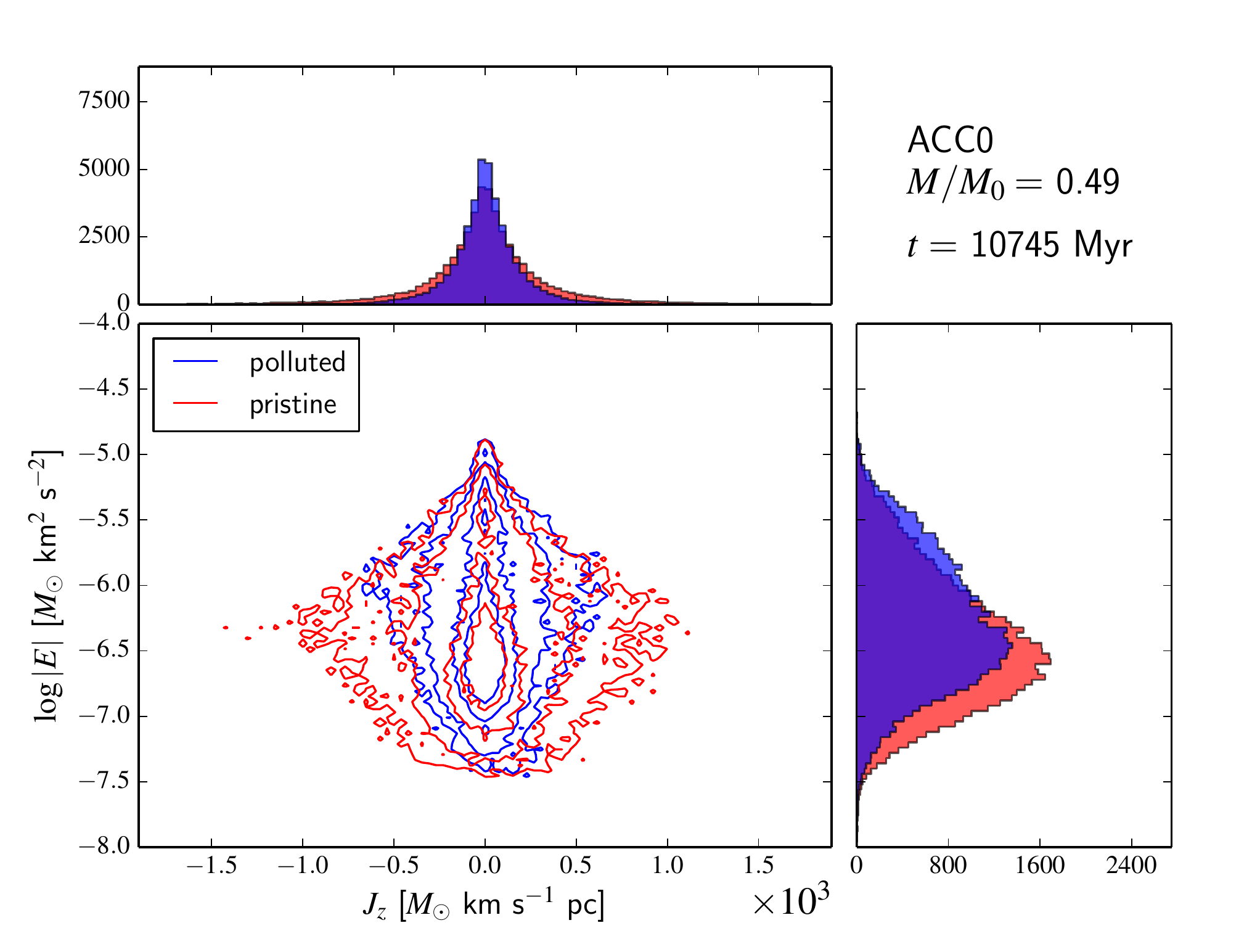}
\caption{Same as Figures \ref{EDA1_mix} and \ref{AGB1_mix} but for model {\tt ACC0}.}
\end{center}
\end{figure}

\begin{figure}
\begin{center}
\includegraphics[width=\columnwidth]{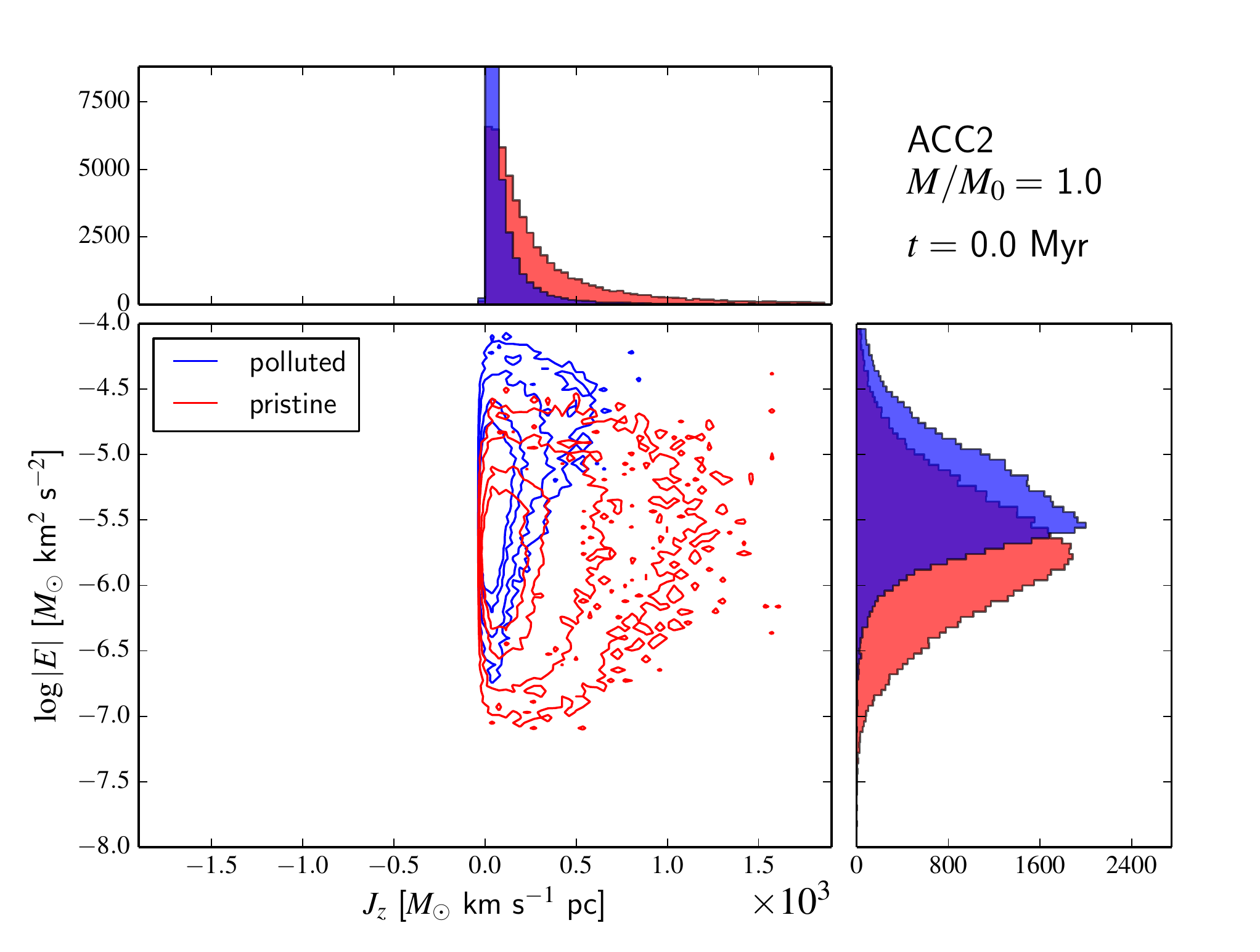}
\includegraphics[width=\columnwidth]{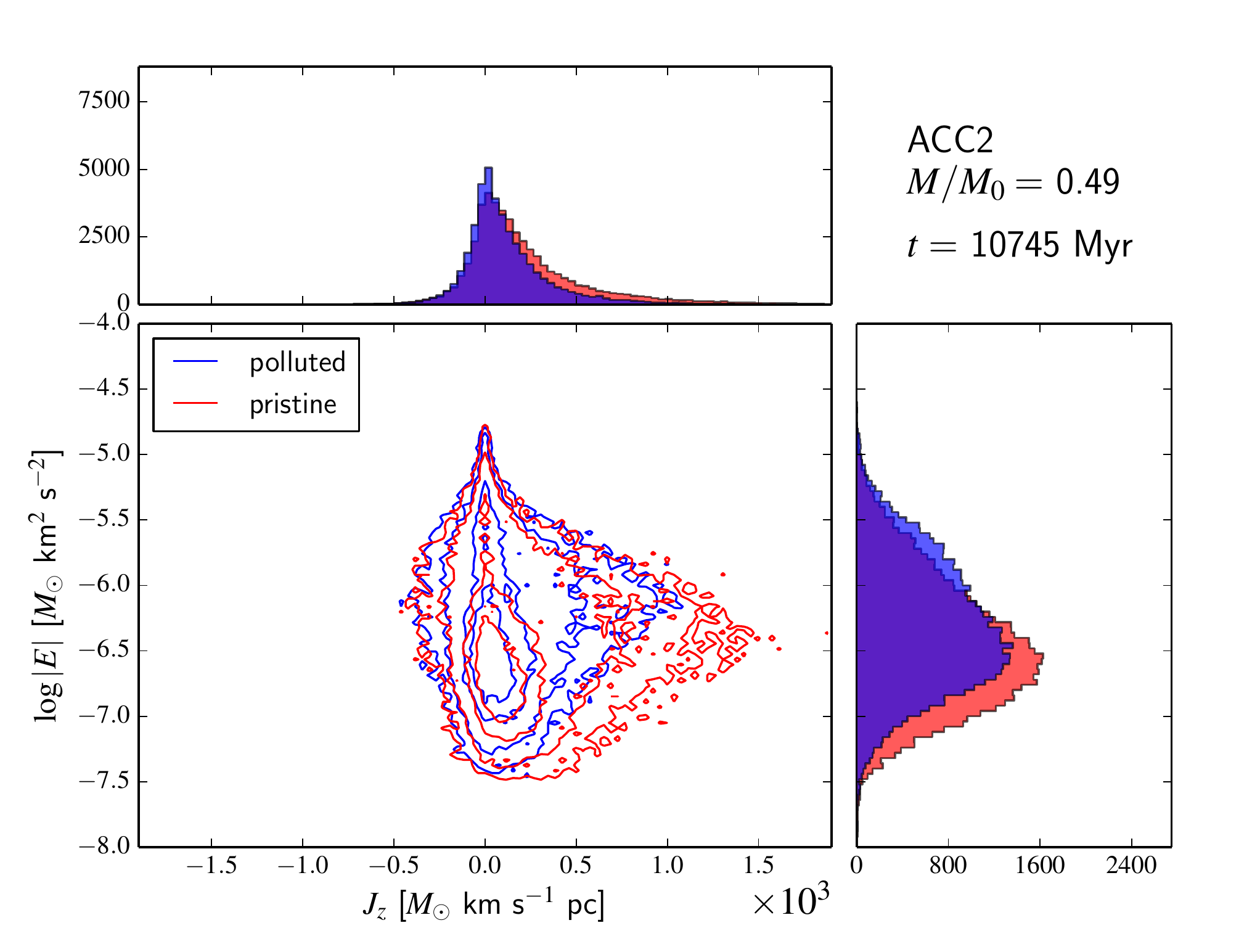}
\caption{Same as Figures \ref{EDA1_mix} and \ref{AGB1_mix} but for model {\tt ACC2}.}
\end{center}
\end{figure}

\begin{figure}
\begin{center}
\includegraphics[width=\columnwidth]{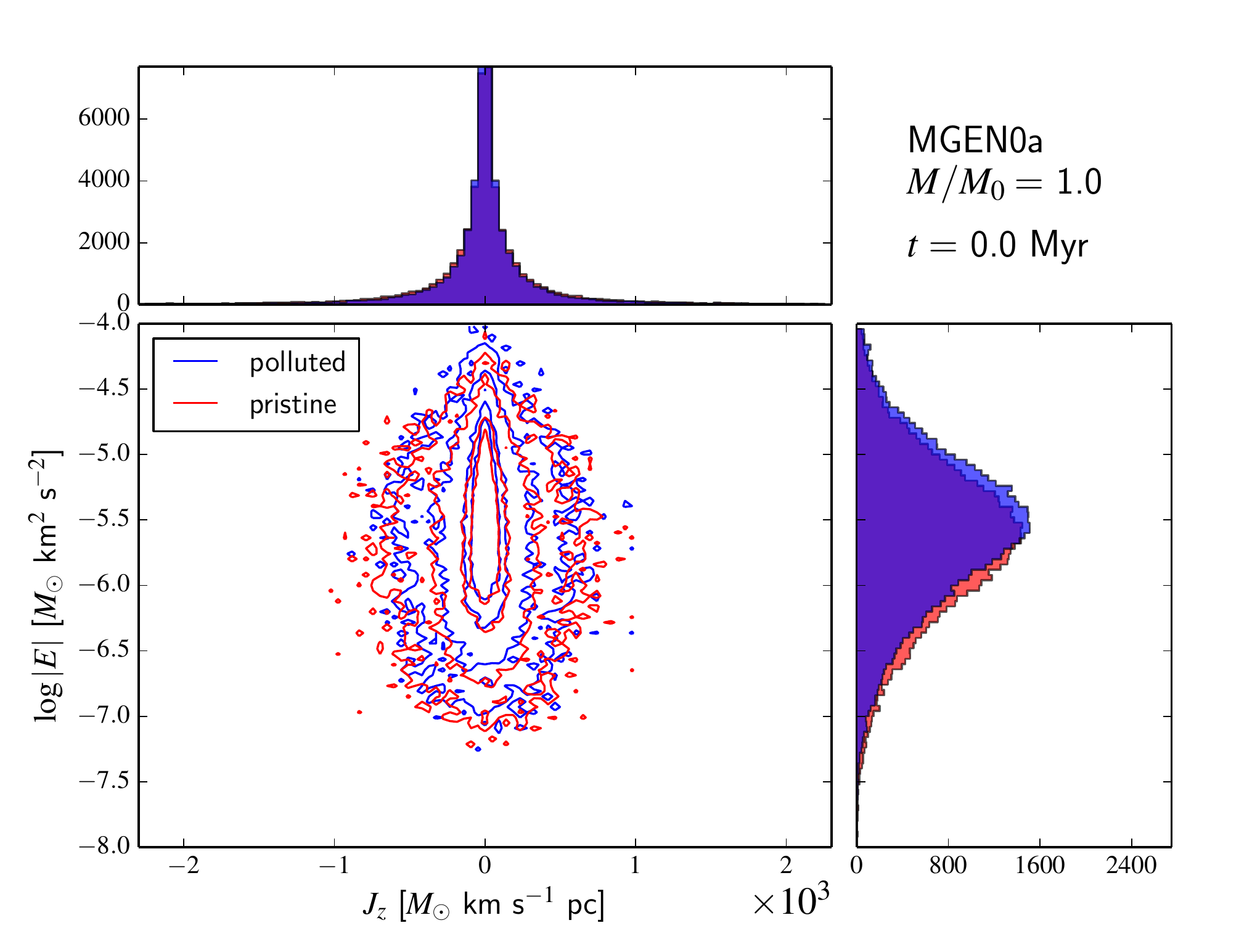}
\includegraphics[width=\columnwidth]{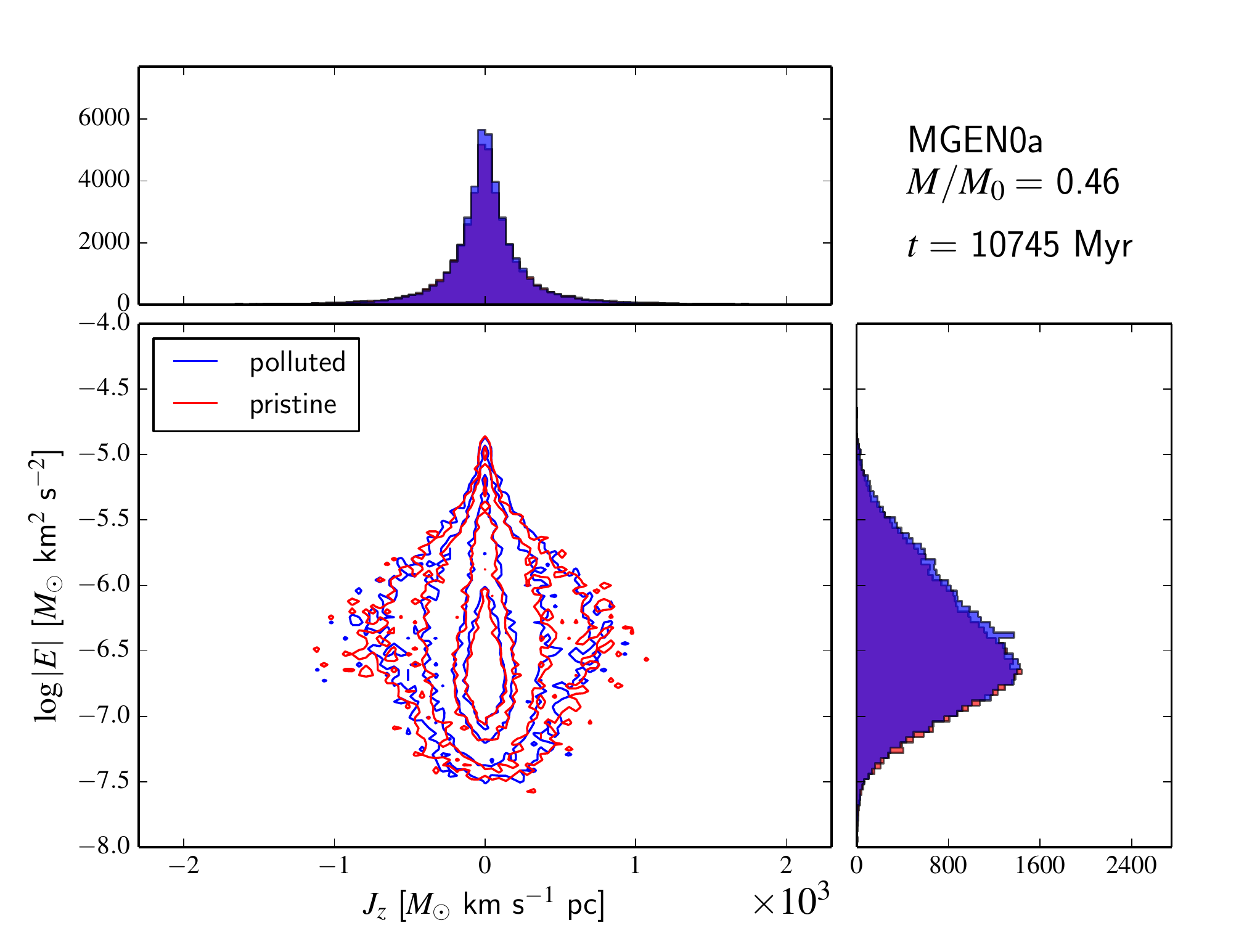}
\caption{Same as Figures \ref{EDA1_mix} and \ref{AGB1_mix} but for model {\tt MGEN0a}.}
\end{center}
\end{figure}

\begin{figure}
\begin{center}
\includegraphics[width=\columnwidth]{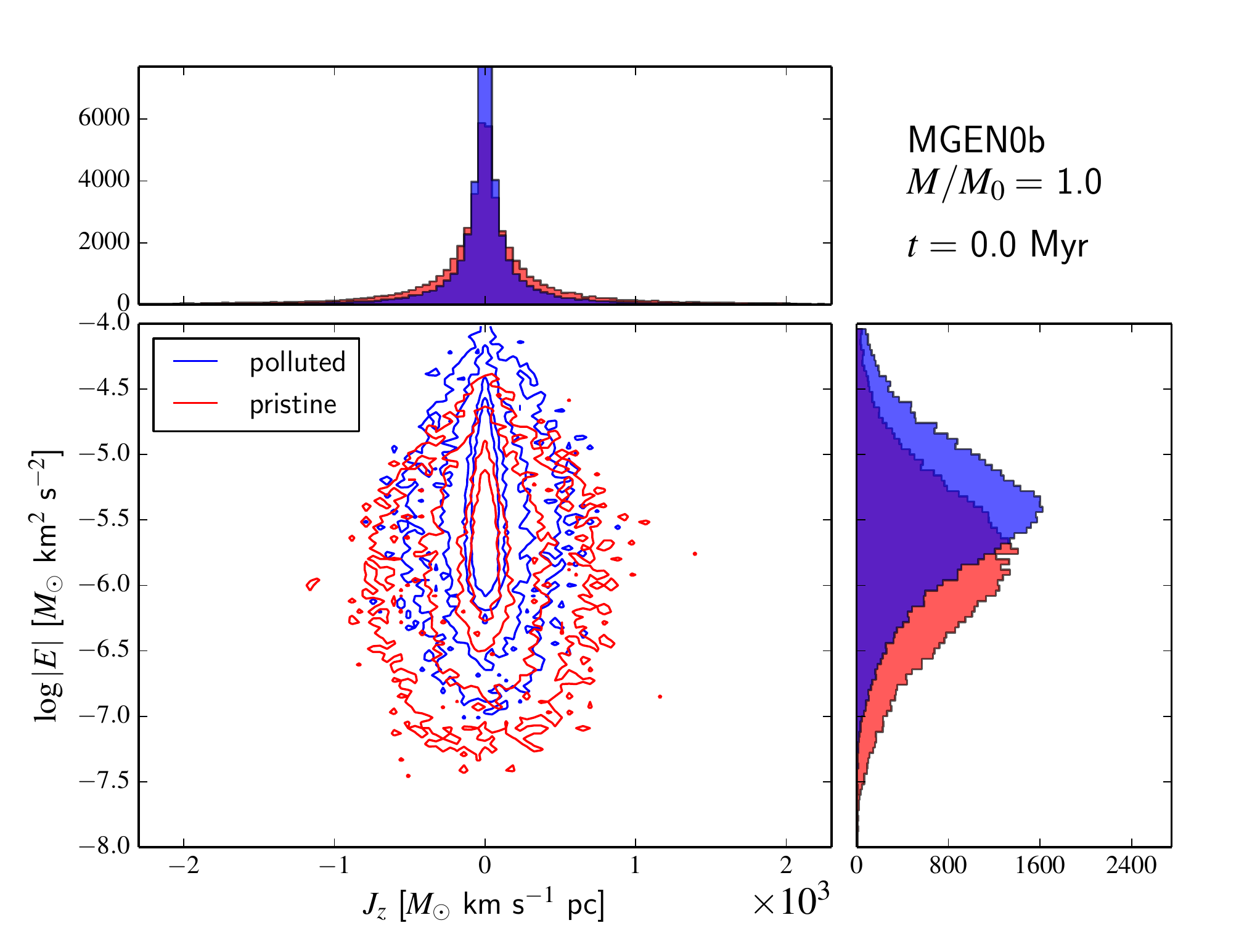}
\includegraphics[width=\columnwidth]{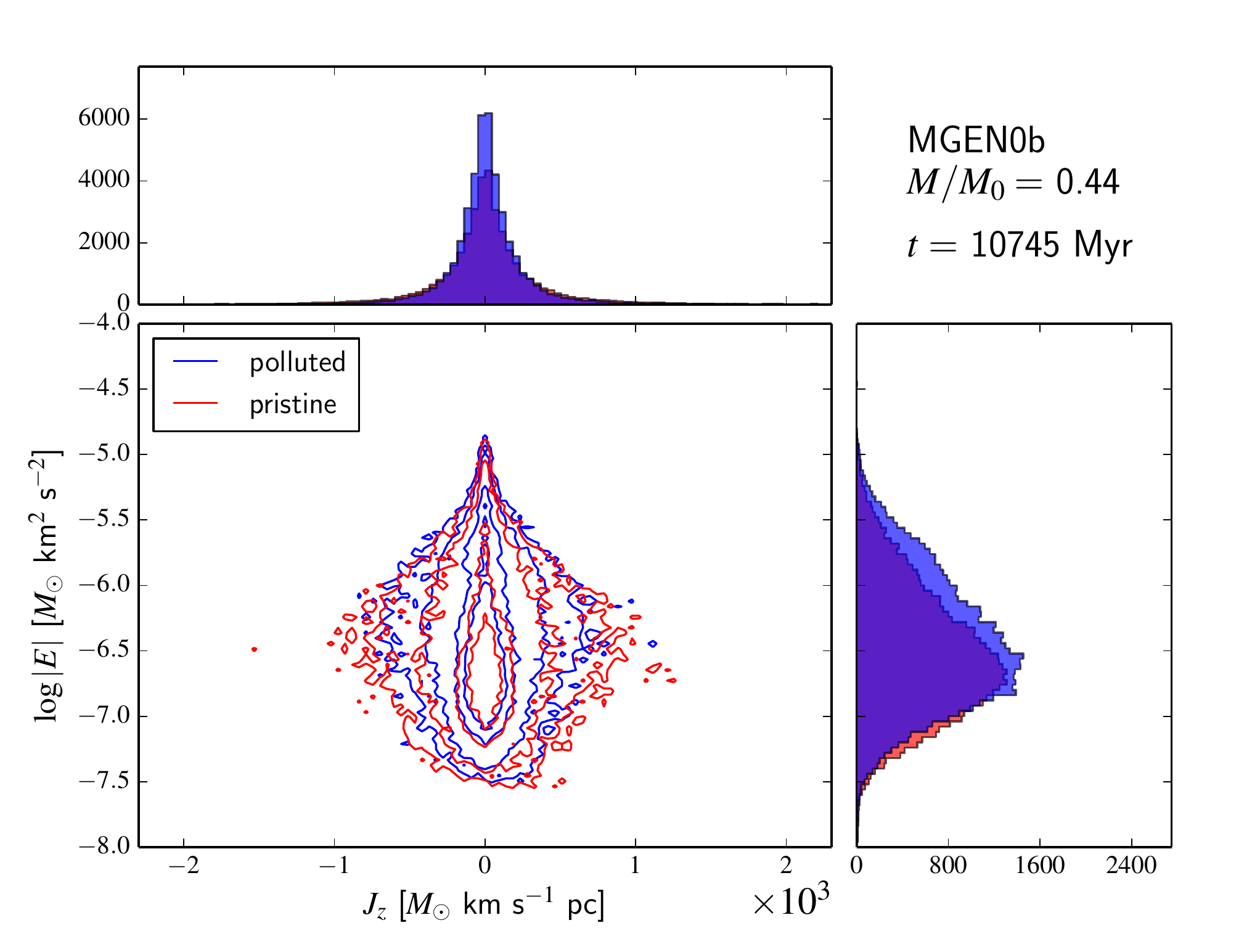}
\caption{Same as Figures \ref{EDA1_mix} and \ref{AGB1_mix} but for model {\tt MGEN0b}.}
\end{center}
\end{figure}

\begin{figure}
\begin{center}
\includegraphics[width=\columnwidth]{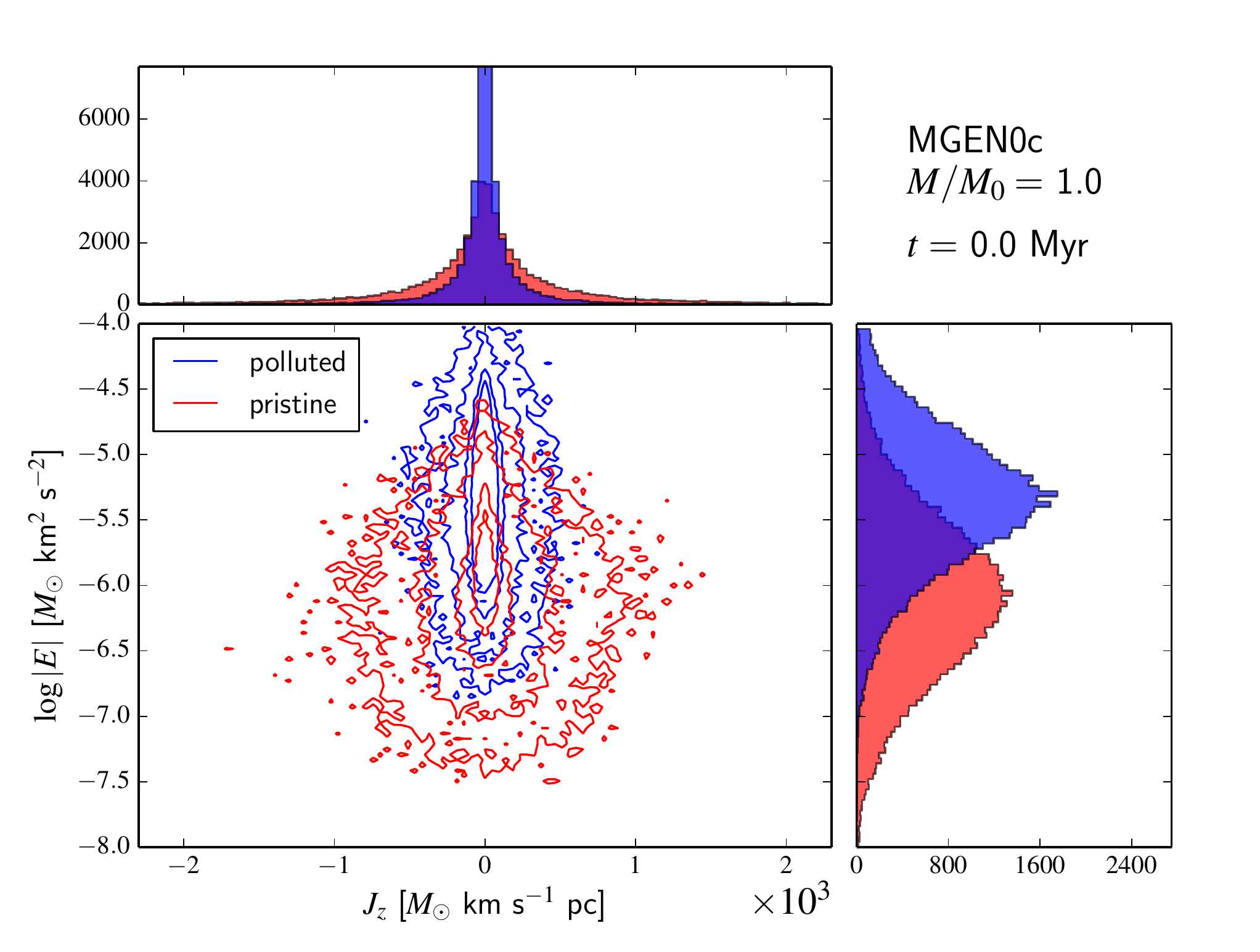}
\includegraphics[width=\columnwidth]{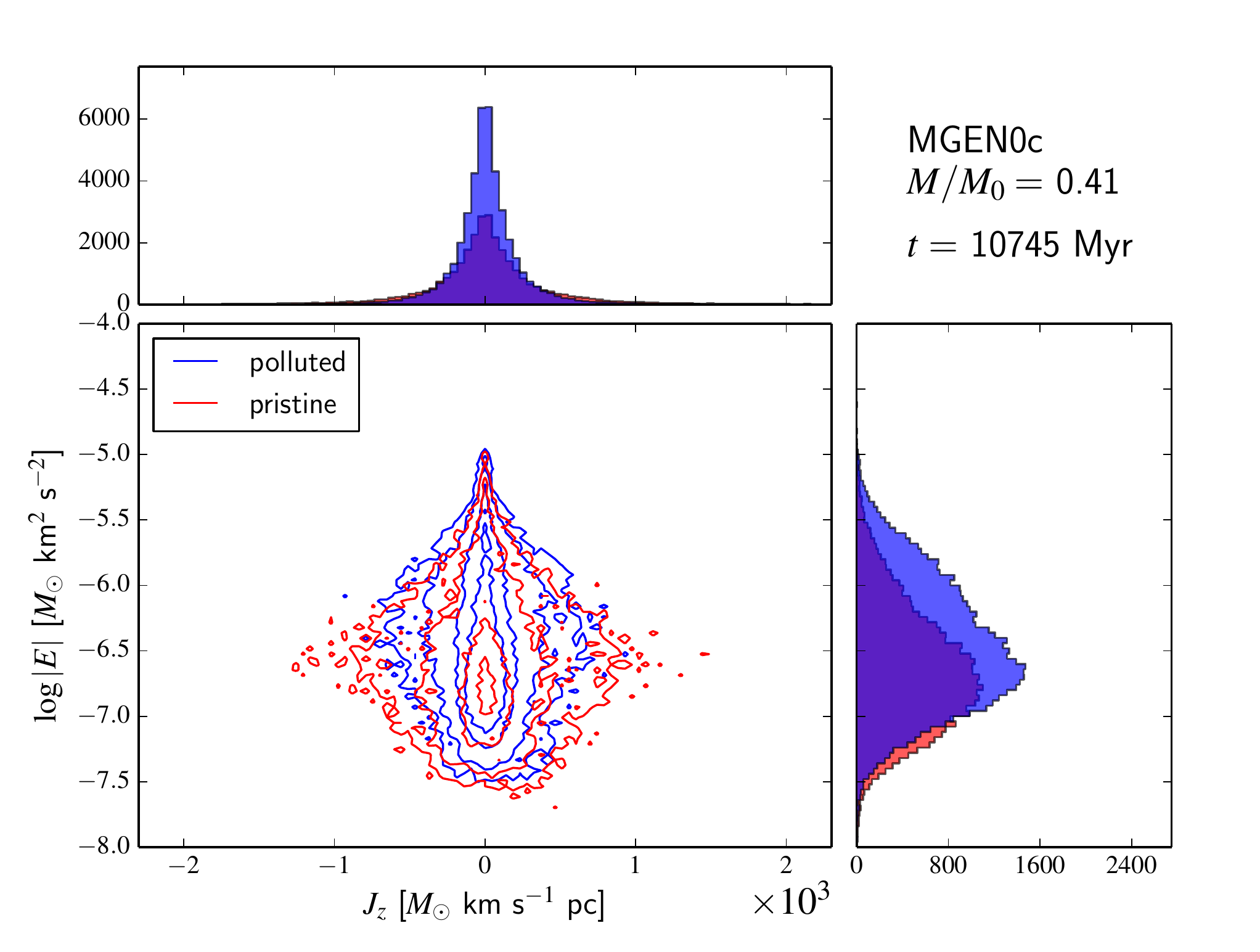}
\caption{Same as Figures \ref{EDA1_mix} and \ref{AGB1_mix} but for model {\tt MGEN0c}.}
\end{center}
\end{figure}

\begin{figure}
\begin{center}
\includegraphics[width=\columnwidth]{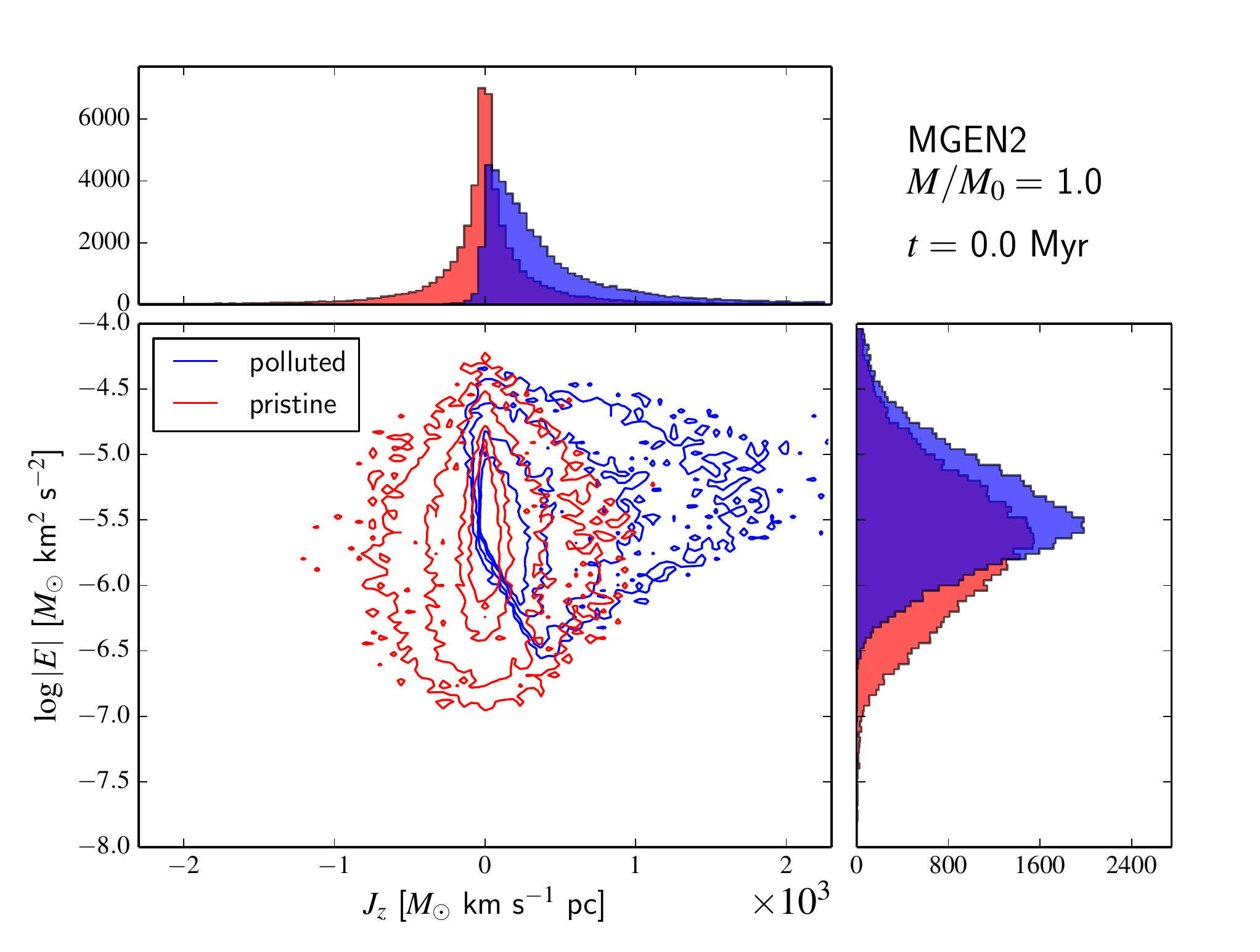}
\includegraphics[width=\columnwidth]{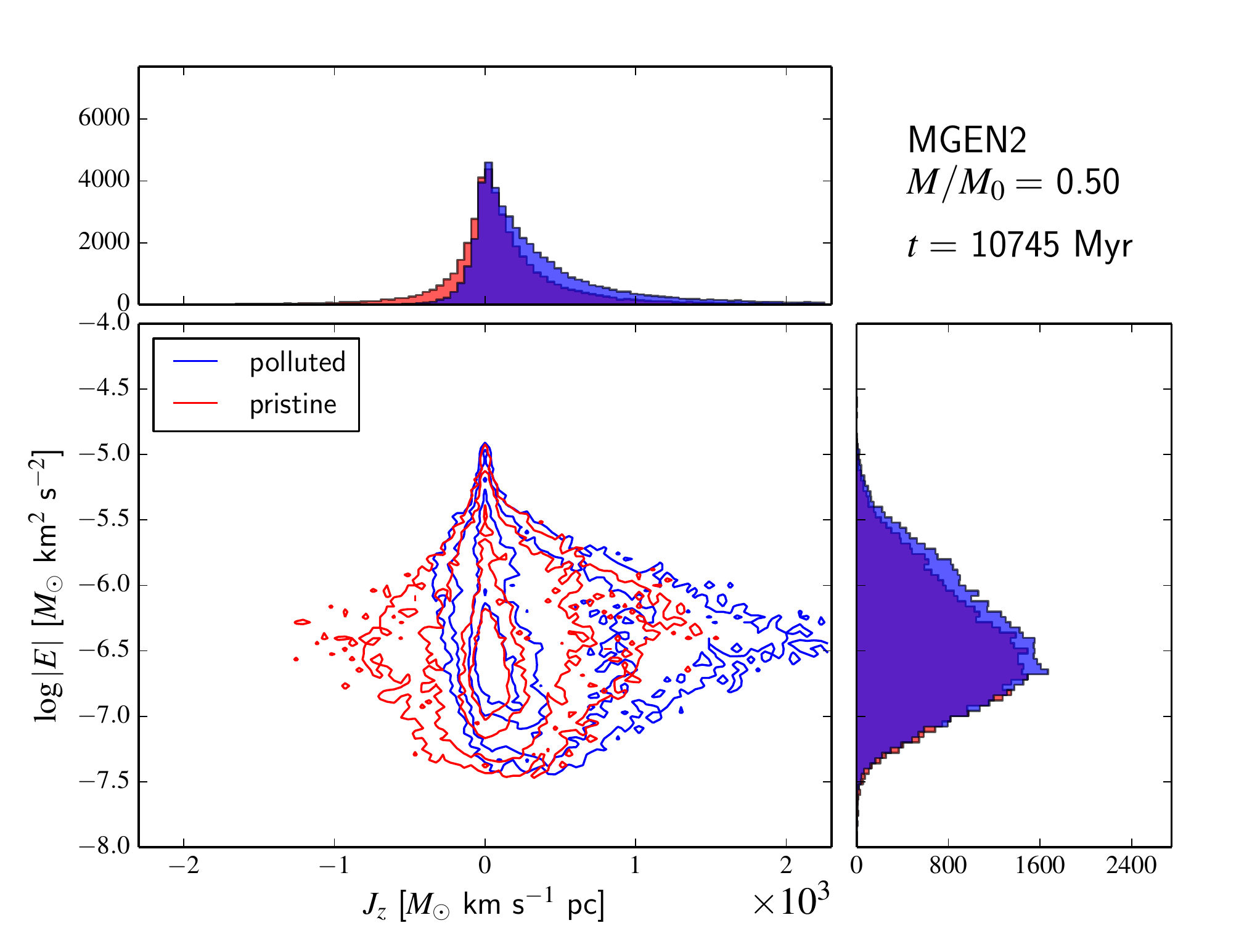}
\caption{Same as Figures \ref{EDA1_mix} and \ref{AGB1_mix} but for model {\tt MGEN2}.}
\end{center}
\end{figure}

\clearpage

\section{Supplementary figures (velocity dispersion)}
\label{appendix_sigma}

This section presents four additional figures to complement \S \ref{sigma_section}, Figure~\ref{sigmaz_evol} and Figure~\ref{sigmax_evol}. These illustrate the evolution of the $x$ and $z$ velocity dispersion profiles for models without net angular momentum initially ({\tt MGEN0c} and {\tt ACC0}; Figure~\ref{sigmaz_evol_0} and Figure~\ref{sigmax_evol_0}), and for models with a larger initial amount of angular momentum ({\tt MGEN2} and {\tt ACC2}; Figure~\ref{sigmaz_evol_2} and Figure~\ref{sigmax_evol_2}).

\begin{figure*}
\begin{center}
\includegraphics[width=6.0in]{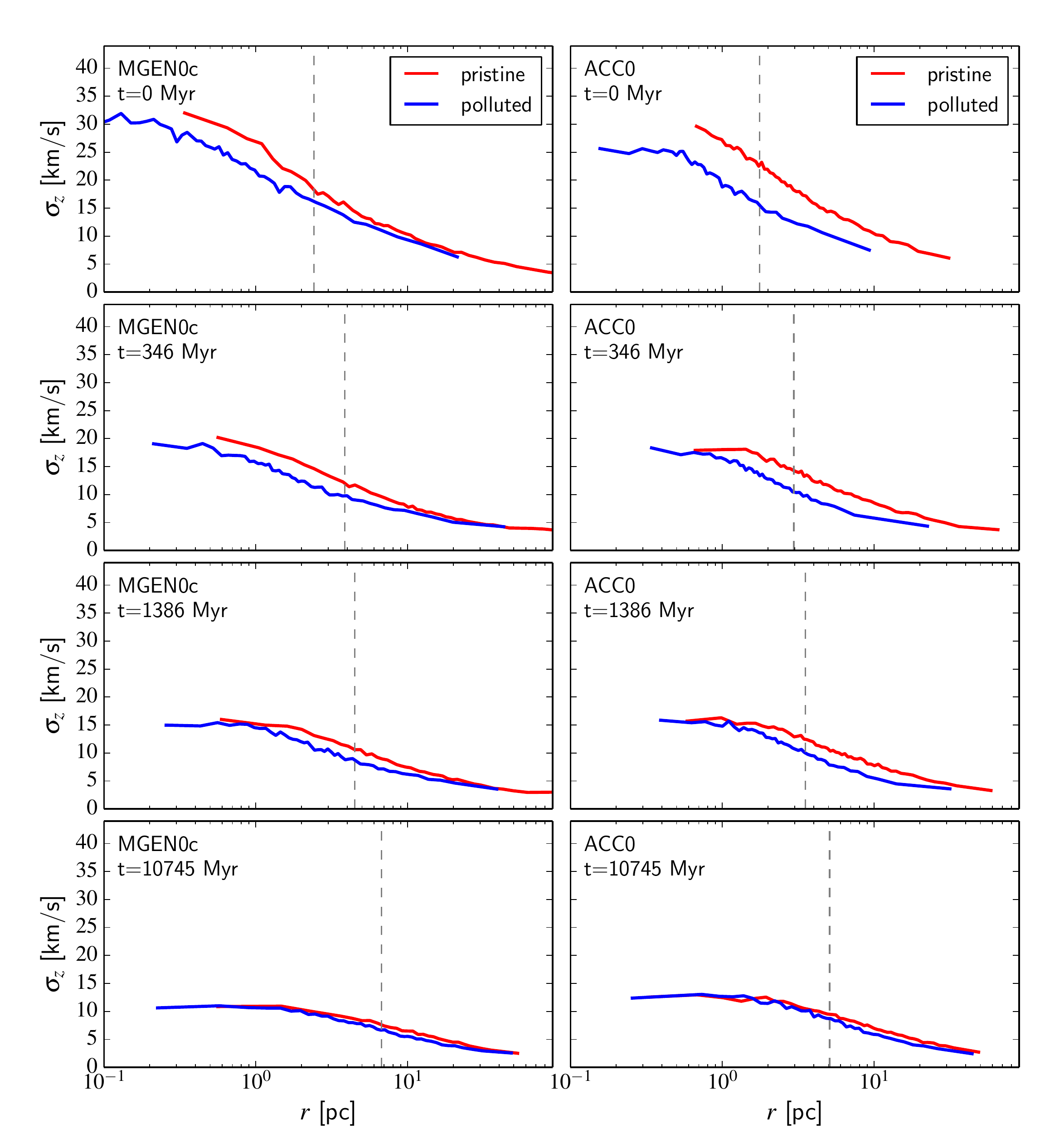}
\caption{\label{sigmaz_evol_0} Same as Figure~\ref{sigmaz_evol} but for models {\tt MGEN0c} and {\tt ACC0}.}
\end{center}
\end{figure*}

\begin{figure*}
\begin{center}
\includegraphics[width=6.0in]{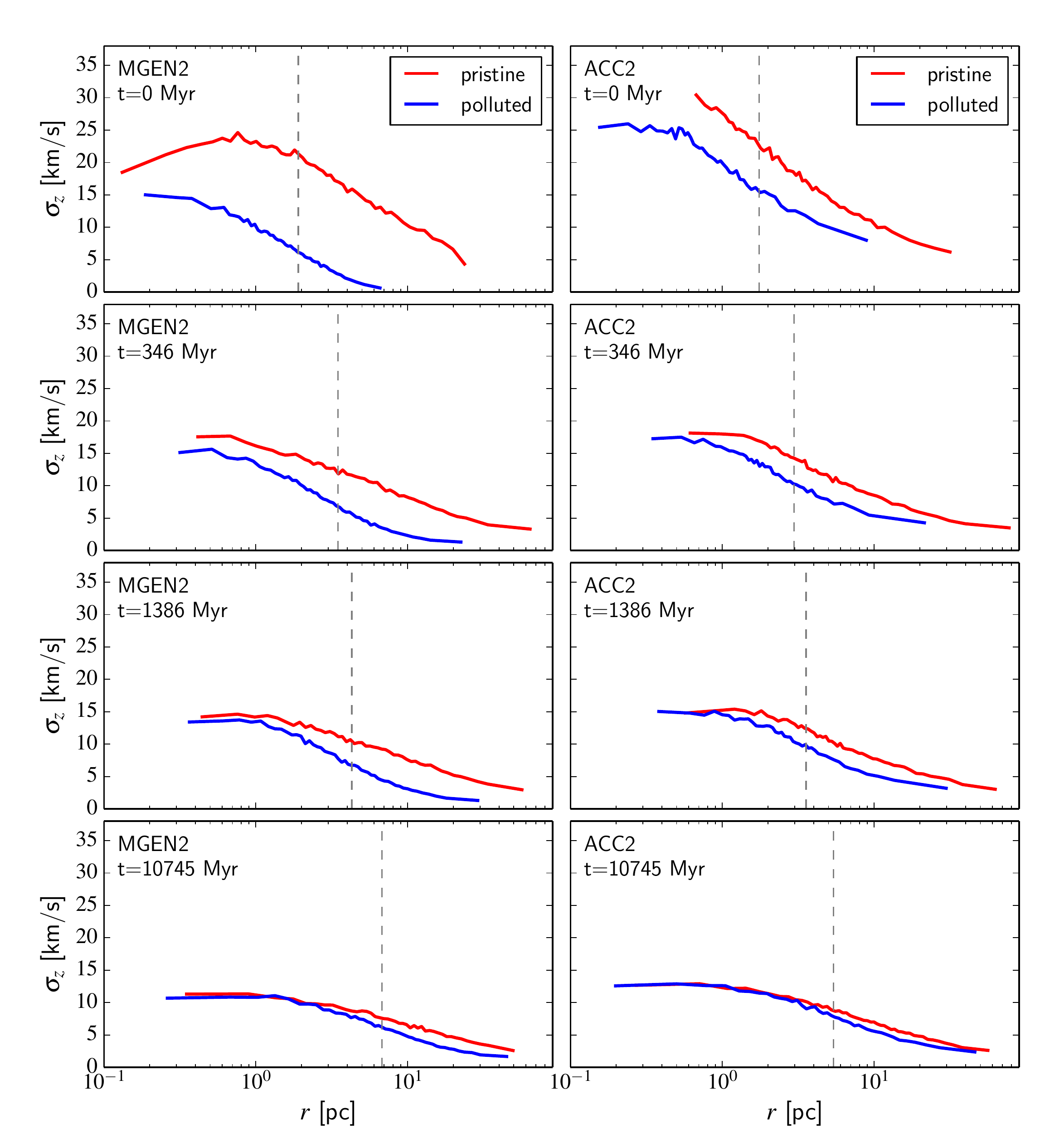}
\caption{\label{sigmaz_evol_2} Same as Figure~\ref{sigmaz_evol} but for models {\tt MGEN2} and {\tt ACC2}.}
\end{center}
\end{figure*}

\begin{figure*}
\begin{center}
\includegraphics[width=6.0in]{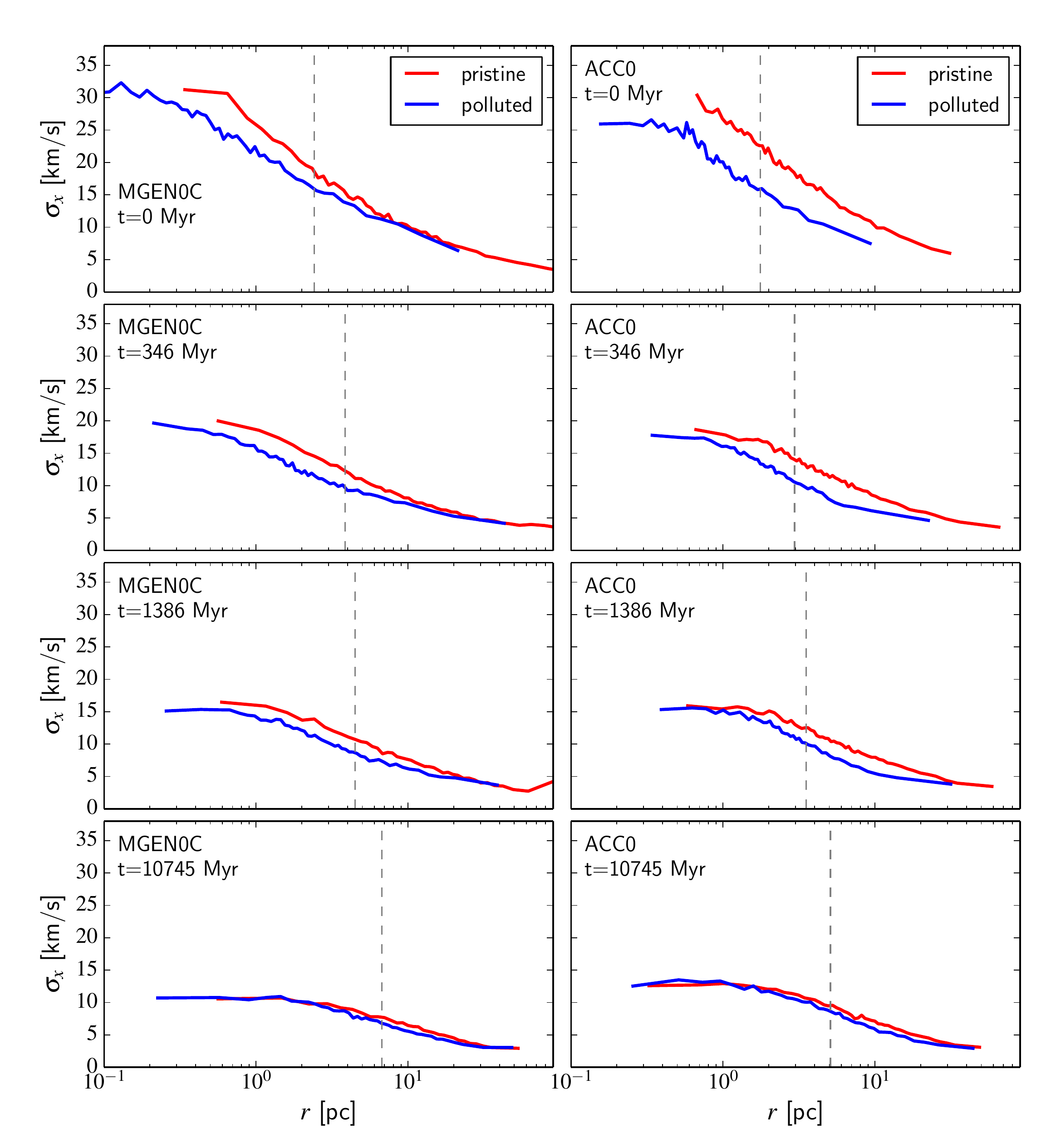}
\caption{\label{sigmax_evol_0} Same as Figure~\ref{sigmax_evol} but for models {\tt MGEN0c} and {\tt ACC0}.}
\end{center}
\end{figure*}

\begin{figure*}
\begin{center}
\includegraphics[width=6.0in]{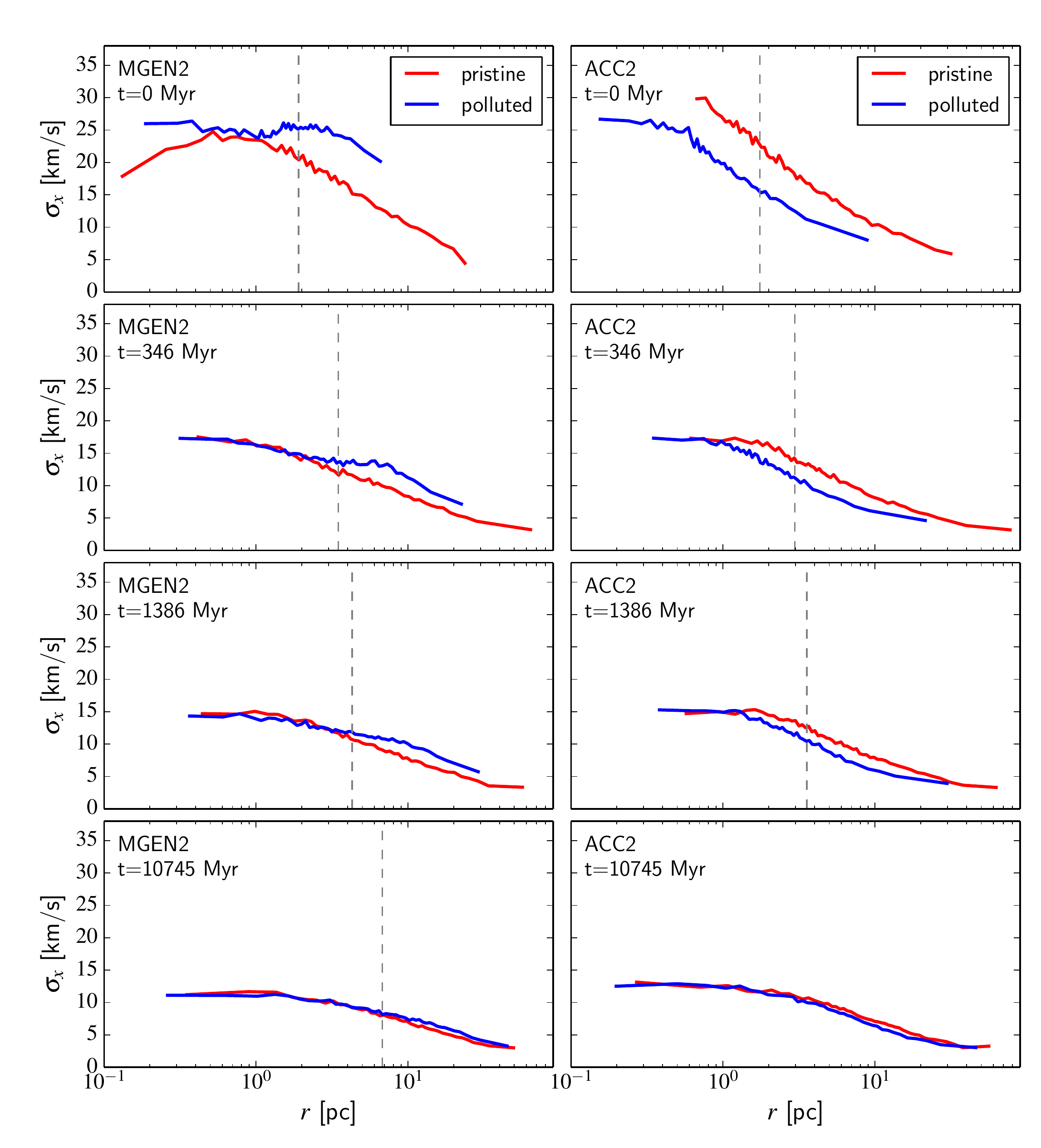}
\caption{\label{sigmax_evol_2} Same as Figure~\ref{sigmax_evol} but for models {\tt MGEN2} and {\tt ACC2}.}
\end{center}
\end{figure*}

\clearpage

\section{Supplementary figures (rotation)}
\label{appendix_rotation}

This section presents two additional figures to complement \S \ref{rot_section} and Figure~\ref{vphi_evol_1} and illustrate the evolution of the rotation curve for models without net angular momentum initially ({\tt MGEN0c} and {\tt ACC0}; Figure~\ref{vphi_evol_0}), and for models with a larger initial amount of angular momentum ({\tt MGEN2} and {\tt ACC2}; Figure~\ref{vphi_evol_2}).

\clearpage

\begin{figure*}
\begin{center}
\includegraphics[width=6.0in]{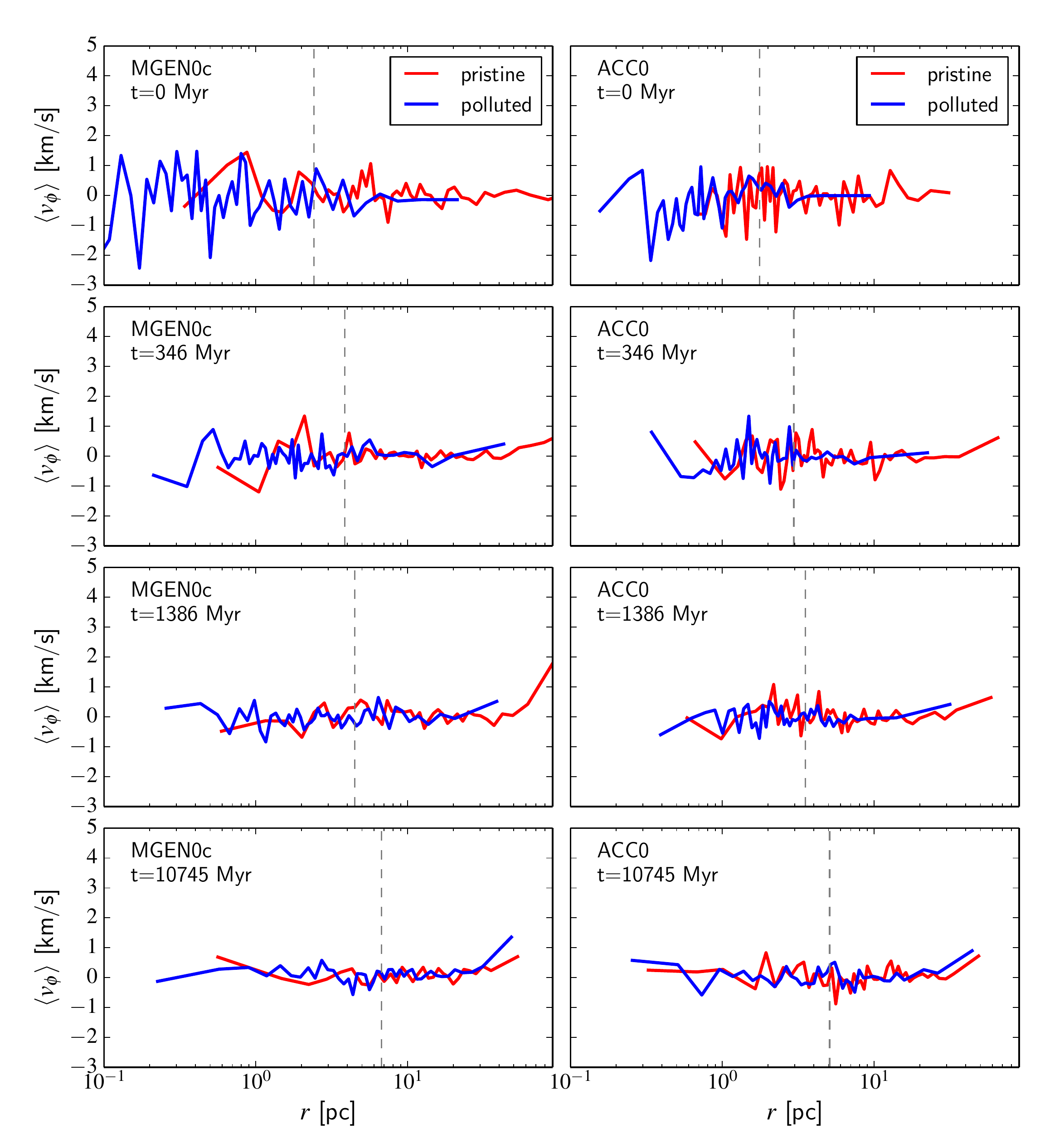}
\caption{Same as Figure~\ref{vphi_evol_1} but for models {\tt MGEN0c} and {\tt ACC0}.}  \label{vphi_evol_0}
\end{center}
\end{figure*}

\begin{figure*}
\begin{center}
\includegraphics[width=6.0in]{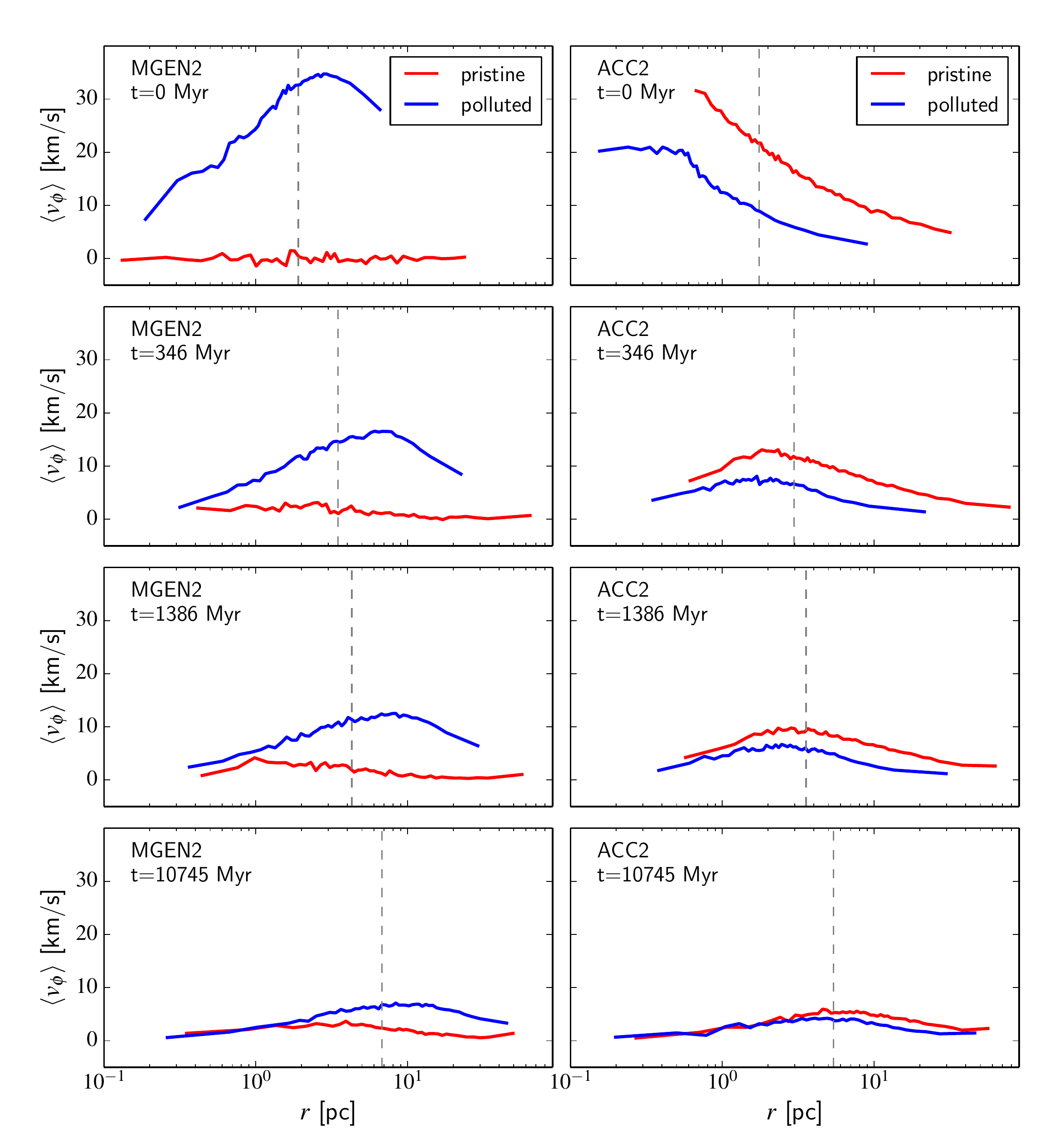}
\caption{Same as Figure~\ref{vphi_evol_1} but for models {\tt MGEN2} and {\tt ACC2}.} \label{vphi_evol_2}
\end{center}
\end{figure*}

\end{document}